\documentclass[aps,preprintnumbers,showpacs,twocolumn,superscriptaddress,floatfix,nofootinbib]{revtex4}
\usepackage{amsmath}
\usepackage{amssymb}
\usepackage{color}
\usepackage{dcolumn}
\usepackage{graphicx}
\usepackage{epstopdf}           
\usepackage{latexsym}
\usepackage{times}
\usepackage{ifpdf}

\input{epsf.sty}

\newcommand{\tg}{\tilde\gamma}
\newcommand{\tG}{\tilde\Gamma}
\newcommand{\tA}{\tilde A}

\newcommand{\dt}{(\partial_t - {\cal L}_\beta)\;}
\DeclareMathOperator{\tr}{\ensuremath{\mathrm{tr}}}

\newcommand{\bea}{\begin{eqnarray}}
\newcommand{\eea}{\end{eqnarray}}
\newcommand{\beq}{\begin{equation}}
\newcommand{\eeq}{\end{equation}}

\newcommand{\cf}{\textit{cf.}}
\newcommand{\ie}{\textit{i.e.}~}
\newcommand{\eg}{\textit{e.g.}~}
\newcommand{\ms}{{\rm ms}}
\newcommand{\km}{{\rm km}}
\newcommand{\hz}{{\rm Hz}}
\newcommand{\khz}{{\rm kHz}}
\newcommand{\mpc}{{\rm Mpc}}

\begin{document}

\title{Accurate evolutions of inspiralling neutron-star binaries:
  prompt and delayed collapse to black hole}

\author{Luca Baiotti}
\affiliation{
Graduate School of Arts and Sciences, University of Tokyo,
Komaba, Meguro-ku, Tokyo, 153-8902, Japan 
}
\affiliation{
  Max-Planck-Institut f\"ur Gravitationsphysik,
  Albert-Einstein-Institut,
  Potsdam-Golm, Germany
}

\author{Bruno Giacomazzo}
\affiliation{
  Max-Planck-Institut f\"ur Gravitationsphysik,
  Albert-Einstein-Institut,
  Potsdam-Golm, Germany
}

\author{Luciano Rezzolla}
\affiliation{
  Max-Planck-Institut f\"ur Gravitationsphysik,
  Albert-Einstein-Institut,
  Potsdam-Golm, Germany
}
\affiliation{
  Department of Physics and Astronomy,
  Louisiana State University,
  Baton Rouge, LA, USA
}
\affiliation{
  INFN, Department of Physics, University of Trieste, Trieste, Italy
}

\date{\today}


\begin{abstract}
  Binary neutron-star systems represent primary sources for the
  gravitational-wave detectors that are presently operating or are
  close to being operating at the target sensitivities. We present a
  systematic investigation in full general relativity of the dynamics
  and gravitational-wave emission from binary neutron stars which
  inspiral and merge, producing a black hole surrounded by a
  torus. Our results represent the state of the art from several
  points of view: \textit{(i)} We use high-resolution shock-capturing
  methods for the solution of the hydrodynamics equations and
  high-order finite-differencing techniques for the solution of the
  Einstein equations; \textit{(ii)} We employ adaptive mesh-refinement
  techniques with ``moving boxes'' that provide high-resolution around
  the orbiting stars; \textit{(iii)} We use as initial data accurate
  solutions of the Einstein equations for a system of binary neutron
  stars in irrotational quasi-circular orbits; \textit{(iv)} We
  exploit the isolated-horizon formalism to measure the properties of
  the black holes produced in the merger; \textit{(v)} Finally, we use
  two approaches, based either on gauge-invariant perturbations or on
  Weyl scalars, to calculate the gravitational waves emitted by the
  system. Within our idealized treatment of the matter, these
  techniques allow us to perform accurate evolutions on timescales
  never reported before (\ie $\sim 30\,\ms$) and to provide the first
  complete description of the inspiral and merger of a neutron-star
  binary leading to the \textit{prompt} or \textit{delayed} formation
  of a black hole and to its ringdown. We consider either a polytropic
  equation of state or that of an ideal fluid and show that already
  with this idealized treatment a very interesting phenomenology can
  be described. In particular, we show that while higher-mass polytropic
  binaries lead to the \textit{prompt} formation of a rapidly rotating
  black hole surrounded by a dense torus, lower-mass binaries give
  rise to a differentially rotating star, which undergoes large
  oscillations and emits large amounts of gravitational
  radiation. Eventually, also the hypermassive neutron star collapses
  to a rotating black hole surrounded by a torus. Finally, we also
  show that the use of a non-isentropic equation of state leads to
  significantly different evolutions, giving rise to a
  \textit{delayed} collapse also with high-mass binaries, as well as
  to a more intense emission of gravitational waves and to a
  geometrically thicker torus. 
\end{abstract}

\maketitle

\section{Introduction} 
\label{sec:intro}

Little is required to justify the efforts in the study of binary
systems. Despite the simplicity of its formulation, the relativistic
two-body problem is, in fact, one of the most challenging problems in
classical general relativity. Furthermore, binary systems of compact
objects are considered one of the most important sources for
gravitational-wave emission and are thought to be at the origin of
some of the most violent events in the Universe. While some of the
numerical difficulties involved in the simulations of such highly
dynamical systems have been overcome in the case of binary black holes
(BHs), numerical simulations of binary neutron stars (NSs) in general
relativity have so far provided only rudimentary descriptions of the
complex physics accompanying the inspiral and merger. Simulations of
this type are the focus of this paper.

Binary NSs are known to exist and for some of the systems in our own
galaxy general-relativistic effects in the binary orbit have been
measured to high precision~\cite{Weisberg89,Stairs98,Kramer04}. The
inspiral and merger of two NSs in binary orbit is the inevitable fate
of close-binary evolution, whose main dissipation mechanism is the
emission of gravitational radiation. An important part of the interest
in the study of coalescing systems of compact objects comes from the
richness of general-relativistic effects that accompany these
processes and, most importantly, from the gravitational-wave
emission. Detection of gravitational waves from NS binaries
will indeed provide a wide variety of physical information on the component
stars, including their mass, spin, radius and equations of state
(EOS)~\cite{Oechslin07a,Oechslin07b}.
Furthermore, NS binary systems are expected to produce signals of amplitude 
large enough to be relevant for Earth-based
gravitational-wave detectors and to be sufficiently frequent sources to
be detectable over the timescale in which the detectors are
operative. Recent improved extrapolations to the local group of the
estimated galactic coalescence rates predict $1$ event per $3-10$
years for the first-generation of interferometric detectors and of
$10-500$ events per year, for the generation of advanced
detectors~\cite{Kalogera04}.

There are three possible characteristic gravitational-wave frequencies
related to the inspiral and merger of binary systems. The first one is
the frequency of the orbital motion of the stars in the last stages of
the inspiral, before tidal distortions become important. The second
characteristic frequency is associated with the fundamental
oscillation modes of the merged massive object formed after the onset
of the merger. Numerical simulations in the frameworks of
Newtonian~\cite{Rasio92}, post-Newtonian (PN)~\cite{Ruffert96b},
semi-relativistic~\cite{Oechslin02} and fully general-relativistic
gravity~\cite{Shibata01a} have shown that, if a BH is not produced
promptly, the frequency of the fundamental oscillation modes of the
merged object is between $2$ and $3\,\khz$, depending on the EOS and
on the initial compactness of the progenitor NSs. Finally, the third
frequency is that of the quasi-normal modes (QNMs) of the BH, which is
eventually formed after the merger.

The study of NS binary systems goes beyond the impact it has on
gravitational-wave astronomy and is also finalized to the
understanding of the origin of some type of $\gamma$-ray bursts
(GRBs), whose short rise times suggest that their central sources have
to be highly relativistic objects~\cite{Piran99}. After the
observational confirmation that GRBs have a cosmological origin, it
has been estimated that the central sources powering these bursts must
provide a large amount of energy ($\sim 10^{51}$ ergs) in a very short
timescale, going from one millisecond to one second (at least for a
subclass of them, called ``short'' GRBs). It has been suggested that
the merger of NS binaries could be a likely candidate for the powerful
central source of a subclass of short GRBs. The typical scenario is
based on the assumption that a system composed of a rotating BH and a
surrounding massive torus is formed after the merger. If the disc had
a mass $\gtrsim 0.1\,M_{\odot}$, it could supply the large amount of
energy by neutrino processes or by extracting the rotational energy of
the BH.

The understanding of GRBs is therefore an additional motivation to
investigate the final fate of binaries after the merger. The total
gravitational masses of the known galactic NS binary systems are in a
narrow range $\sim 2.65-2.85\,M_{\odot}$ and the present observational
evidence indicates that the masses of the two stars are nearly
equal. If this is the general situation, NSs in binary systems will
not be tidally disrupted before the merger. As a result, the mass loss
from the binary systems is expected to be small during the evolution
and the mass of the merged object will be approximately equal to the
initial mass of the binary system. Since the maximum allowed
gravitational mass for spherical NSs is in the range $\sim
1.5-2.3\,M_{\odot}$, depending on the EOS, the compact objects formed
just after the merger of these binary systems are expected to collapse
to a BH, either promptly after the merger or after a certain
``delay''. Indeed, if the merged object rotates differentially, the
final collapse may be prevented on a timescale over which dissipative
effects like viscosity, magnetic fields or gravitational-wave emission
bring the star towards a configuration which is unstable to
gravitational collapse. During this process, if the merged object has
a sufficiently high ratio of rotational energy to the gravitational
binding energy, it could also become dynamically unstable to nonlinear
instabilities, such as the bar-mode
instability~\cite{Baiotti06b,Manca07}. It is quite clear, therefore,
that while the asymptotic end state of a binary NS system is a
rotating BH, the properties of the intermediate product of the merger
are still pretty much an open question, depending not only on the
nuclear EOS for high-density neutron matter, but also on the
rotational profile of the merged object and on the physical processes
through which the object can lose angular momentum and energy.

Several different approaches have been developed over the years to
tackle the binary NS problem. One of these approaches attempts to
estimate the properties of the binary evolution by considering
sequences of quasi-equilibrium configurations, that is by neglecting
both gravitational waves and wave-induced deviations from a circular
orbit; this is expected to be a very good approximation if the stars
are well
separated~\cite{Lombardi97,Baumgarte98b,Uryu98,Bonazzola97,Bonazzola98b,
  Bonazzola99a,Gourgoulhon01,Uryu00, Marronetti99,Taniguchi01,
  Taniguchi02a,Taniguchi02b,Taniguchi03}. Other approaches have tried
to simplify some aspects of the coalescence, by solving, for instance,
the Newtonian or PN version of the hydrodynamics equations
(see~\cite{Nakamura99a,Rasio92,New97,Zhuge94,Zhuge96,Ruffert96a,Ruffert96b,Ruffert97,Ruffert01}
and references therein). At the same time, alternative treatments of
the gravitational fields, such as the conformally-flat approximation,
have been developed and coupled to the solution of the relativistic
hydrodynamical equations~\cite{Wilson95,Mathews98}, either in the
fluid approximation or in its smooth-particle hydrodynamics (SPH)
variant~\cite{Oechslin06}. Special attention has also been paid to the
role played in these calculations by the EOS and progress has been
made recently with SPH calculations~\cite{Oechslin07a,Oechslin07b}.

While all of the above-mentioned works have provided insight into the
coalescence process and some of them represent the state of the art
for their realistic treatment of the matter
properties~\cite{Oechslin07a,Oechslin07b}, they are nevertheless
only approximations to the full general-relativistic solution. The
latter is however required for quantitatively reliable coalescence
waveforms and to determine those qualitative features of the final
merger which can only result from strong-field effects.

Several groups have launched efforts to solve the equations of
relativistic hydrodynamics together with the Einstein equations and to
model the coalescence and merger of NS
binaries~\cite{Oohara97,Baumgarte99d,Font00,Font02c}. The first
successful simulations of binary NS mergers were those of Shibata and
Uryu \cite{Shibata99a,Shibata99d,Shibata02a}. Later on, Shibata and
Taniguchi have extended their numerical studies to unequal-mass
binaries providing a detailed and accurate discussion of simulations
performed with realistic EOSs (see \cite{Shibata06a} and references
therein). More recently, Anderson {\it et al.} ~\cite{Anderson2007}
have made an important technical progress by presenting results of
binary NS evolutions using an adaptive-mesh-refinement (AMR)
code. However, despite the high resolution available with the use of
AMR, no waveforms from the BH formation were reported in
ref.~\cite{Anderson2007} over the timescales discussed for the
evolutions. Finally, a set of simulations involving (among other
compact binaries) also binary neutron stars have been recently
presented by Yamamoto {\it et al.} using AMR techniques with moving
boxes~\cite{Yamamoto2008}.

An aspect common to all the above-mentioned simulations is that, while
they represent an enormous progress with respect to what was possible
to calculate only a few years ago, they provide a description of the
dynamics which is limited to a few $\ms$ after the merger.  The work
presented here aims at pushing this limit further and to provide a
systematic investigation of the inspiral, but also of the merger and
of the (possibly) delayed collapse to a BH. Despite the fact that we
do not account for the transport of radiation or
neutrinos, our results benefit from a number of technical
advantages (some of which are also shared by other groups):
\textit{(i)} The use of high-resolution shock-capturing methods for
the solution of the hydrodynamics equations and high-order (\ie fourth
order in space and third order in time) finite-differencing techniques
for the solution of the Einstein equations; \textit{(ii)} The use of
adaptive mesh-refinement techniques that provide higher resolution
around the orbiting stars; \textit{(iii)} The use of consistent
initial data representing a system of binary NSs in irrotational
quasi-circular orbits; \textit{(iv)} The use of the isolated-horizon
formalism to measure the properties of the BHs produced in the merger;
\textit{(v)} The use of two complementary approaches for the
extraction of the gravitational waves produced. Most importantly,
however, our simulations can rely on unprecedented evolution
timescales spanning more than $30\,\ms$. 

Exploiting these features we provide, within an idealized treatment of
the matter, the first complete description of the inspiral and merger
of a NS binary leading to the \textit{prompt} or \textit{delayed}
formation of a BH and to its ringdown. While our treatment of the
matter is simplified with the use of analytic EOSs, we show that this
does not prevent us from reproducing some of the most salient aspects
that a more realistic EOSs would yield. In particular, we show that an
isentropic (\ie polytropic) EOS leads either to the \textit{prompt}
formation of a rapidly rotating BH surrounded by a dense torus if the
binary is sufficiently massive, or, if the binary is not very massive,
to a differentially rotating star, which undergoes oscillations,
emitting large amounts of gravitational radiation and experiencing a
\textit{delayed} collapse to BH. In addition, we show that the use of
non-isentropic (\ie ideal-fluid) EOS inevitably leads to a further
delay in the collapse to BH, as a result of the larger pressure
support provided by the temperature increase via shocks.

Our interest also goes to the small-scale hydrodynamics of the merger
and to the possibility that dynamical instabilities develop. In
particular, we show that, irrespective of the EOS used, coalescing
irrotational NSs form a vortex sheet when the outer layers of the
stars come into contact. This interface is Kelvin-Helmholtz unstable
on all wavelengths (see, \eg \cite{Rasio99} and references therein)
and, exploiting the use of AMR techniques, we provide a first
quantitative description of this instability in general-relativistic
simulations.

Special attention in this work is obviously dedicated to the analysis
of the waveforms produced and to their properties for the different
configurations. In particular, we find that the largest loss rates of
energy and angular momentum via gravitational radiation develop at the
time of the collapse to BH and during the first stages of the subsequent
ringdown. Nevertheless, the configurations which emit the highest
amount of energy and angular momentum are those with lower mass,
because they do not collapse promptly to a BH, but instead produce a
violently oscillating transient object, which produces copious
gravitational radiation while rearranging its angular-momentum
distribution. We also show that although the gravitational-wave
emission from NS binaries has spectral distributions with large powers
at high frequencies (\ie $f \gtrsim 1\,\khz$), a signal-to-noise ratio
(SNR) as large as $3$ can be estimated for a source at $10\,\mpc$ if
using the sensitivity of currently operating gravitational-wave
interferometric detectors.

Many aspects of the simulations reported here could be improved and
probably the most urgent among them is the inclusion of magnetic
fields. Recent calculations have shown that the corrections produced
by strong magnetic fields could be large and are very likely to be
present [see ref.~\cite{Price06} for Newtonian magnetohydrodynamical
  (MHD) simulations and ref.~\cite{Anderson2008,Etienne08} for a
  recent general-relativistic extension]. While we have already
developed the numerical code that would allow to perform the study of
such binaries in the ideal-MHD limit~\cite{Giacomazzo:2007ti}, our
analysis is here limited to unmagnetized NSs.

The paper is organized as follows. In Section II we first summarize
the formalism we adopt for the numerical solution of the Einstein and
of the relativistic-hydrodynamics equations; we then describe briefly
the numerical methods we implemented in the {\tt Whisky}
code~\cite{Baiotti03a,Baiotti04}, we outline our mesh-refined grid
setup, and we finally describe the quasi-equilibrium initial data we
use. In Sections III A and B we describe binaries evolved with the
polytropic EOS and having a comparatively ``high'' or ``low'' mass,
respectively. In Sections III C and D we instead discuss the dynamics
of the same initial models when evolved with the ideal-fluid EOS,
while Section III E is dedicated to our analysis of the
Kelvin-Helmholtz instability. In Sections IV A and B we characterise
the gravitational-wave emission for the case of the polytropic and
ideal-fluid EOS, respectively. Finally in Sections IV C and D we
report about the energy and angular momentum carried by the
gravitational waves and their power spectra. In the Appendix, further
comments on numerical and technical issues are discussed.

We here use a spacelike signature $(-,+,+,+)$ and a system of units in
which $c=G=M_\odot=1$ (unless explicitly shown otherwise for
convenience). Greek indices are taken to run from $0$ to $3$, Latin
indices from $1$ to $3$ and we adopt the standard convention for the
summation over repeated indices.

\section{Mathematical and Numerical Setup}
\label{sec:NumericalMethods}

\subsection{Evolution system for the fields}

We evolve a conformal-traceless ``$3+1$'' formulation of the Einstein
equations~\cite{Nakamura87, Shibata95, Baumgarte99, Alcubierre99d}, in
which the spacetime is decomposed into three-dimensional spacelike
slices, described by a metric $\gamma_{ij}$, its embedding in the full
spacetime, specified by the extrinsic curvature $K_{ij}$, and the gauge
functions $\alpha$ (lapse) and $\beta^i$ (shift) that specify
a coordinate frame (see Sect.~\ref{sec:Gauges} for details on how we
treat gauges and~\cite{York79} for a general description of the $3+1$
split). The particular
system which we evolve transforms the standard ADM variables as
follows. The three-metric $\gamma_{ij}$ is conformally transformed via
\begin{equation}
  \label{eq:def_g}
  \phi = \frac{1}{12}\ln \det \gamma_{ij}, \qquad
  \tilde{\gamma}_{ij} = e^{-4\phi} \gamma_{ij}
\end{equation}
and the conformal factor $\phi$ is evolved as an independent variable,
whereas $\tilde{\gamma}_{ij}$ is subject to the constraint
$\det \tilde{\gamma}_{ij} = 1$. The extrinsic curvature is
subjected to the same conformal transformation and its trace
$\tr K_{ij}$ is evolved as an independent variable. That is, in place of
$K_{ij}$ we evolve:
\begin{equation}
  \label{eq:def_K}
  K \equiv \tr K_{ij} = g^{ij} K_{ij}, \qquad
  \tilde{A}_{ij} = e^{-4\phi} (K_{ij} - \frac{1}{3}\gamma_{ij} K),
\end{equation}
with $\tr\tilde{A}_{ij}=0$. Finally, new evolution variables
\begin{equation}
  \label{eq:def_Gamma}
  \tilde{\Gamma}^i = \tilde{\gamma}^{jk}\tilde{\Gamma}^i_{jk}
\end{equation}
are introduced, defined in terms of the Christoffel symbols of
the conformal three-metric.

The Einstein equations specify a well-known set of evolution equations
for the listed variables and are given by
\begin{align}
  \label{eq:evolution}
  &\dt \tg_{ij} = -2 \alpha \tA_{ij}\,,  \\ \nonumber \\
  &\dt \phi = - \frac{1}{6} \alpha K\,, \\ \nonumber \\ 
  &\dt \tA_{ij} = e^{-4\phi} [ - D_i D_j \alpha 
   + \alpha (R_{ij} - 8 \pi S_{ij}) ]^{TF} \nonumber\\
   & \hskip 2.0cm + \alpha (K \tA_{ij} - 2 \tA_{ik} \tA^k{}_j), \\
  &\dt K  = - D^i D_i \alpha \nonumber \\
   & \hskip 2.0cm + \alpha \Big [\tA_{ij} \tA^{ij} + \frac{1}{3} K^2 + 
  4\pi (\rho_{_{\rm  ADM}}+S)\Big ], \\ \nonumber \\
& 
  \partial_t \tG^i  = \tilde\gamma^{jk} \partial_j\partial_k \beta^i
    + \frac{1}{3} \tilde\gamma^{ij}  \partial_j\partial_k\beta^k
    + \beta^j\partial_j \tilde\Gamma^i
   - \Tilde\Gamma^j \partial_j \beta^i \nonumber \\
   & \hskip 1.0cm
   + \frac{2}{3} \tilde\Gamma^i \partial_j\beta^j 
   - 2 \tilde{A}^{ij} \partial_j\alpha
   + 2 \alpha ( 
   \tilde{\Gamma}^i{}_{jk} \tilde{A}^{jk} + 6 \tilde{A}^{ij}
   \partial_j \phi \nonumber\\
   & \hskip 1.0cm - \frac{2}{3} \tg^{ij} \partial_j K - 8 \pi \tg^{ij} S_j),
\end{align}
where $R_{ij}$ is the three-dimensional Ricci tensor, $D_i$ the
covariant derivative associated with the three metric $\gamma_{ij}$,
``TF'' indicates the trace-free part of tensor objects and $
{\rho}_{_{\rm ADM}}$, $S_j$ and $S_{ij}$ are the matter source terms
defined as
\begin{align}
\rho_{_{\rm ADM}}&\equiv n_\alpha n_\beta T^{\alpha\beta}, \nonumber \\ 
S_i&\equiv -\gamma_{i\alpha}n_{\beta}T^{\alpha\beta}, \\
S_{ij}&\equiv \gamma_{i\alpha}\gamma_{j\beta}T^{\alpha\beta}, \nonumber
\end{align}
where $n_\alpha\equiv (-\alpha,0,0,0)$ is the future-pointing four-vector
orthonormal to the spacelike hypersurface and $T^{\alpha\beta}$ is the
stress-energy tensor for a perfect fluid ({\it cf.} eq. \ref{hydro
  eqs}). The Einstein equations also lead to a set of physical
constraint equations that are satisfied within each spacelike slice,
\begin{align}
  \label{eq:einstein_ham_constraint}
  \mathcal{H} &\equiv R^{(3)} + K^2 - K_{ij} K^{ij} - 16\pi\rho_{_{\rm ADM}} = 0, \\
  \label{eq:einstein_mom_constraints}
  \mathcal{M}^i &\equiv D_j(K^{ij} - \gamma^{ij}K) - 8\pi S^i = 0,
\end{align}
which are usually referred to as Hamiltonian and momentum constraints.
Here $R^{(3)}=R_{ij} \gamma^{ij}$ is the Ricci scalar on a
three-dimensional timeslice.  Our specific choice of evolution
variables introduces five additional constraints,
\begin{align}
  \det \tilde{\gamma}_{ij} & = 1, 
    \label{eq:gamma_one}\\
  \tr \tilde{A}_{ij} & = 0,
    \label{eq:trace_free_A}\\
  \tilde{\Gamma}^i & = \tilde{\gamma}^{jk}\tilde{\Gamma}^i_{jk}.
   \label{eq:Gamma_def}
\end{align}
Our code actively enforces the algebraic
constraints~(\ref{eq:gamma_one}) and~(\ref{eq:trace_free_A}).  The
remaining constraints, $\mathcal{H}$, $\mathcal{M}^i$,
and~(\ref{eq:Gamma_def}), are not actively enforced and can be used
as monitors of the accuracy of our numerical solution.
See~\cite{Alcubierre02a} for a more comprehensive discussion of the
these points.\\

Among the diagnostic quantities, we compute the angular momentum as a
volume integral with the expression~\cite{Shibata99c}:

\begin{align}
\label{AngMomFormula}
J^i_{\rm vol} &=& \varepsilon^{ijk}\int_V \bigg(\frac{1}{8\pi}\tilde A_{jk}
+ x_j S_k + \frac{1}{12\pi}x_j K_{,k} + \nonumber \\
&&-\frac{1}{16\pi}x_j\tilde\gamma^{lm}{}_{,k}\tilde A_{lm}\bigg)
e^{6\phi} d^3x\ .
\end{align}

\subsection{Gauges}
\label{sec:Gauges}

We specify the gauge in terms of the standard ADM lapse
function, $\alpha$, and shift vector, $\beta^i$~\cite{misner73}.
We evolve the lapse according to the ``$1+\log$'' slicing
condition~\cite{Bona94b}:
\begin{equation}
  \partial_t \alpha - \beta^i\partial_i\alpha 
    = -2 \alpha (K - K_0),
  \label{eq:one_plus_log}
\end{equation}
where $K_0$ is the initial value of the trace of the extrinsic
curvature and
equals zero for the maximally sliced initial data we consider here.
The shift is evolved using the hyperbolic $\tilde{\Gamma}$-driver
condition~\cite{Alcubierre02a},
\begin{eqnarray}
\label{shift_evol}
  \partial_t \beta^i - \beta^j \partial_j  \beta^i & = & \frac{3}{4} \alpha B^i\,,
  \\
  \partial_t B^i - \beta^j \partial_j B^i & = & \partial_t \tilde\Gamma^i 
    - \beta^j \partial_j \tilde\Gamma^i - \eta B^i\,,
\end{eqnarray}
where $\eta$ is a parameter which acts as a damping coefficient which
we set to be constant ($\eta=1$). The advection terms on the
right-hand sides of these equations have been suggested
in~\cite{Baker05a, Baker:2006mp, Koppitz-etal-2007aa}.

All the equations discussed above are solved using the
\texttt{CCATIE} code, a three-dimensional finite-differencing code
based on the Cactus Computational Toolkit~\cite{Goodale02a}. A
detailed presentation of the code and of its convergence properties
have been recently presented in ref.~\cite{Pollney:2007ss}.

\subsection{Evolution system for the matter}
\label{hd_eqs}

An important feature of the {\tt Whisky} code is the implementation of
a \textit{flux-conservative} formulation of the hydrodynamics
equations~\cite{Marti91,Banyuls97,Ibanez01}, in which the set of
conservation equations for the stress-energy tensor $T^{\mu\nu}$ and
for the matter current density $J^\mu$, namely
\begin{equation}
\label{hydro eqs}
\nabla_\mu T^{\mu\nu} = 0\;,\;\;\;\;\;\;
\nabla_\mu J^\mu = 0\, ,
\end{equation}
is written in a hyperbolic, first-order and flux-conservative form of
the type
\begin{equation}
\label{eq:consform1}
\partial_t {\mathbf q} + 
        \partial_i {\mathbf f}^{(i)} ({\mathbf q}) = 
        {\mathbf s} ({\mathbf q})\ ,
\end{equation}
where ${\mathbf f}^{(i)} ({\mathbf q})$ and ${\mathbf s}({\mathbf q})$
are the flux vectors and source terms, respectively~\cite{Font03}. Note
that the right-hand side (the source terms) does not depend
on derivatives of the stress-energy tensor. Furthermore,
while the system (\ref{eq:consform1}) is not strictly hyperbolic,
strong hyperbolicity is recovered in a flat spacetime, where ${\mathbf s}
({\mathbf q})=0$.

        As shown by~\cite{Banyuls97}, in order to write system
(\ref{hydro eqs}) in the form of system (\ref{eq:consform1}), the
\textit{primitive} hydrodynamical variables ({\it i.e.} the
rest-mass density $\rho$, the pressure $p$ measured in the
rest-frame of the fluid, the fluid three-velocity $v^i$ measured by a local
zero-angular momentum observer, the specific internal energy $\epsilon$
and the Lorentz factor $W$) are mapped to the so-called \textit{conserved}
variables \mbox{${\mathbf q} \equiv (D, S^i, \tau)$} via the relations

\vspace{-0.2 cm}
\begin{eqnarray}
  \label{eq:prim2con}
   D &\equiv& \sqrt{\gamma}W\rho\ , \nonumber\\
   S^i &\equiv& \sqrt{\gamma} \rho h W^2 v^i\ ,  \\
   \tau &\equiv& \sqrt{\gamma}\left( \rho h W^2 - p\right) - D\ , \nonumber
\end{eqnarray}
where $h \equiv 1 + \epsilon + p/\rho$ is the specific enthalpy and
\hbox{$W \equiv (1-\gamma_{ij}v^i v^j)^{-1/2}$}. Note that only five of
the seven primitive variables are independent.

\begin{table*}[t]
  \caption{Properties of the initial data: proper separation between
    the centers of the stars $d/M_{_{\rm ADM}}$; baryon mass $M_{b}$
    of each star in solar masses; total ADM mass $M_{_{\rm ADM}}$ in
    solar masses, as measured on the finite-difference grid; total
    ADM mass $\tilde{M}_{_{\rm ADM}}$ in solar masses, as provided by
    the Meudon initial data; angular momentum $J$, as measured on the
    finite-difference grid; angular momentum $\tilde{J}$, as
    provided by the Meudon initial data; initial orbital angular
    velocity $\Omega_0$; mean coordinate equatorial radius of each
    star $r_e$ along the line connecting the two stars; ratio
    $r_{e'}/r_e$ of the equatorial coordinate radius of a star in the
    direction orthogonal to the line connecting the two stars and
    $r_e$; ratio of the polar to the equatorial coordinate radius of
    each star $r_p/r_e$; maximum rest-mass density of a star
    $\rho_{\rm max}$.  The initial data for the evolutions with
    polytropic and ideal-fluid EOS are the same. Note that the
    asterisk in the model denomination will be replaced by ``${\rm
      P}$'' or by ``${\rm IF}$'' according to whether the binary is
    evolved using a polytropic or an ideal-fluid EOS.}
\begin{ruledtabular}
\begin{tabular}{l|ccccccccccc}
Model &
\multicolumn{1}{c}{$d/M_{_{\rm ADM}}$} &
\multicolumn{1}{c}{$M_{b}$} &
\multicolumn{1}{c}{$M_{_{\rm ADM}}$} &
\multicolumn{1}{c}{$\tilde{M}_{_{\rm ADM}}$} &
\multicolumn{1}{c}{$J$} &
\multicolumn{1}{c}{$\tilde{J}$} &
\multicolumn{1}{c}{$\Omega_0$} &
\multicolumn{1}{c}{$r_e$} &
\multicolumn{1}{c}{$r_{e'}/r_e$}&
\multicolumn{1}{c}{$r_p/r_e$}&
\multicolumn{1}{c}{$\rho_{\rm max}$} \\
~ &
\multicolumn{1}{c}{$~$} &
\multicolumn{1}{c}{$(M_{\odot})$} &
\multicolumn{1}{c}{$(M_{\odot})$} &
\multicolumn{1}{c}{$(M_{\odot})$} &
\multicolumn{1}{c}{$({\rm g\, cm^2/s})$} &
\multicolumn{1}{c}{$({\rm g}\, {\rm cm}^2{\rm /s})$} &
\multicolumn{1}{c}{$({\rm rad/ms})$} &
\multicolumn{1}{c}{$({\rm km})$} &
\multicolumn{1}{c}{$~$}&
\multicolumn{1}{c}{$~$}&
\multicolumn{1}{c}{$({\rm g/cm}^3)$} \\
\hline
$1.46$-$45$-$*$ & $14.3$ & $1.456$ & $2.681$ & $2.694$ & $6.5077\times10^{49}$ & $6.5075\times10^{49}$ & $1.78$ & $15$ & $0.890$ & $0.899$ & $4.58\times10^{14}$\\
$1.62$-$45$-$*$ & $12.2$ & $1.625$ & $2.982$ & $2.998$ & $7.7833\times10^{49}$ & $7.7795\times10^{49}$ & $1.85$ & $14$ & $0.923$ & $0.931$ & $5.91\times10^{14}$\\
$1.62$-$60$-$*$ & $16.8$ & $1.625$ & $2.987$ & $3.005$ & $8.5548\times10^{49}$ & $8.5546\times10^{49}$ & $1.24$ & $13$ & $0.972$ & $0.977$ & $5.93\times10^{14}$\\
\end{tabular}
\end{ruledtabular}
\vskip -0.25cm
\label{table:ID}
\end{table*}

In this approach, all variables ${\bf q}$ are represented on the
numerical grid by cell-integral averages. The functions the ${\bf q}$ represent are then {\it
  reconstructed} within each cell, usually by piecewise polynomials, in
a way that preserves conservation of the variables ${\bf
  q}$~\cite{Toro99}. This operation produces two values at each cell boundary, which
are then used as initial data for the local Riemann problems, whose (approximate) solution gives the
fluxes through the cell boundaries. A Method-of-Lines
approach~\cite{Toro99}, which reduces the partial differential
equations~\eqref{eq:consform1} to a set of ordinary differential
equations that can be evolved using standard numerical methods, such
as Runge-Kutta or the iterative Cranck-Nicholson
schemes~\cite{Teukolsky00,Leiler_Rezzolla06}, is used to update the
equations in time (see ref.~\cite{Baiotti03a} for further
details). The {\tt Whisky} code implements several reconstruction
methods, such as Total-Variation-Diminishing (TVD) methods,
Essentially-Non-Oscillatory (ENO) methods~\cite{Harten87} and the
Piecewise Parabolic Method (PPM)~\cite{Colella84}. Also, a variety of
approximate Riemann solvers can be used, starting from the
Harten-Lax-van Leer-Einfeldt (HLLE) solver~\cite{Harten83}, over to
the Roe solver~\cite{Roe81} and the Marquina flux
formula~\cite{Aloy99b} (see~\cite{Baiotti03a,Baiotti04} for a more
detailed discussion).  A comparison among different numerical methods
in our binary-evolution simulations is reported in
Appendix~\ref{sec:influence num methods}.

Note that in order to close the system of equations for the
hydrodynamics an EOS which relates the pressure to the rest-mass
density and to the energy density must be specified. The code has been
written to use any EOS, but all the tests so far have been performed
using either an (isentropic) polytropic EOS
\begin{eqnarray}
\label{poly}
p &=& K \rho^{\Gamma}\ , \\
e &=& \rho + \frac{p}{\Gamma-1}\ ,
\end{eqnarray}
or an ``ideal-fluid'' EOS
\begin{equation}
\label{id fluid}
p = (\Gamma-1) \rho\, \epsilon \ . 
\end{equation}
Here, $e$ is the energy density in the rest frame of the fluid, $K$
the polytropic constant (not to be confused with the trace of the
extrinsic curvature defined earlier) and $\Gamma$ the adiabatic
exponent. In the case of the polytropic EOS (\ref{poly}),
$\Gamma=1+1/N$, where $N$ is the polytropic index and the evolution
equation for $\tau$ does not need to be solved. Note that polytropic
EOSs \eqref{poly} do not allow any transfer of kinetic energy to
thermal energy, a process which occurs in physical shocks (shock
heating). It is also useful to stress that by being isentropic, the
polytropic EOS~\eqref{poly}, far from being realistic, can
be considered as unrealistic for describing the merger and the
post-merger evolution. However, it is used here because it provides
three important advantages. Firstly, it provides one ``extreme'' of
the possible binary evolution by being perfectly adiabatic. Secondly,
it allows us to use the same initial data but to evolve it with two
different EOS. This is a unique possibility which is not offered by
other (more realistic) EOSs. As we will comment below, it has allowed
us to highlight subtle properties of the binary dynamics during the
inspiral which were not reported before (see the discussion in
Sec.~\ref{IF_binaries_hm}). Finally, by being isentropic it provides
the most realistic description of the inspiral phase, during which the
neutron stars are expected to interact only between themselves and
only gravitationally.

In contrast to the polytropic EOS, when using the ideal-fluid EOS
(\ref{id fluid}), non-isentropic changes can take place in the fluid
and the evolution equation for $\tau$ needs to be solved. 

\subsection{Adaptive Mesh Refinement and Singularity Handling}  
\label{sec:NumericalSpecifications}

We use the \texttt{Carpet} code that implements a vertex-centered
adaptive-mesh-refinement scheme adopting nested
grids~\cite{Schnetter-etal-03b} with a $2:1$ refinement factor for
successive grid levels. We center the highest resolution level around
the peak in the rest-mass density of each star. This represents our
rather basic form of AMR.

The timestep on
each grid is set by the Courant condition (expressed in terms of the
speed of light) and so by the spatial grid resolution for that level;
the typical Courant coefficient is set to be $0.35$.
The time evolution is carried out using third-order accurate
Runge-Kutta integration steps. Boundary data for finer grids are
calculated with spatial prolongation operators employing third-order
polynomials and with prolongation in time employing second-order
polynomials. The latter allows a significant memory saving, requiring
only three time levels to be stored, with little loss of accuracy due
to the long dynamical timescale relative to the typical grid timestep.

In the results presented below we have used $6$ levels of mesh
refinement with the finest grid resolution of
$h=0.12\,M_{\odot}=0.177\,\km$ and the wave-zone grid resolution of
$h=3.84\,M_{\odot}=5.67\,\km$. Our finest grid has then a resolution
of a factor of two higher compared to the one used
in~\cite{Shibata06a}, where a uniform grid with $h=0.4\,\km$ was
instead used. Each star is covered with two of the finest grids, so
that the high-density regions of the stars are tracked with the
highest resolution available. These ``boxes'' are then moved by
tracking the position of the rest-mass density as the stars orbit and
are merged when they overlap. In addition, a set of refined but fixed
grids is set up at the center of the computational domain so as to
capture the details of the Kelvin-Helmholtz instability ({\it cf.}
Sect.~\ref{vs_and_khi}). The finest of these grids extends to
$r=7.5\,M_{\odot}=11\,\km$. A single grid resolution covers then the
region between $r=150\,M_{\odot}=221.5\,\km$ and
$r=250\,M_{\odot}=369.2\,\km$, in which our wave extraction is carried
out. A reflection symmetry condition across the $z=0$ plane and a
$\pi$-symmetry condition\footnote{Stated differently, we evolve only
the region $\{x>0,\,z>0\}$ applying a
180-degrees--rotational-symmetry boundary condition across the plane
at $x=0$.} across the $x=0$ plane are used. 

We have performed extensive tests to ensure that both the hierarchy of
the refinement levels described above and the resolutions used yield
results that are numerically consistent although not always in a
convergent regime. The initial data used for these tests refer to a
binary evolved with an ideal-fluid EOS from an initial separation of
$45\,\km$ and in which each star has a mass of $1.8\,M_{\odot}$; such a mass is larger than the one
used for the rest of our analysis (\cf Table~\ref{table:ID}) and it has been
employed because it leads to a prompter formation of a black hole,
thus saving computational costs.

The $2$-norm of the typical violation of the Hamiltonian constraint
grows from $\lesssim 10^{-6}$ at the beginning of the simulations to
$\lesssim 10^{-4}$ at the end of the simulations. The convergence rate
measured in the $2$-norm of the violation of the Hamiltonian
constraint is $\simeq 1.7$ before the merger (\ie the same convergence
rate measured in the evolution of isolated stars~\cite{Baiotti04}),
but it then drops to $\simeq 1.2$ during the merger. It is still
unclear whether this difference is due to the generation of a
turbulent regime at the merger (see the discussion in
Sect.~\ref{vs_and_khi} and in ref.~\cite{Fromang2007}) or to a
resolution which is close but not yet in a fully convergent
regime. Tests showing the convergence rate, the conservation of the
mass (baryonic and gravitational) and angular momentum, as well as the
consistency in the gravitational waves have been validated by the
referee but, for compactness, are not reported here.

The apparent horizon (AH) formed during the simulation is located
every few timesteps during the
evolution~\cite{Thornburg2003:AH-finding_nourl}.
Exploiting a technique we have first developed when performing
simulations of gravitational collapse to rotating
BHs~\cite{Baiotti06,Baiotti07} and that has now been widely adopted by other codes, we do not
make use of the excision technique~\cite{Baiotti04b}. Rather, we add a
small amount of dissipation to the evolution equations for the metric
and gauge variables only and rely on the singularity-avoiding
gauge~\eqref{eq:one_plus_log} to extend the simulations well past the
formation of the AH (note that no dissipation is added to the
evolution of matter variables). More specifically, we use an
artificial dissipation of the Kreiss--Oliger type~\cite{Kreiss73} on
the right-hand sides of the evolution equations for the spacetime
variables and the gauge quantities. This is needed mostly because all
the field variables develop very steep gradients in the region inside
the AH. Under these conditions, small high-frequency oscillations
(either produced by finite-differencing errors or by small reflections
across the refinement or outer boundaries) can easily be amplified,
leave the region inside the AH and rapidly destroy the solution. In
practice, for any time-evolved quantity $u$, the right-hand side of
the corresponding evolution equation is modified with the introduction
of a term of the type ${\cal L}_{\mbox{\tiny diss}}(u) = -\varepsilon
h^3 \partial^4_i u$, where $h$ is the grid spacing and
$\varepsilon=0.075$ and is kept constant in space (see
ref.~\cite{Baiotti07} for additional information and different
prescriptions).

\subsection{Initial data}
\label{sec:initial_data}

As initial data for relativistic-star binary simulations we use the
ones produced by the group working at the Observatoire de
Paris-Meudon~\cite{Gourgoulhon01,Taniguchi02b}. These data, which we
refer to also as the {\it ``Meudon data''}, are obtained under the
simplifying assumptions of quasi-equilibrium and of conformally-flat
spatial metric. The initial data used in the simulations shown here
were produced with the additional assumption of irrotationality of the
fluid flow, {\it i.e.}  the condition in which the spins of the stars
and the orbital motion are not locked; instead, they are defined so as
to have vanishing vorticity. Initial data obtained with the
alternative assumption of rigid rotation were not used because,
differently from what happens for binaries consisting of ordinary
stars, relativistic-star binaries are not thought to achieve
synchronisation (or corotation) in the timescale of the
coalescence~\cite{Bildsten92}. The Meudon initial configurations are
computed using a multi-domain spectral-method code, {\tt LORENE}, which
is publicly available~\cite{lorene}. A specific routine is used to transform the
solution from spherical coordinates to a Cartesian grid of the desired
dimensions and shape.

The binaries used as initial-data configurations have been chosen so
as to provide the variety of behaviours that we wanted to illustrate
and some of their physical quantities are reported in
Table~\ref{table:ID}. Furthermore, since it is the least
computationally expensive, we have chosen model $1.62$-$45$-$*$ as our
fiducial initial configuration. For this binary the initial coordinate
distance between stellar centres in terms of the initial gravitational
wavelength is $d = 0.09\,\lambda_{_{\rm GW}}$, where $\lambda_{_{\rm
    GW}} = \pi/\Omega_0$ is the gravitational wavelength for a
Newtonian binary of orbital angular frequency $\Omega_0$. For
evolutions that employ a polytropic EOS, the polytropic exponent is
$\Gamma=2$ and the polytropic coefficient $K=123.6=1.798\times10^5{\rm
  g}^{-1}{\rm cm}^5{\rm s}^{-2}$. (Ref.~\cite{Oechslin07a} has recently
made the useful remark that a choice of $\Gamma=2.75$ and $K=30000$
leads to an EOS that fits well the supernuclear regime of the
Shen-EOS at zero temperature~\cite{Shen98}, as well as yielding
density profiles that are very similar to those obtained with that
realistic EOS; unfortunately no initial data with this adiabatic
exponent is available at the moment.)

\section{Binary dynamics}
\label{sec:bd}

In what follows we describe the matter dynamics of the binary initial
data discussed in the previous Section. To limit the discussion and
highlight some of the most salient aspects we will consider two main
classes of initial data, represented by models $1.62$-$45$-$*$ and
$1.46$-$45$-$*$, respectively. These models differ only in the rest
mass, the first one being composed of stars each having a mass of
$1.625\,M_{\odot}$ (which we refer to as the \textbf{high-mass}
binaries), while the second one is composed of stars of mass
$1.456\,M_{\odot}$ (which we refer to as the \textbf{low-mass}
binaries).

\begin{figure*}[ht]
\begin{center}
   \includegraphics[width=0.45\textwidth]{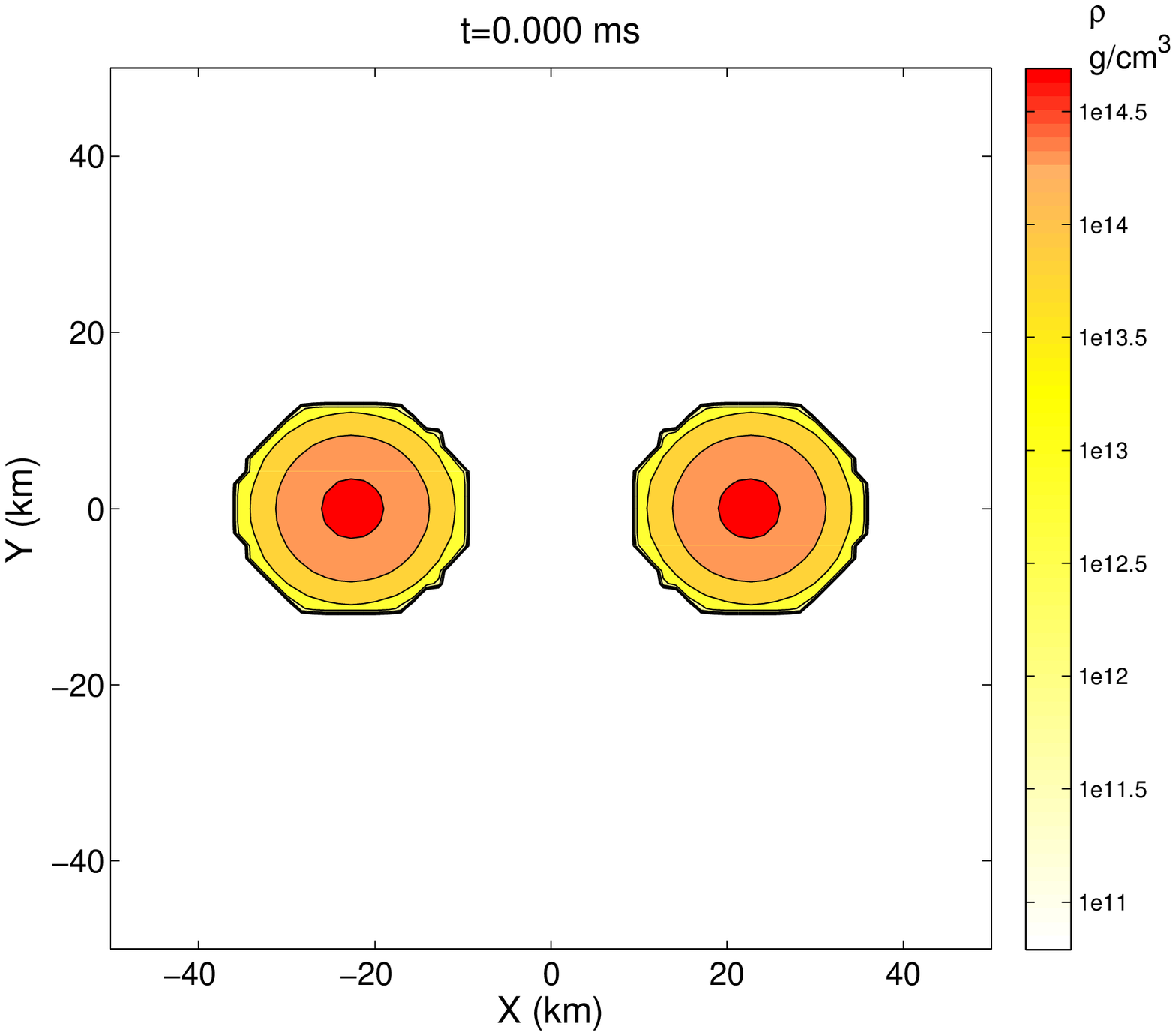}
   \includegraphics[width=0.45\textwidth]{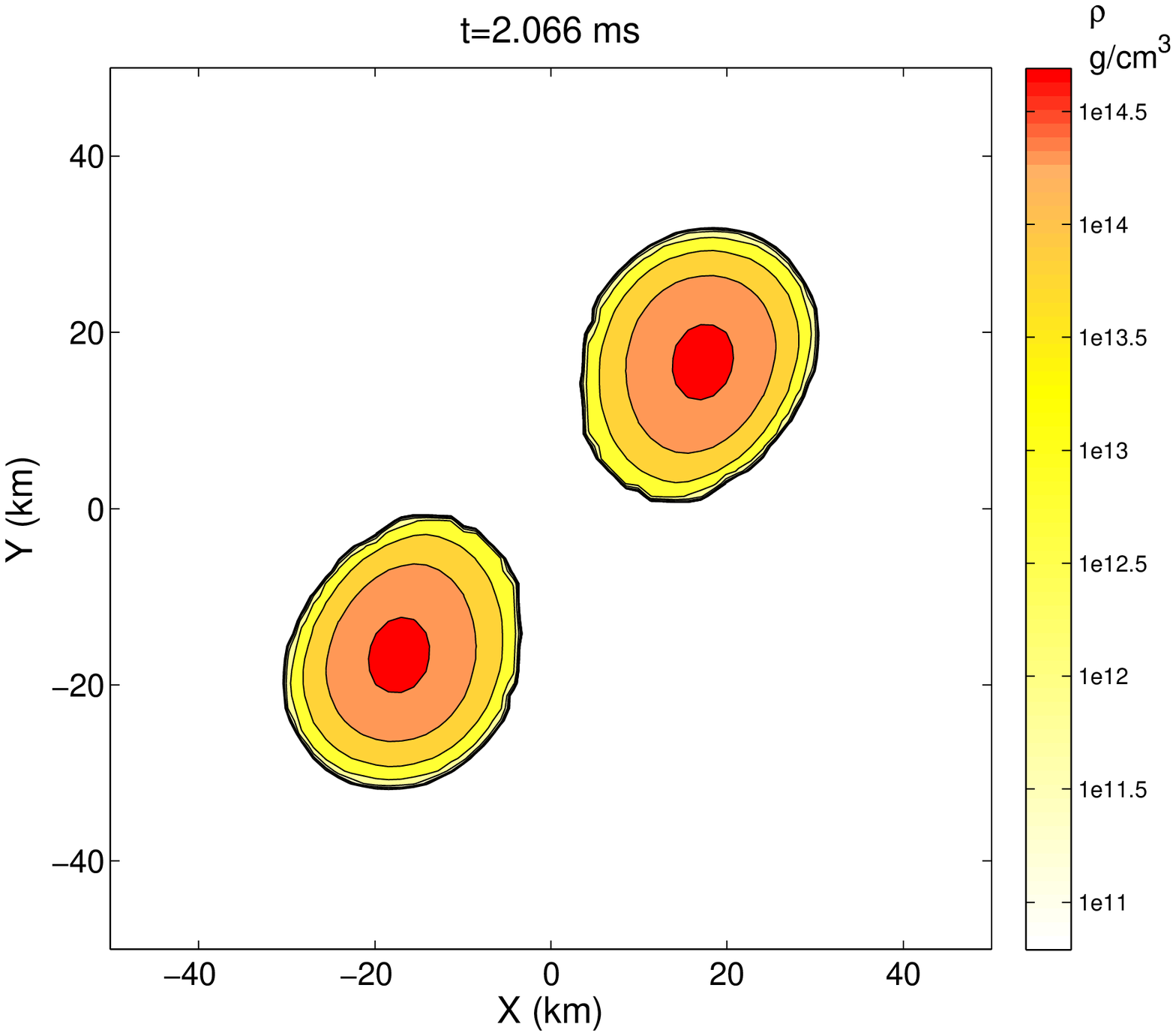}
   \includegraphics[width=0.45\textwidth]{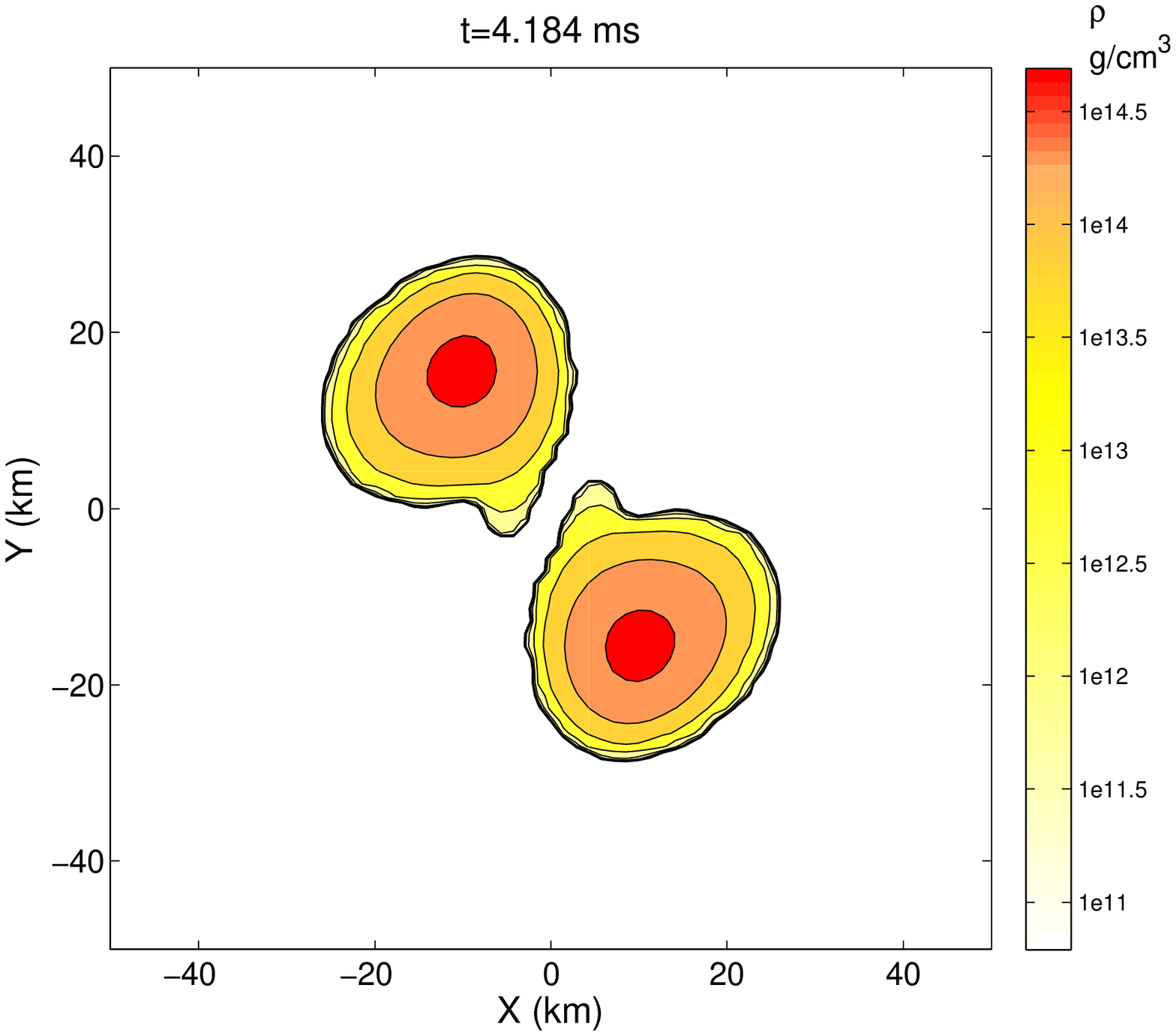}
   \includegraphics[width=0.45\textwidth]{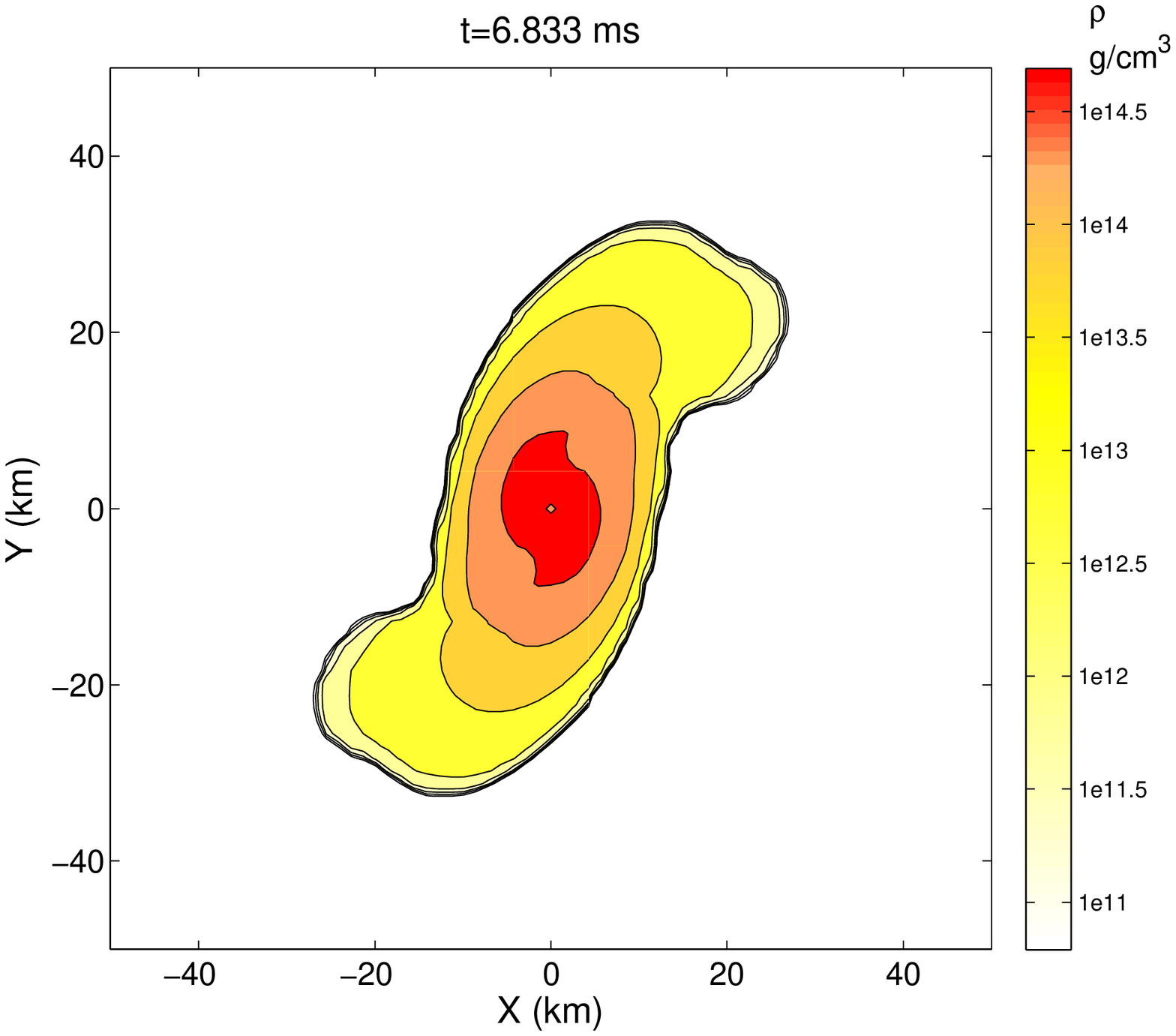}
   \includegraphics[width=0.45\textwidth]{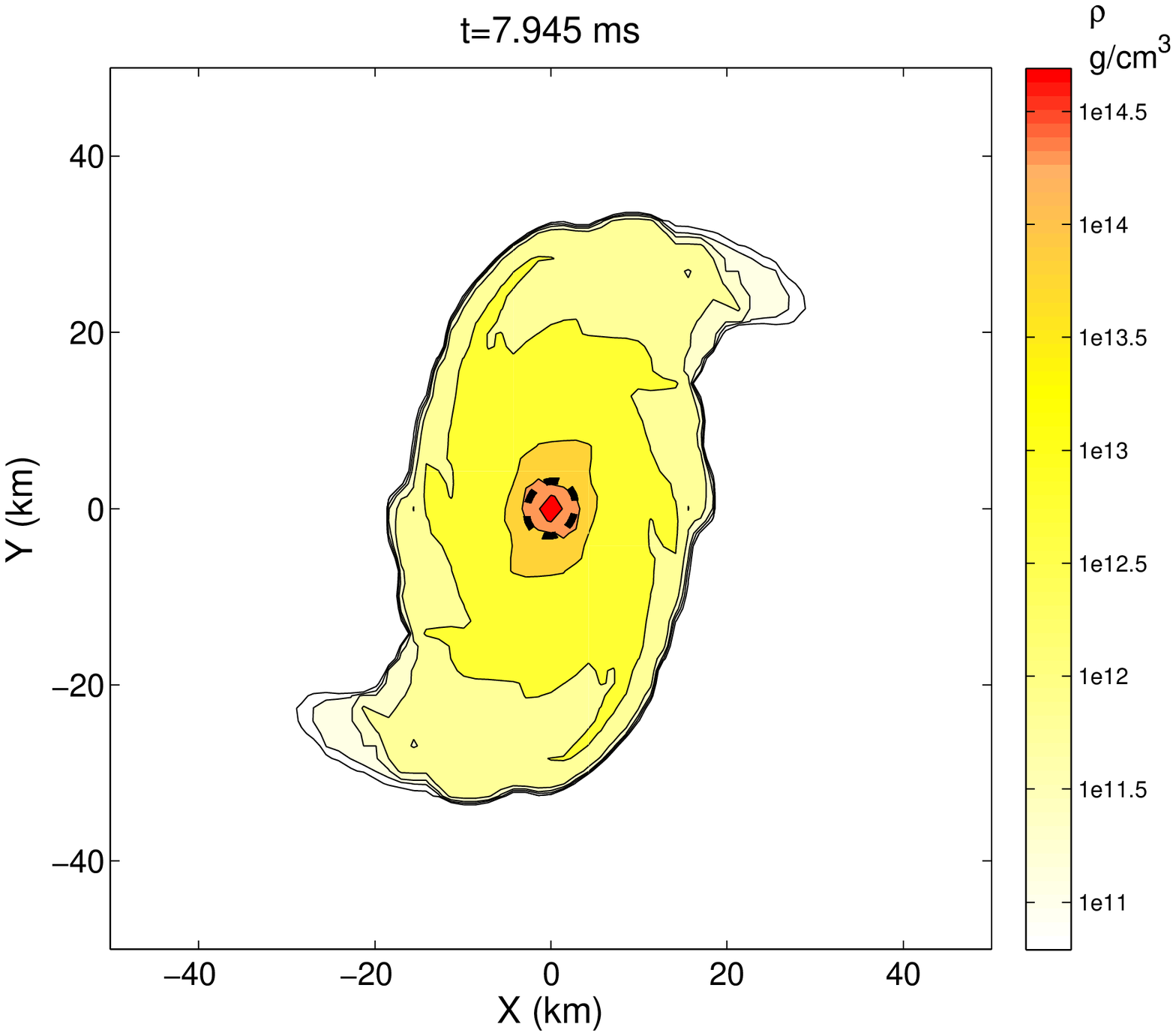}
   \includegraphics[width=0.45\textwidth]{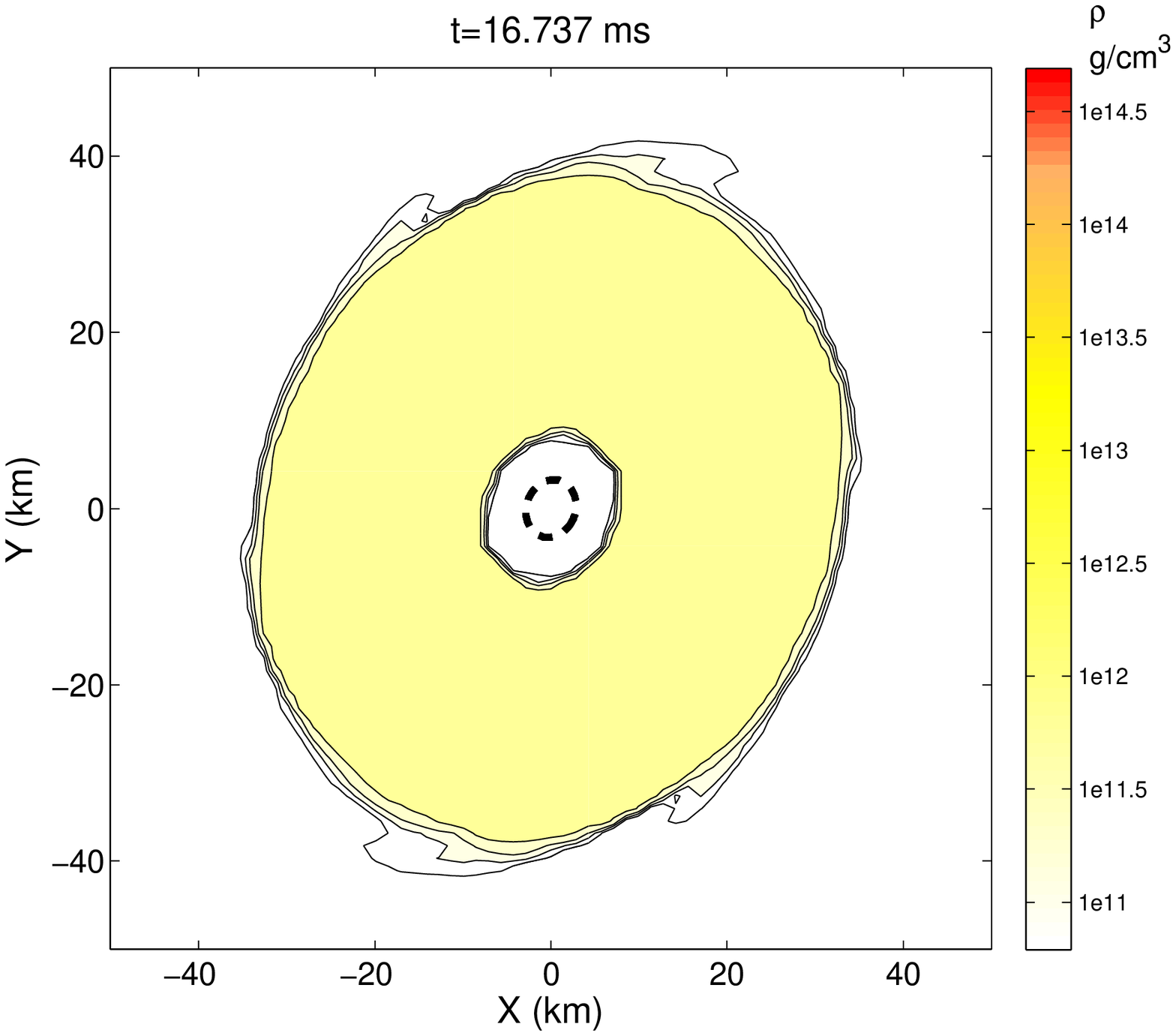}
\end{center}
   \caption{Isodensity contours on the $(x,y)$ (equatorial) plane
     for the evolution of the \textbf{high-mass} binary with the
     \textbf{polytropic} EOS (\ie model $1.62$-$45$-${\rm P}$ in
     Table~\ref{table:ID}). The time stamp in $\ms$ is shown on the
     top of each panel the color-coding bar is shown
     on the right in units of ${\rm g/cm}^3$ and the thick
     dashed line represents the AH. A high-resolution version of this figure can be found at~\cite{weblink}.\label{fig:rho2D_poly_high_xy}}
\end{figure*}

\begin{figure}[ht]
  \begin{center}
   \includegraphics[angle=-0,width=0.45\textwidth]{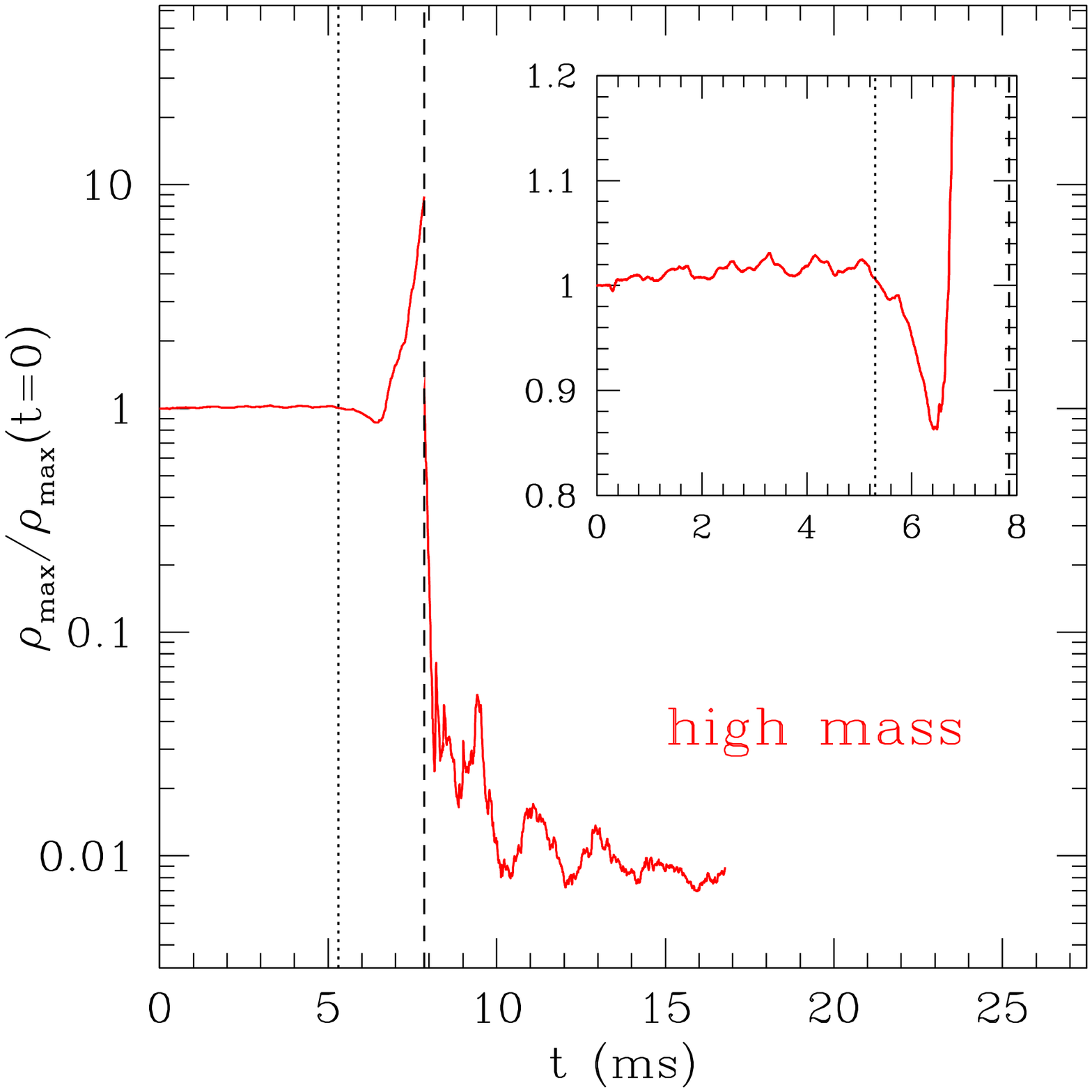} 
  \end{center}
  \vskip -0.5cm
  \caption{Evolution of the maximum rest-mass density normalized to
    its initial value for the \textbf{high-mass} binary. Indicated
    with a dotted vertical line is the time at which the stars merge,
    while a vertical dashed line shows the time at which an AH is first
    found and which is a few ${\rm ms}$ only after the merger in this
    case. After this time, the maximum rest-mass density is computed
    in a region outside the AH and therefore it refers to the density
    of the oscillating torus. It is only a few orders of magnitude
    smaller. Note that the non-normalised value of the maximum
    rest-mass density at $t=0$ is $5.91\times10^{14}\,{\rm g}/{\rm cm^3}$ (see
    Table~\ref{table:ID}). The binary has been evolved using the
    \textbf{polytropic} EOS.}
\label{fig:poly-rho-high}
\end{figure}

Variations of these initial data will also be considered by changing,
for instance, either the initial coordinate separation (\ie $60\,{\rm
  km}$ in place of ~$45\,{\rm km}$) or the EOS (\ie an ideal-fluid EOS
or a polytropic one). Additional variations involving, for instance,
different mass ratios, will be presented elsewhere~\cite{BGR08}.

\subsection{Polytropic EOS: high-mass binary}
\label{sec:poly_hm}

We start by considering the evolution of the high-mass binary evolved
with the polytropic EOS, \ie model $1.62$-$45$-${\rm P}$ in
Table~\ref{table:ID}. Fig.~\ref{fig:rho2D_poly_high_xy}, in
particular, collects some representative isodensity contours
(\ie contours of equal rest-mass density) on the $(x,y)$ (equatorial)
plane, with the time stamp being shown on the top of each panel and
with the color-coding bar being shown on the right in units of
${\rm g/cm}^3$.

The binary has an initial coordinate separation between the maxima in
the rest-mass density of $45\,{\rm km}$ and, as we will discuss more
in detail later on, a certain amount of coordinate eccentricity and
tidal coordinate deformation is introduced by the initial choice for
the shift vector. As the stars inspiral, their orbital angular velocity increases
and after about $2.2$ orbits, or equivalently after about $5.3\ {\rm
  ms}$ from the beginning of the simulation, they merge, producing an
object which has a mass well above the maximum one for uniformly
rotating stars, but which supports itself against gravitational
collapse by a large differential rotation. Such an object is usually
referred to as a hyper-massive neutron star or HMNS\footnote{We recall
  that a HMNS is a star whose mass is larger than the maximum one
  allowed for a uniformly rotating model with the same EOS (\ie the
  supramassive limit). For the $\Gamma=2$ polytropes considered here
  this maximum mass is $M=2.324\ M_{\odot}$ (with an equivalent baryon
  mass of $M_b= 2.559\ M_{\odot}$), while the maximum mass for a
  nonrotating model is $M=2.027\ M_{\odot}$ (with an equivalent baryon
  mass of $M_b= 2.225\ M_{\odot}$). Clearly, all the models considered
  in Table~\ref{table:ID} lead to a HMNS after the merger.}. As the
inspiral proceeds and the two NSs progressively approach each other,
tidal waves produced by the tidal interaction become visible
(\cf~first and second rows of panels in
Fig.~\ref{fig:rho2D_poly_high_xy}) and these are particularly large,
\ie of $\sim 5\%$, for the high-mass binary and considerably smaller
for the low-mass one (\cf~Fig.~\ref{fig:poly-rho-low}).

\begin{figure}[ht]
\begin{center}
   \includegraphics[width=0.45\textwidth]{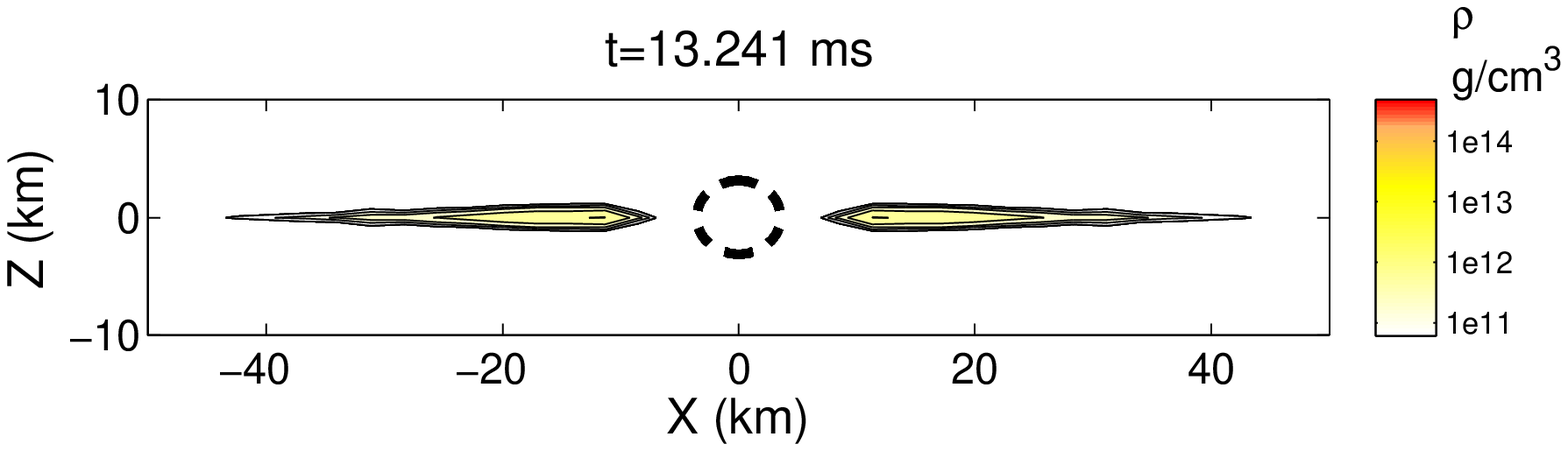}
   \includegraphics[width=0.45\textwidth]{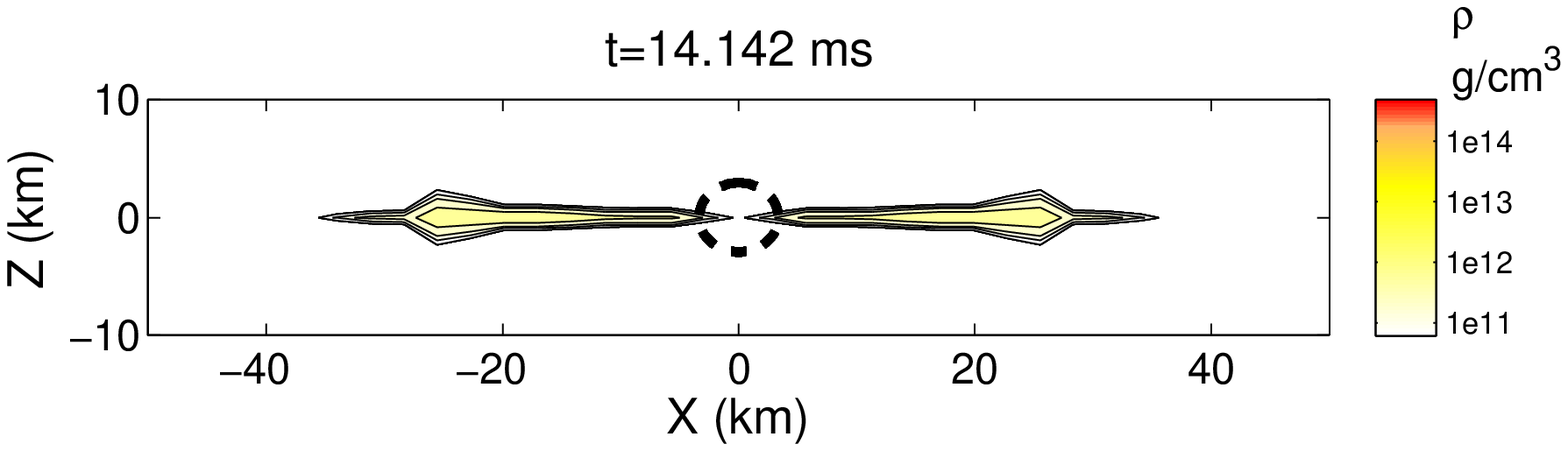}
   \includegraphics[width=0.45\textwidth]{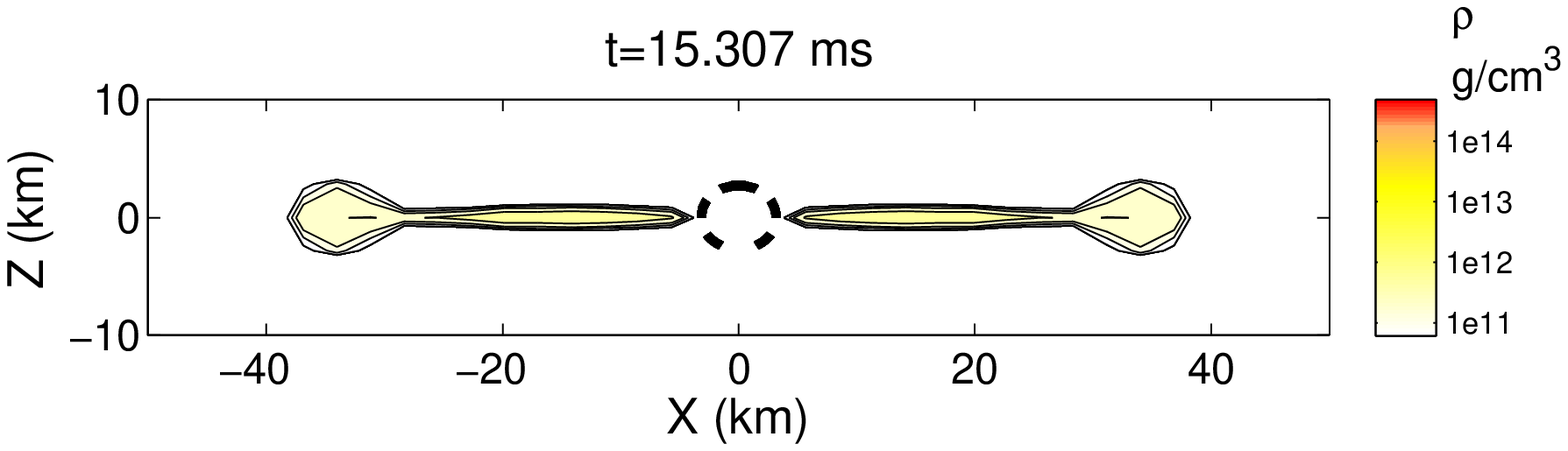}
   \includegraphics[width=0.45\textwidth]{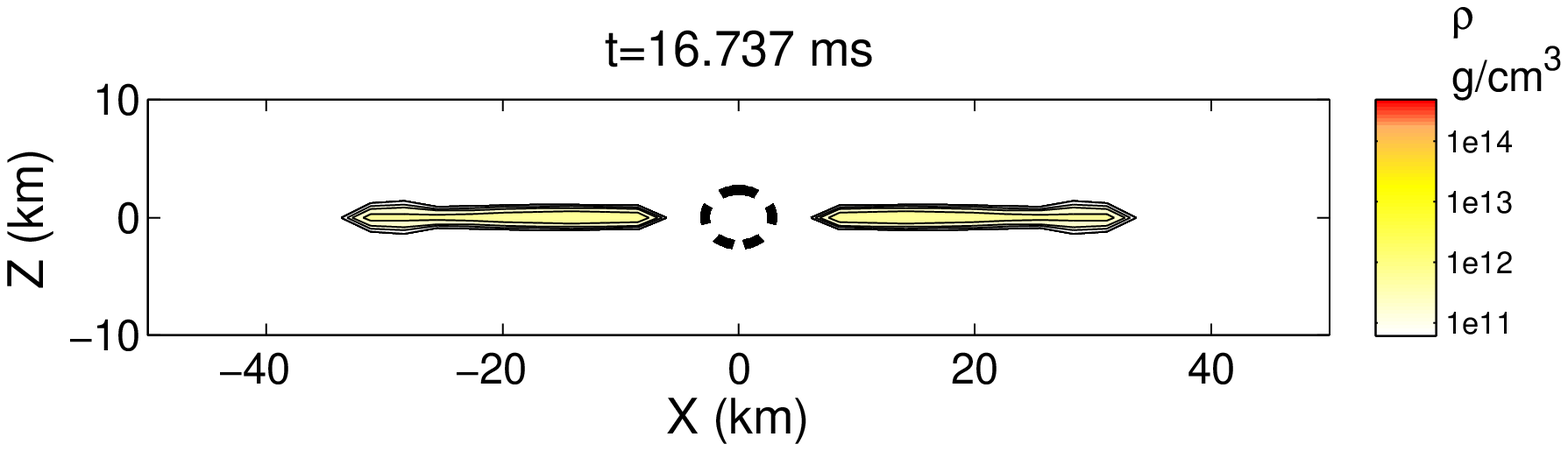}
\end{center}
\vskip -0.5cm
   \caption{Isodensity contours on the $(x,z)$ plane highlighting the
     formation of a torus surrounding the central BH, whose AH
     is indicated with a thick dashed line. The data
     refers to the \textbf{high-mass} binary evolved with the
     \textbf{polytropic} EOS. A high-resolution version of this figure can be found at~\cite{weblink}.\label{fig:BH_short_torus_poly_high}}
\end{figure}

This is shown in Fig.~\ref{fig:poly-rho-high}, which reports the
evolution of the maximum rest-mass density normalized to its initial
value. Indicated with a dotted vertical line is the time at which the
stars merge (which we define as the time at which the outer layers of
the stars enter in contact), while a vertical dashed line shows the
time at which an AH is found. After this time the maximum rest-mass
density is computed in a region outside the AH and therefore it refers
to the density of the oscillating torus. It is only a few orders of
magnitude smaller. Note that before the merger the central rest-mass
density not only oscillates but it also increases secularly, although
at a much smaller rate (\cf~also
Fig.~\ref{fig:rho_pol_high_45_vs_60}).

As mentioned above, the merger takes place after about $5\,\ms$ and
the two NSs collide with a rather large impact parameter. This reduces
significantly the strength of the shocks which have been computed in
the case of head-on collisions~\cite{Evans03a}, but it also produces a
considerable amount of shear, which could then lead to a series of
interesting dynamical instabilities (see also the discussion in
Sect.~\ref{vs_and_khi}). Because of the adiabatic nature of the EOS,
which prevents the formation of physical shocks (\ie discontinuities
where entropy is increased locally)\footnote{Note that very large
  gradients can nevertheless be produced because of the nonlinear
  nature of the hydrodynamic equations. These gradients, however, are
  not physical shocks in that no entropy is increased locally.}, the
HMNS produced at the merger is beyond the stability limit for
gravitational collapse, so that despite the high amount of angular
momentum and the large degree of differential rotation\footnote{Note
  that the HMNS is not axisymmetric and hence it is difficult to
  provide a unique measure of the degree of differential rotation. On
  average, however, the angular velocity decreases of about one order
  of magnitude between the rotation axis and the surface.}, it rapidly
collapses to produce a rotating BH, at about $8\,\ms$.

More specifically, soon after the merger, the two massive and
high-density cores of the NSs coalesce and during this rapid infall
they experience a considerable decompression of $\sim 15\%$ or more as
shown in the small inset of Fig.~\ref{fig:poly-rho-high}. However,
after $t\sim 6\,\ms$, the maximum rest-mass density is seen to
increase exponentially, a clear indication of the onset of a
quasi-radial dynamical instability, and this continues through the
formation of an AH, which is first found at time $t=7.85\,\ms$ (see
the last row of panels in Fig.~\ref{fig:rho2D_poly_high_xy} where the
AH is shown with a thick dashed line, or
Fig.~\ref{fig:poly-rho-high}, where the time of appearance is marked
by a dashed vertical line).

This complex general behavior, namely the very small secular increase
in the central rest-mass density accompanied by small tidal
oscillations, and the final decompression as the two NS cores merge
into one, should help to clarify a long-standing debate on whether the
NSs experience a compression prior to the merger which leads them to
collapse to a BH~\cite{Wilson95,Mathews98,Mathews00} or, rather, a
decompression~\cite{Baumgarte97,Bonazzola97,Baumgarte98c}, as a result
of the dynamical instability leading to the merger. Clearly, for the
rather restricted set of stellar models which are close to the
stability limit to BH collapse, the small secular increase
could lead to the formation of two BHs prior to the merger.

\label{footnote1}
After an AH is first found, the amount of matter outside of it is
still quite large and, most importantly, it is the one with the
largest amount of angular momentum. This leads to the formation of an
accretion torus with an average density between $10^{12}$ and
$10^{13}\ {\rm g/cm^3}$, a vertical size of a few ${\rm km}$ but a
horizontal one between $20$ and $30\,{\rm km}$ (see evolution of
$\rho_{\rm max}$ in Fig.~\ref{fig:poly-rho-high} after the AH). The
torus has an \textit{initial} rest mass of $(M_{_{\rm T}})_0 \simeq
0.04\, M_{\odot}$\footnote{We define the initial mass of the torus as
  the rest mass outside the AH soon after the AH is first found. Note
  that such a measure could be ambiguous since the time of the first
  AH detection depends also on the frequency with which the AH has
  been searched for and on the initial guess for the AH radius. 
  To improve this notion and to give a measure that
  is indicatively comparable for different simulations, we take the
  values of the rest mass of the torus at the time at which the AH
  mean radius has reached the arbitrarily chosen value of $2.1$. This
  mass should really be taken as an upper limit for the torus
  rest mass, since its value decreases considerably as the evolution
  proceeds and the torus accretes onto the BH.}, which however
decreases rapidly to become $(M_{_{\rm T}})_{3\,{\rm ms}} = 0.0117\,
M_{\odot}$ only $3\,\ms$ later.

The dynamics of the torus are summarized in
Fig.~\ref{fig:BH_short_torus_poly_high}, which shows the isodensity
contours on the $(x,z)$ plane; also in this case the time stamp is
shown on the top of each panel, while the color-coding bar is shown on
the right in units of ${\rm g/cm}^3$. Note that the panels refer to
times between $13.2\ {\rm ms}$ and $16.7\ {\rm ms}$ and thus to a
stage in the evolution which is between the last two panels of
Fig.~\ref{fig:rho2D_poly_high_xy}. Other information on the properties
of the merged object can be found in Table~\ref{table:FD}.

\begin{figure}[t]
\begin{center}
   \includegraphics[angle=-0,width=0.45\textwidth]{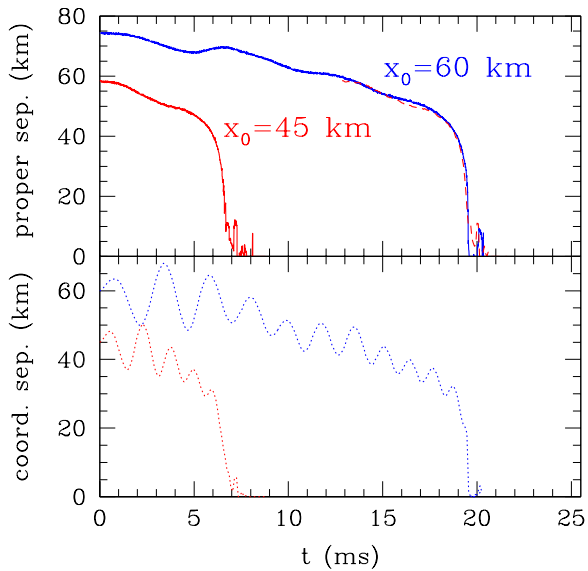}
\end{center}
\vskip -0.5cm
\caption{Evolution of the proper separation (top part) and of the
  coordinate separation (bottom part) for binaries with initial
  coordinate separation of either $45$ or $60\,{\rm km}$ (\ie models
  $1.62$-$45$-${\rm P}$ and $1.62$-$60$-${\rm P}$ in
  Table~\ref{table:ID}). Indicated with a dashed line is the proper
  separation for the binary starting at $45\,{\rm km}$ and suitably
  shifted in time.}
  \label{fig:pol_high_45_vs_60}
\end{figure}

Overall, the torus has a dominant $m=0$ (axisymmetric) structure but,
because of its violent birth, it is very far from an equilibrium.  As
a result, it is subject to large oscillations, mostly in the radial
direction, as it tries to compensate between the excess angular
momentum and the intense gravitational field produced by the BH.
In doing so, it triggers quasi-periodic oscillations with a
period of $\sim 2\ {\rm ms}$, during which the torus moves in towards
the BH, accreting part of its mass. A behavior very similar to this
one has been studied in detail in a number of related
works~\cite{Zanotti02,Zanotti05,Rubio_Lee05a,Rubio_Lee05b,Montero07},
in which the torus was treated as a test fluid. While the above-mentioned 
studies represent an idealization of the dynamics simulated
here, they have highlighted that the harmonic dynamics of the torus
represent a generic response of the fluid to a quasi-radial
oscillation with a frequency reminiscent of the epicyclic frequency
for point-like particles in a gravitational
field~\cite{Rezzolla_qpo_03a,Rezzolla_qpo_03b}. Furthermore, because
of the large quadrupole moment possessed by the torus and its large
variations produced by the oscillations, a non-negligible amount of
gravitational radiation can be produced as a result of this process
(see also the discussion of Fig.~\ref{fig:psi4_pol_high}).

As mentioned in the Introduction, the existence of a massive torus
around the newly formed rotating BH is a key ingredient in the
modeling of short GRBs and the ability of reproducing this feature
through a fully nonlinear simulation starting from consistent initial
data is a measure of the maturity of simulations of this type. For
compactness, we cannot present here a detailed study of the dynamics of
the torus, of the variation of its mass and of the consequent
accretion onto the BH. Such an analysis will be presented
elsewhere~\cite{BGR08}, but it is sufficient to remark here that the
choice of suitable gauge conditions and the use of artificial
viscosity for the field variables allow for a stable evolution of the
system well past BH formation and for all of the time we could afford
computationally.

Using the isolated-horizon formalism~\cite{Ashtekar:2004cn} and its
numerical implementation discussed in ref.~\cite{Dreyer02a}, we have
measured the final BH to have a mass $M_{_{\rm BH}}=2.99\,M_{\odot}$
and spin $J_{_{\rm BH}} = 7.3\,M^2_{\odot}=6.4\times10^{49}{\rm g}\,{\rm cm}^2{\rm s}^{-1}$, 
thus with a dimensionless spin $a\equiv J_{_{\rm BH}}/M^2_{_{\rm BH}} =0.82$
(\cf~Table~\ref{table:FD}). This is a rather surprising result when
compared to the equivalent measure made in the inspiral and merger of
an equal-mass BH binary. In that case, it has been found that
the final dimensionless spin is $a_{\rm fin}\simeq 0.68$ for BHs that
are initially non-spinning and increasing/decreasing for BHs that have
spins parallel/anti-parallel with the orbital angular momentum (see,
\eg \cite{Campanelli:2006gf,Rezzolla-etal-2007,Rezzolla-etal-2007b,Rezzolla-etal-2007c}). More
specifically, the two initial BHs need to have a substantial spin,
with $a_{\rm initial} \simeq 0.45$, in order to produce a final BH
with $a_{\rm final} \simeq 0.82$. On the other hand, the NSs have
here little initial spin (they are essentially spherical besides the
tidal deformation) and the little they have is anti-parallel to the
orbital angular momentum (\ie they counter-spin with respect to the
orbital angular momentum). Yet, they are able to produce a rapidly
spinning BH. It is apparent therefore that the merger of two equal-mass
NSs is considerably less efficient in losing the orbital angular
momentum (or equivalently more efficient in transferring the orbital
angular momentum to the final BH), thus producing a BH which is
comparatively more rapidly spinning.

An important validation of the accuracy of the simulations presented
here can be appreciated when comparing the evolution of the same
binary when evolved starting from different initial separations. More
specifically, we have considered high-mass binaries with initial
coordinate separation of either $45$ or $60\,{\rm km}$ (\ie models
$1.62$-$45$-${\rm P}$ and $1.62$-$60$-${\rm P}$ in
Table~\ref{table:ID}) and evolved them with a polytropic EOS. The
results of this verification are summarized in
Fig.~\ref{fig:pol_high_45_vs_60}, with the upper part reporting the
evolution of the proper separation (continuous lines) and the lower
one that of the coordinate separation (dotted lines) for binaries with
initial coordinate separation of either $45$ or $60\,{\rm km}$
(\ie models $1.62$-$45$-${\rm P}$ and $1.62$-$60$-${\rm P}$ in
Table~\ref{table:ID}). It should be remarked that the evolution of the
latter binary is computationally much more challenging, with an
inspiral phase that is about three times as long when compared with
the small-separation binary. In particular, the stars merge at $t\sim
18\ {\rm ms}$, corresponding to $\sim 5.5$ orbits. This is to be
compared with the $\sim 2.2$ orbits of $1.62$-$45$-${\rm P}$ and it is
close to the limit of what is computationally feasible at these
resolutions.

\begin{figure}[t]
\begin{center}
   \includegraphics[angle=-0,width=0.45\textwidth]{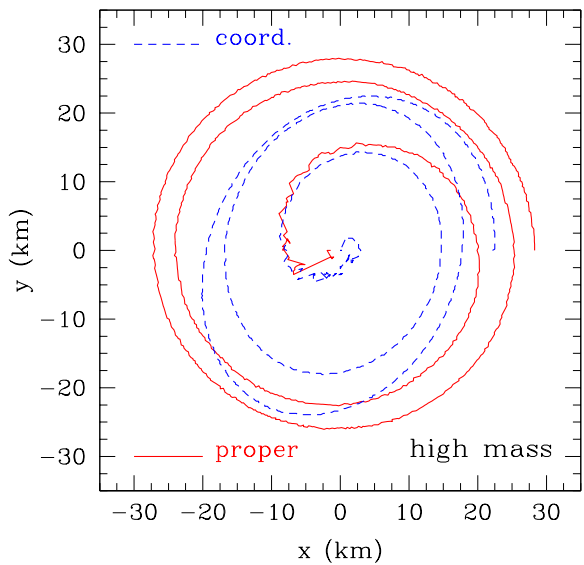}
\end{center}
\vskip -0.5cm
\caption{Coordinate (dashed line) and proper (solid line) trajectory of
  one of the stars for the \textbf{high-mass} binary from a coordinate
  distance of $45\,{\rm km}$ (about $2.2$ orbits).}
  \label{fig:traj_pol_high_45_vs_60}
\end{figure}

The first thing to note in Fig.~\ref{fig:pol_high_45_vs_60} is the
remarkable difference between the coordinate separation, which shows
very large oscillations, and the proper separation, which instead
shows only very little variations superposed to the secular decrease.
These are probably associated to a small but nonzero residual
eccentricity like the one observed in binary BH
simulations~\cite{Pollney:2007ss}. The oscillations in the coordinate
separation, which have been reported also in ref.~\cite{Anderson2007}
(\cf~their Figs.~5 and~7), are in our case clearly related to the
gauge choice, as demonstrated by the evolution of the proper separation. 
This is also apparent when looking at
Fig.~\ref{fig:traj_pol_high_45_vs_60}, which shows the coordinate
trajectory (dashed line) and the proper trajectory (solid line) of one
of the two NSs in the high-mass binary starting from a coordinate
separation of $45\,{\rm km}$. While a certain amount of eccentricity is
present also in the proper trajectory, this is rather small.

A more careful analysis has revealed that the large oscillations in
the coordinate separation are simply the result of non-optimal gauge
conditions. As mentioned in Sect.~\ref{sec:initial_data}, we
import the initial data from the solution of the Meudon group,
adopting the same shift vector $\beta^i$ computed for the
quasi-circular solution. While this may seem like a reasonable thing
to do, it actually introduces the oscillations commented above. We
have also performed alternative simulations in which the shift vector
has been set to be zero initially and then evolved with the gauge
conditions~\eqref{shift_evol}. We have found that in this case also
the coordinate separation is much better behaved and only very small
oscillations are present (see also the discussion in
Appendix~\ref{sec:0shift}).

When concentrating on the evolution of the proper separation it is
clear that the binary starting at a large separation has a larger
eccentricity, but also that most of it is lost by the time the stars
merge. Indicated with a dashed line in
Fig.~\ref{fig:pol_high_45_vs_60} is also the evolution of the
proper separation for the binary starting at $45\,{\rm km}$ when this
is suitably shifted in time of $\sim 13\ {\rm ms}$; the very good
overlap between the two curves is what one expects for a binary system
that is simply translated in time and it gives a measure of our
ability of accurately evolving binary NSs for a large number of
orbits. This provides us with sufficiently long waveforms to perform a
first match with the PN expectations and also to establish the role
played by the tidal interaction between the two NSs as they
inspiral. Both of these studies will be presented
elsewhere~\cite{BGR08}.

A comparison of the waveforms produced in these two simulations will
be discussed in Sect.~\ref{gws_pol}, but we show in
Fig.~\ref{fig:rho_pol_high_45_vs_60} the evolution of the maximum
rest-mass density normalized to its initial value for high-mass
binaries with initial coordinate separation of either $45$ or
$60\,{\rm km}$. For the large-separation binary, we observe a behavior
very similar to that of the small-separation binary, as discussed for
Fig.~\ref{fig:poly-rho-high}, namely the very small secular increase
with superposed small tidal oscillations, the decompression as the two
NS cores merge and the final exponential growth produced by the
collapse to a BH. Note, however, that the two evolutions are not
exactly the same and that small differences are appreciable both in
the decompression phase and in the post-collapse phase which is
dominated by the dynamics of the torus around the BH (Note that the
different final values in $\rho_{\rm max}$ are due to the different
times at which the AH is first found and which do not coincide for the
two runs; see also the footnote on page~\pageref{footnote1}).

\begin{figure}[t]
\begin{center}
   \includegraphics[angle=-0,width=0.45\textwidth]{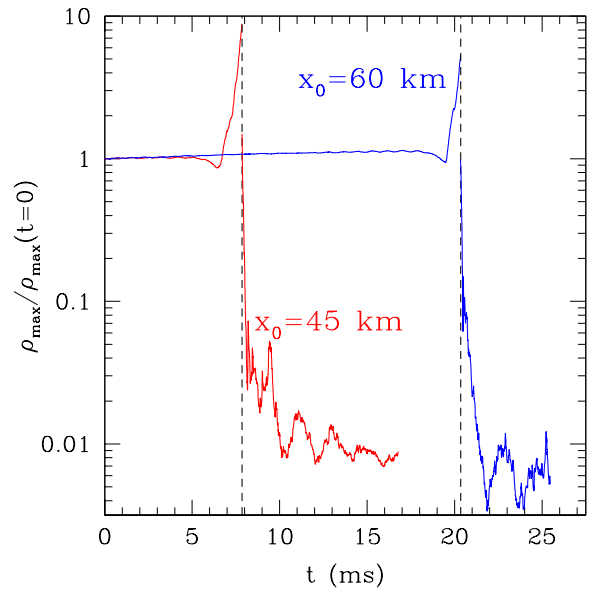}
\end{center}
\vskip -0.5cm
\caption{Evolution of the maximum rest-mass density normalized to its
  initial value for {\bf high-mass} binaries with initial coordinate
  separation of either $45$ or $60\,{\rm km}$. The vertical dashed
  lines denote the time at which an AH was found. The {\bf polytropic}
  EOS was used for the evolutions.}
  \label{fig:rho_pol_high_45_vs_60}
\end{figure}

\begin{figure*}[ht]
\begin{center}
   \includegraphics[width=0.45\textwidth]{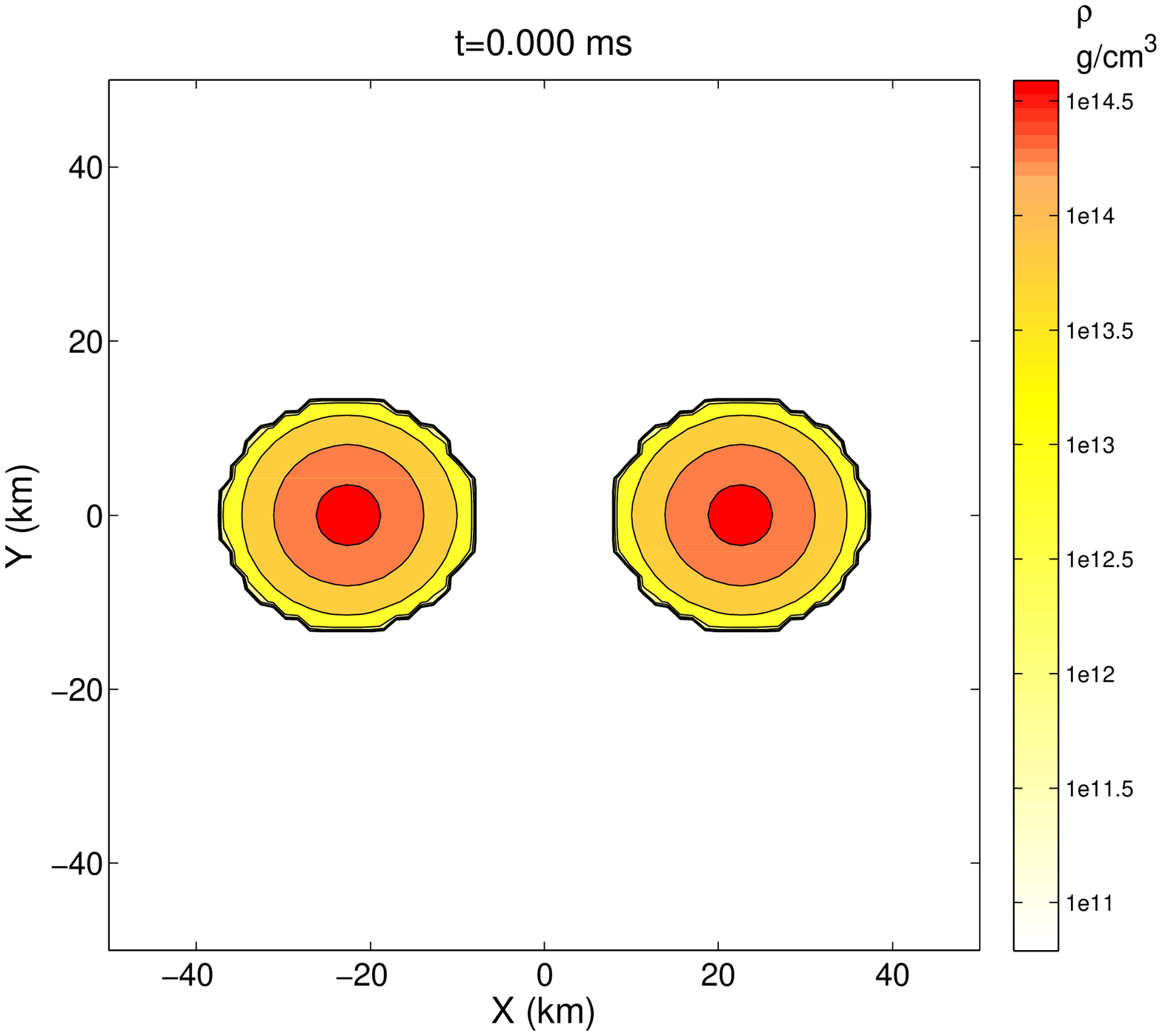}
   \hskip 1.0cm
   \includegraphics[width=0.45\textwidth]{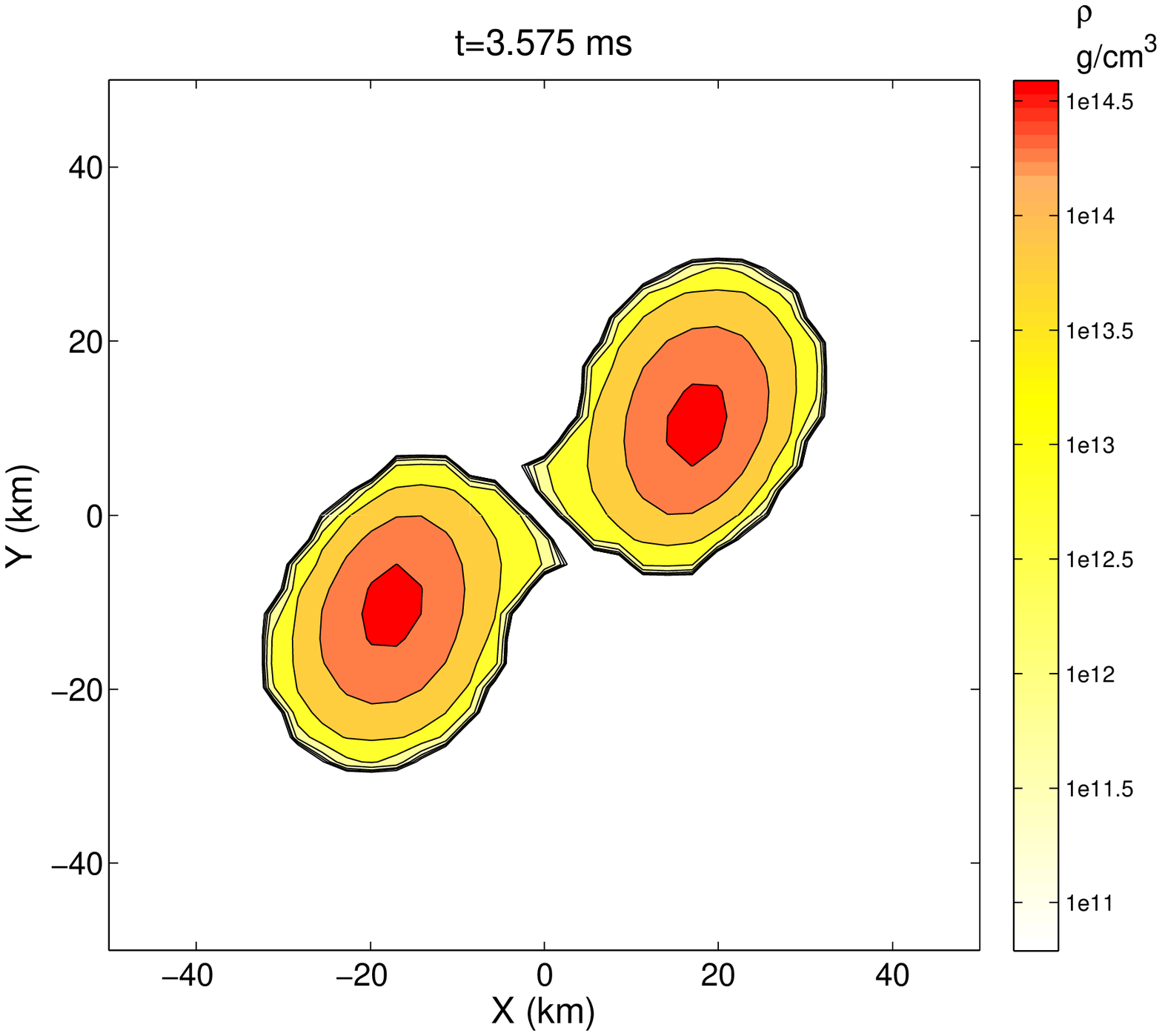}
   \includegraphics[width=0.45\textwidth]{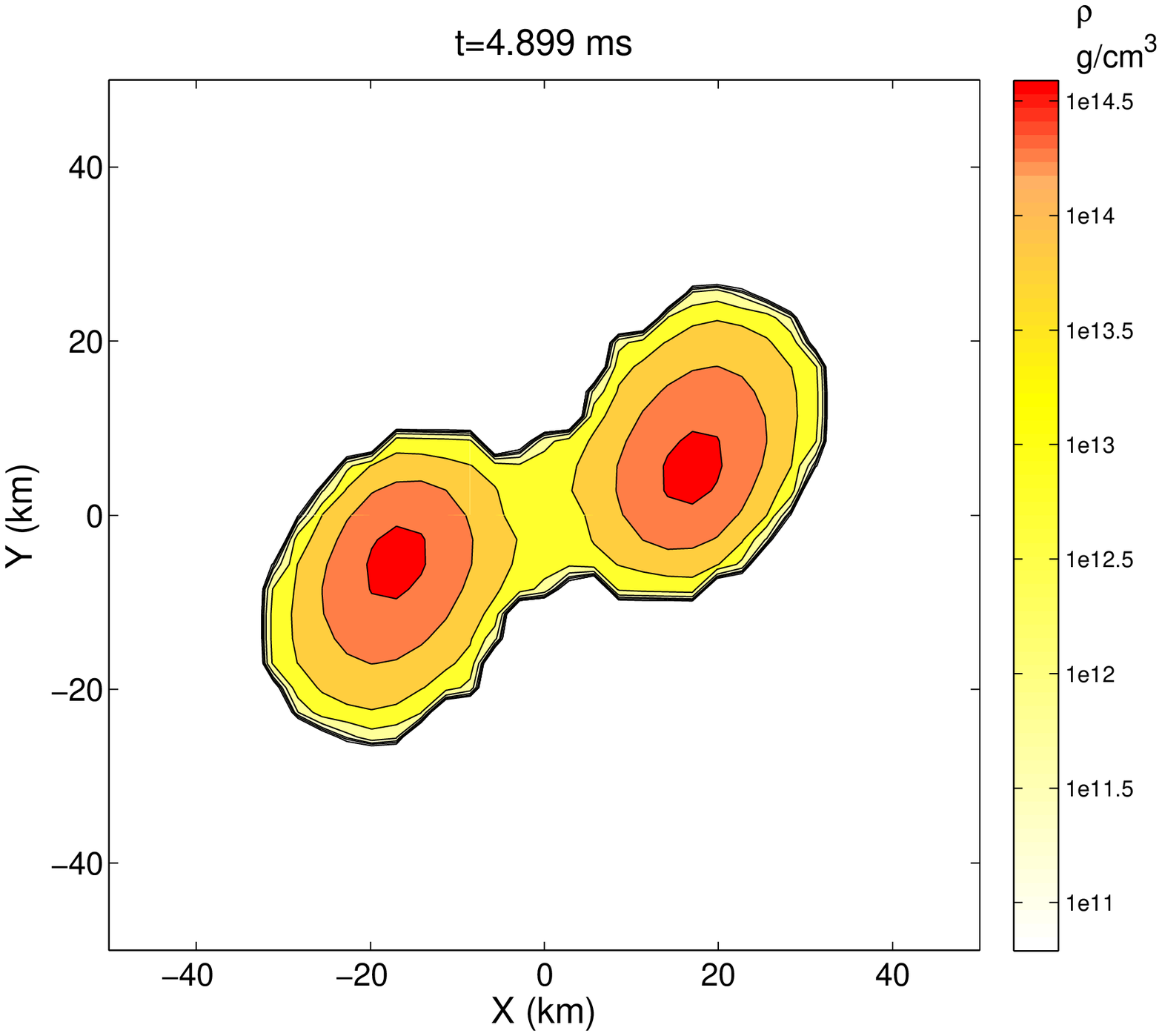}
   \hskip 1.0cm
   \includegraphics[width=0.45\textwidth]{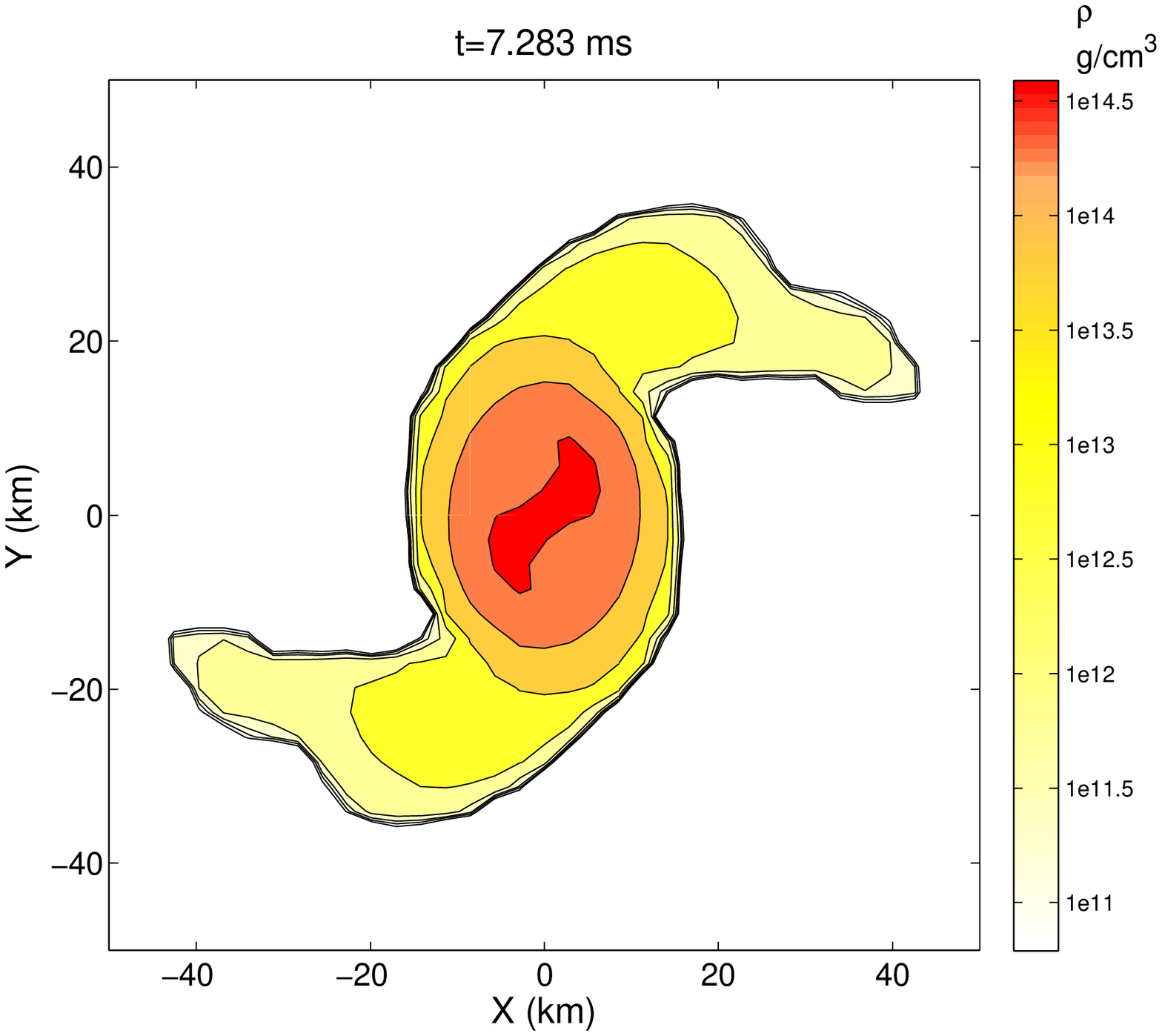}
   \includegraphics[width=0.45\textwidth]{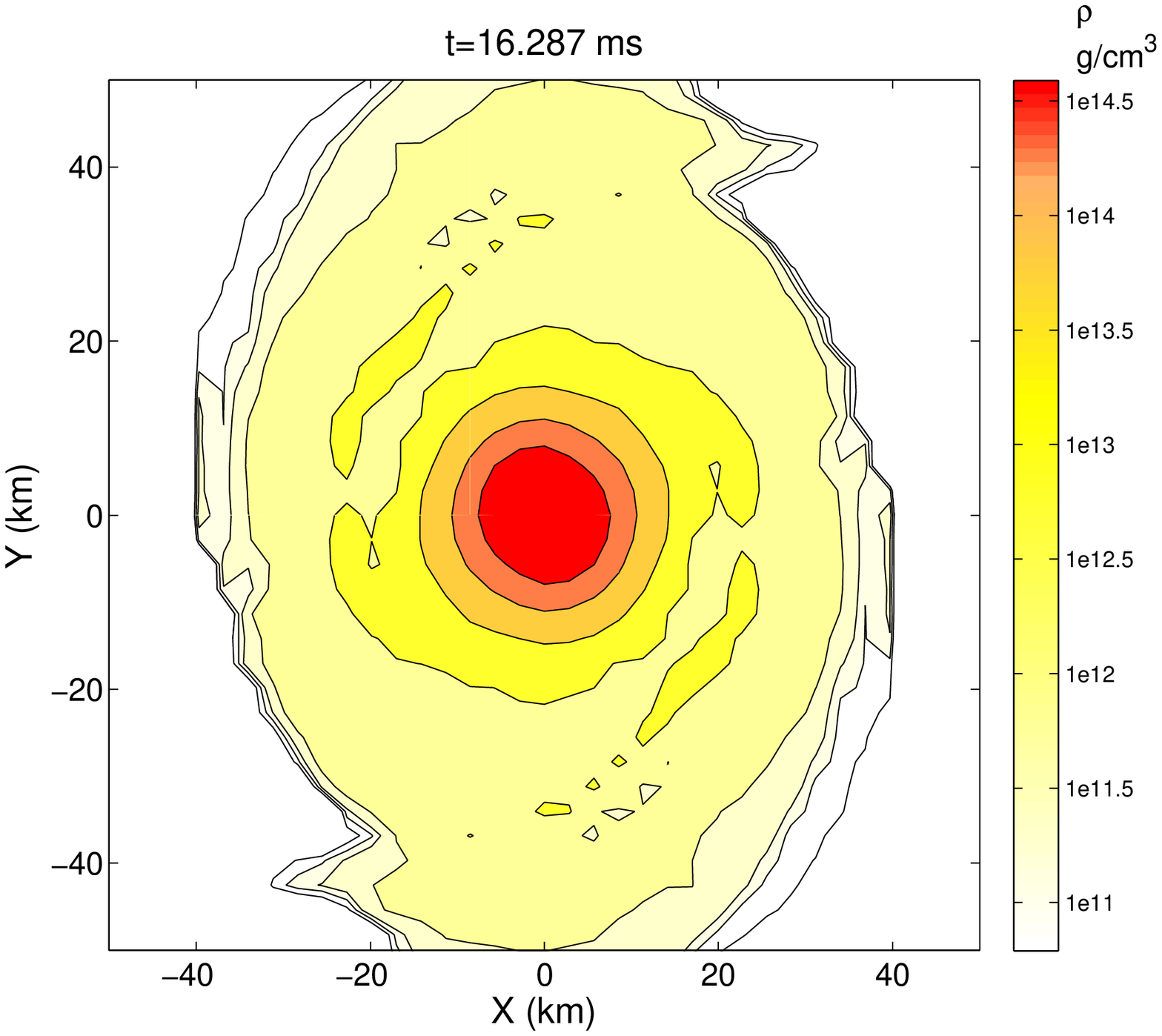}
   \hskip 1.0cm
   \includegraphics[width=0.45\textwidth]{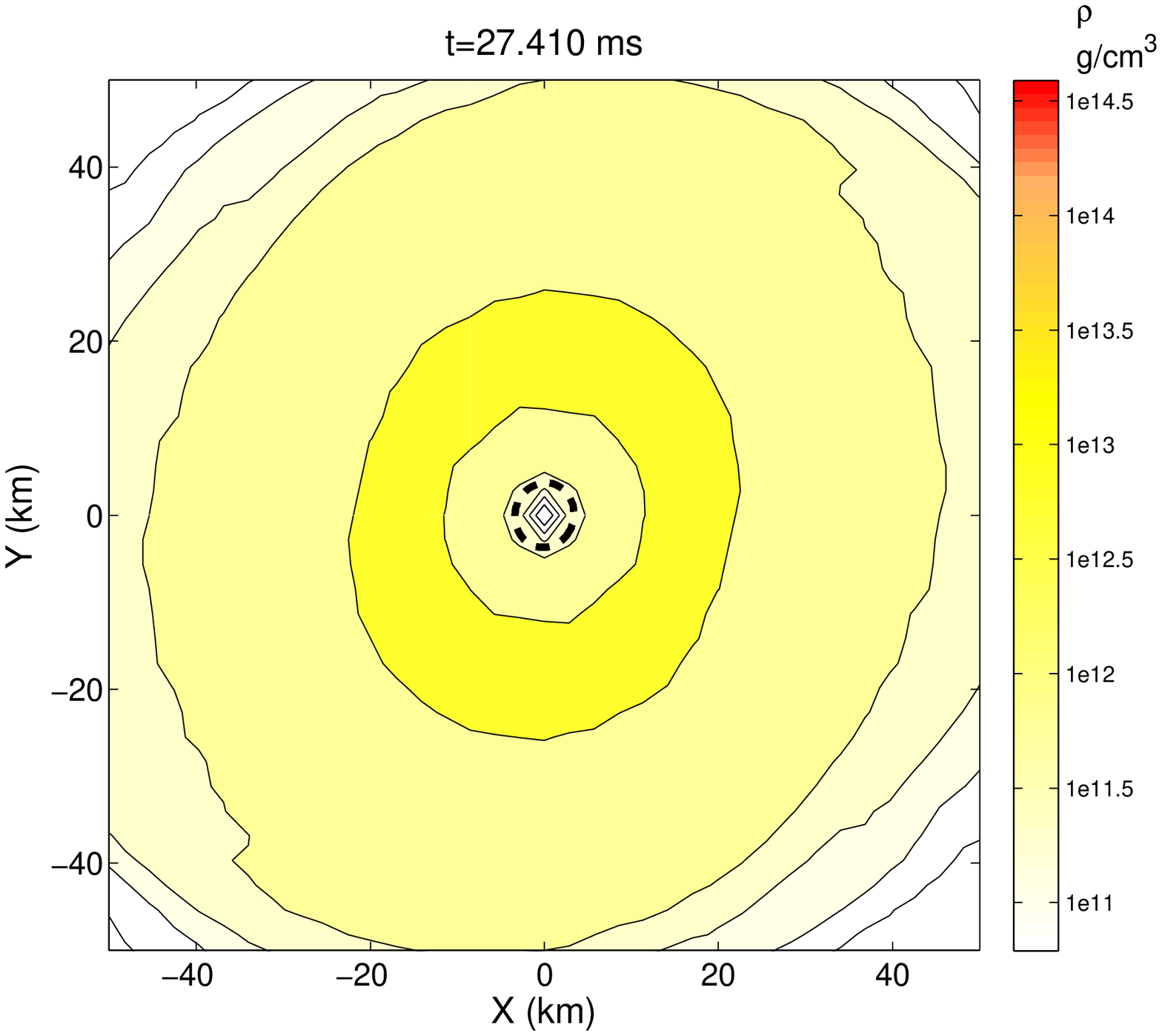}
\end{center}
   \caption{Isodensity contours on the $(x,y)$ (equatorial) plane
     for the evolution of the \textbf{low-mass} binary with the
     \textbf{polytropic} EOS (\ie model $1.46$-$45$-${\rm P}$ in
     Table~\ref{table:ID}). The time stamp in $\ms$ is shown on the
     top of each panel, the color-coding bar is shown
     on the right in units of ${\rm g/cm}^3$ and the thick
     dashed line represents the AH. A high-resolution version of this figure can be found at~\cite{weblink}.\label{fig:rho2D_poly_low_xy}}
\end{figure*}

Although a larger truncation error is to be expected in the case of
the large-separation binary simply because of the larger integration
time, we believe these differences are genuine and reflect the fact
that the initial data used are not invariant under time
translation. Stated differently, the large-separation binary
$1.62$-$60$-${\rm P}$, when evolved down to a separation of $45\ {\rm
  km}$, will not coincide with the equilibrium solution
$1.62$-$45$-${\rm P}$ computed for a quasi-circular binary in
equilibrium at $45\,{\rm km}$. Because these differences are mostly in
the internal structure, the deviations in the evolution become evident
only at and after the merger and are essentially absent in the
pre-merger evolution of both the central-density
(\cf~Fig.~\ref{fig:rho_pol_high_45_vs_60}) and of the waveforms
(\cf~Fig.~\ref{fig:psi4_amp_pol_45_vs_60} in Sect.~\ref{gws_pol}).

Since considerations of this type have never been made before in the
literature and we are not aware of careful comparative studies of this
type, our conclusions require further validation. Work is now in
progress to perform similar simulations with different polytropic
indices. If the differences reported in
Fig.~\ref{fig:rho_pol_high_45_vs_60} are indeed physical, they will
also show variations in case stiffer or softer EOS are considered and
they should indeed disappear for perfectly incompressible stars. The
results of this analysis will be reported in a future
work~\cite{BGR08}.

\subsection{Polytropic EOS: low-mass binary}

We next consider the evolution of the low-mass binary evolved with the
polytropic EOS, \ie model $1.46$-$45$-${\rm P}$ in
Table~\ref{table:ID}. As for the high-mass binary, we first show in
Fig.~\ref{fig:rho2D_poly_low_xy}, the representative isodensity
contours on the $(x,y)$ plane, with the time stamp being shown on the
top of each panel and with the color-coding bar being shown on the
right in units of ${\rm g/cm}^3$. Note that because the evolution is
different in this case, the times at which the isodensity contours are
shown are different from those in Fig.~\ref{fig:rho2D_poly_high_xy}.

\begin{figure}[t]
\begin{center}
   \includegraphics[angle=-0,width=0.45\textwidth]{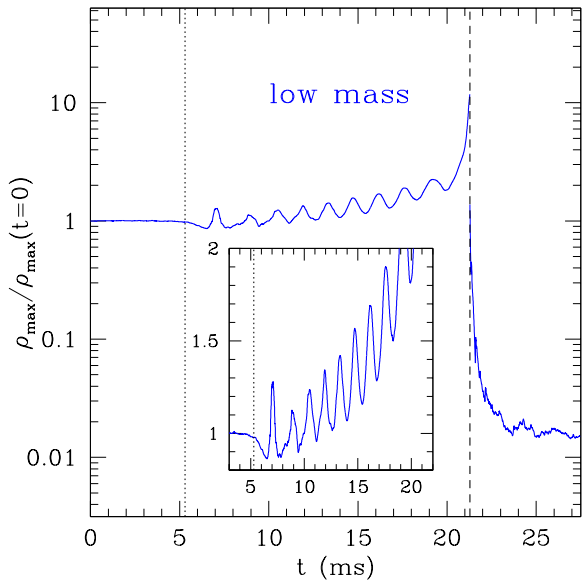}
\end{center}
\vskip -0.5cm
    \caption{The same as in Fig.~\ref{fig:poly-rho-high} but for the
      \textbf{low-mass} binary. Note that the merger time is
      essentially the same as for the high-mass binary but there is a
      long delay in the collapse and the onset of quasi-harmonic
      oscillations in the HMNS. The binary has been evolved using the
      \textbf{polytropic} EOS. Note that the non-normalised value of the maximum
      rest-mass density at $t=0$ is $4.58\times10^{14}\,{\rm g}/{\rm cm^3}$ (see
      Table~\ref{table:ID}).}
 \label{fig:poly-rho-low}
\end{figure}

Since the mass difference with model $1.62$-$45$-${\rm P}$ is less
than $10\%$, one expects that the orbital dynamics \textit{before} the
merger are essentially the same. Indeed this is what our simulations
indicate and differences appear only as higher-order effects, such as
in the strength of the tidal waves (see
Fig.~\ref{fig:rho2D_poly_low_xy}). However, despite the small
difference in mass, the evolution \textit{after} the merger is
considerably different. This is nicely summarized in
Fig.~\ref{fig:poly-rho-low} which shows that the merger time is
essentially the same as for the high-mass binary (\ie $\sim
5.3\,\ms$), but the subsequent evolution does not lead to the prompt
formation of a BH. Rather, the HMNS is still quite far from the
instability threshold and undergoes a number of quasi-periodic
oscillations (\cf~Fig.~\ref{fig:poly-rho-low}), which have almost
constant amplitude in the central rest-mass density.

\begin{figure}[t]
\begin{center}
   \includegraphics[angle=-0,width=0.45\textwidth]{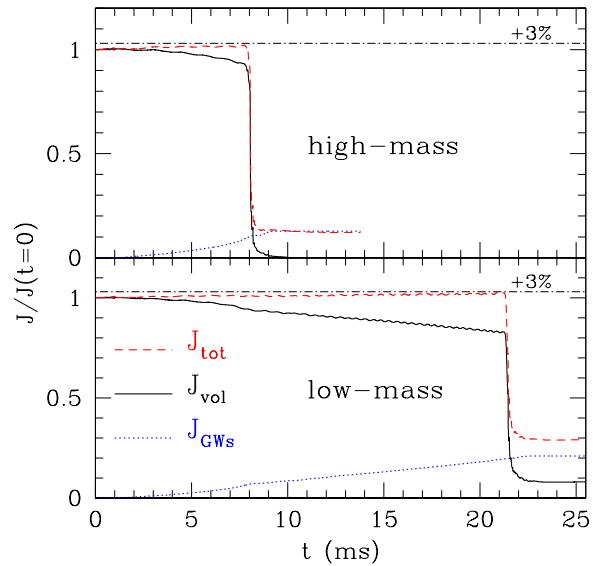}
\end{center}
\vskip -0.5cm
   \caption{Conservation of the total angular momentum for the
     \textbf{high-mass} binary (upper plot) and the \textbf{low-mass}
     one (lower plot). Indicated with different lines are the computed
     values of the volume-integrated angular momentum (solid line), of
     the angular momentum lost to gravitational waves (dotted line)
     and of their sum (dashed line). The dot-dashed line marks a $3\%$
     error.}
  \label{fig:ang_mom_pol}
\end{figure}

A more careful analysis reveals that the core of the HMNS undergoes
violent non-axisymmetric oscillations, with the development of an
overall $m=2$ deformation, \ie a bar, as the system tries to reach a
configuration which is energetically favourable through the
rearrangement of the angular-momentum distribution. The bar-deformed
star has an initial value of the ratio of the kinetic energy to the
binding energy $T/|W|\simeq 0.22$, which remains roughly constant and
only slightly decreases to $0.19$ until the time of the collapse. As
the bar rotates, it also loses large amounts of angular momentum
through gravitational radiation and this is reported in
Fig.~\ref{fig:ang_mom_pol}, which shows the evolution of the angular
momentum as normalized to the initial value. The top panel, in
particular, refers to the high-mass binary, while the bottom one to
the low-mass binary (a more detailed discussion of the losses of
energy and angular momentum will also be presented in
Sect.~\ref{sec:eam_losses}). Indicated with different lines are the
computed values of the volume-integrated angular momentum [solid line,
  computed with the integral (\ref{AngMomFormula})], of the angular
momentum lost to gravitational waves (dotted line) and of their sum
(dashed line). In both cases the slight secular increase is due to the
truncation error and is at most of $3\%$ over more than $20\ {\rm ms}$
(\cf~dot-dashed line). A very similar figure can be made for the one of
the ADM mass, whose conservation is even higher (the error is below
$1\%$). For compactness we will not show such a figure here.

Note that the loss of angular momentum is of $\sim 5\%$ of the total
initial angular momentum during the inspiral and merger, but becomes
much larger once the HMNS has been produced and while the bar-deformed core
rotates. Indeed, in the case of the large-mass binary this
loss increases to $\sim 13\%$ after the BH quasi-normal ringing, while
it becomes as large as $\sim 22\%$ for the low-mass binary. Overall,
the post-merger evolution for the low-mass binary is rather long and
spans over $\sim 16\,\ms$. The inset in Fig.~\ref{fig:poly-rho-low}
shows that during this time the maximum rest-mass density oscillates
but it also increases secularly of a factor of about $2$. This is due
to the fact that as the HMNS loses angular momentum, its centrifugal
support is also decreased and thus it reaches more and more compact
configurations. At one point the HMNS is past the threshold of the
quasi-radial instability for the collapse to a BH, which takes place
at $\sim 20\,\ms$ (\cf~Fig.~\ref{fig:poly-rho-low}), with an AH being
found at $t=21.3\,\ms$.

\begin{figure}[t]
\begin{center}
   \includegraphics[width=0.45\textwidth]{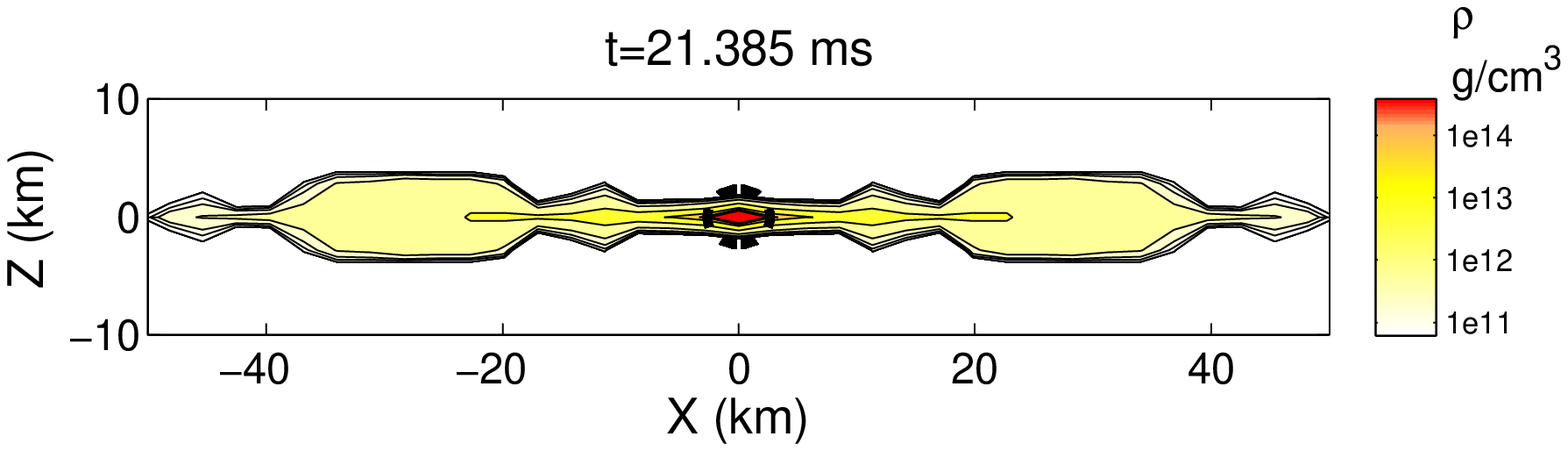}
   \includegraphics[width=0.45\textwidth]{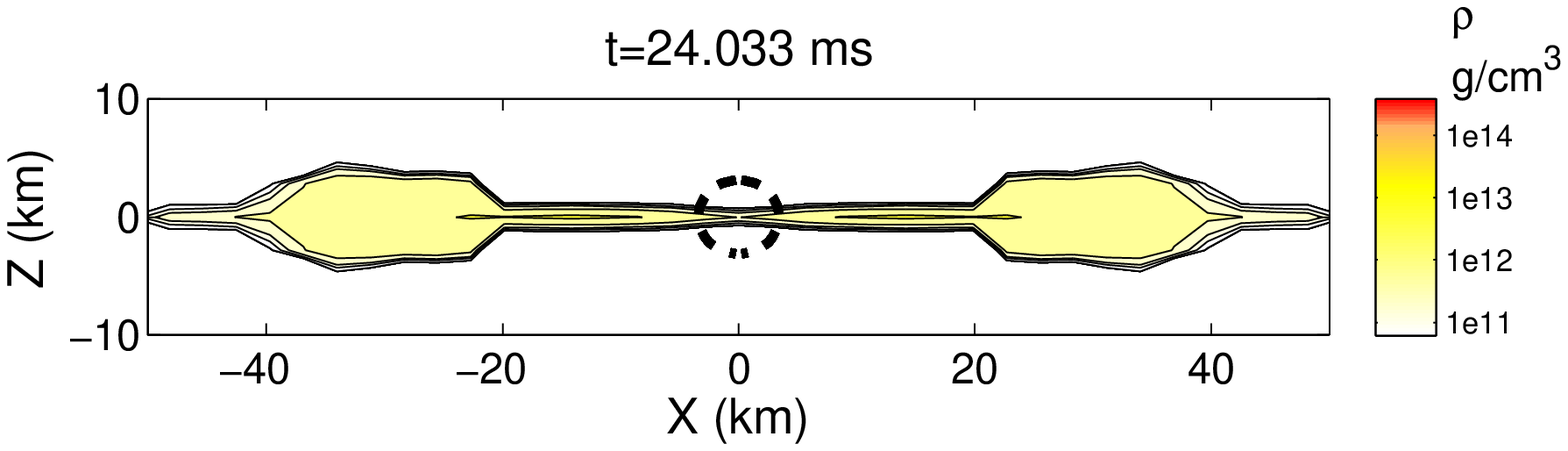}
   \includegraphics[width=0.45\textwidth]{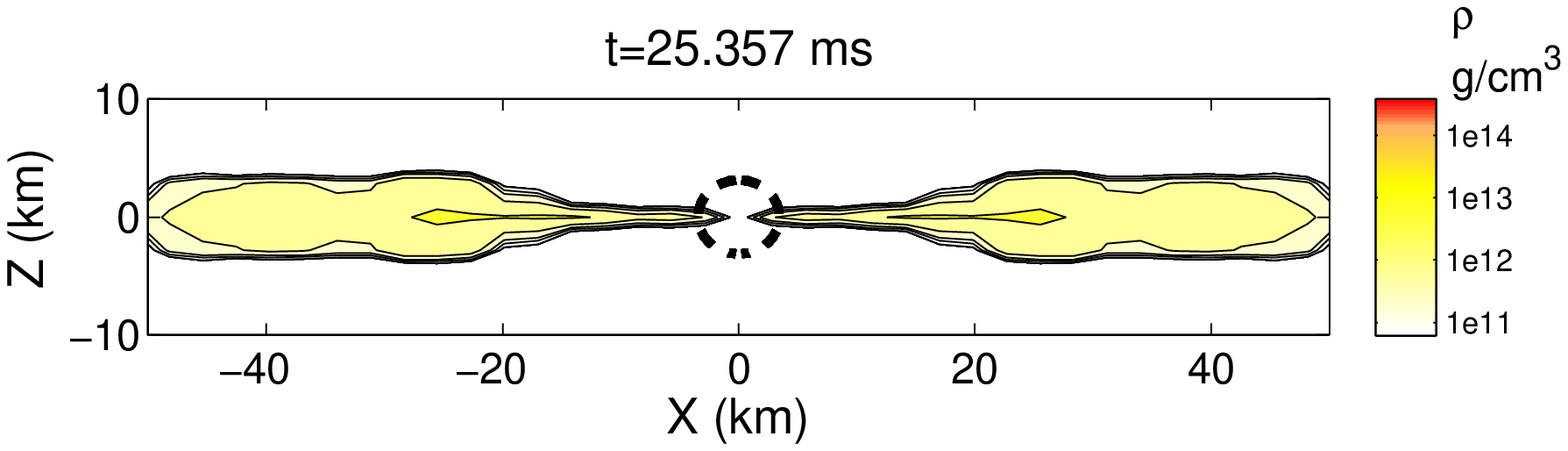}
   \includegraphics[width=0.45\textwidth]{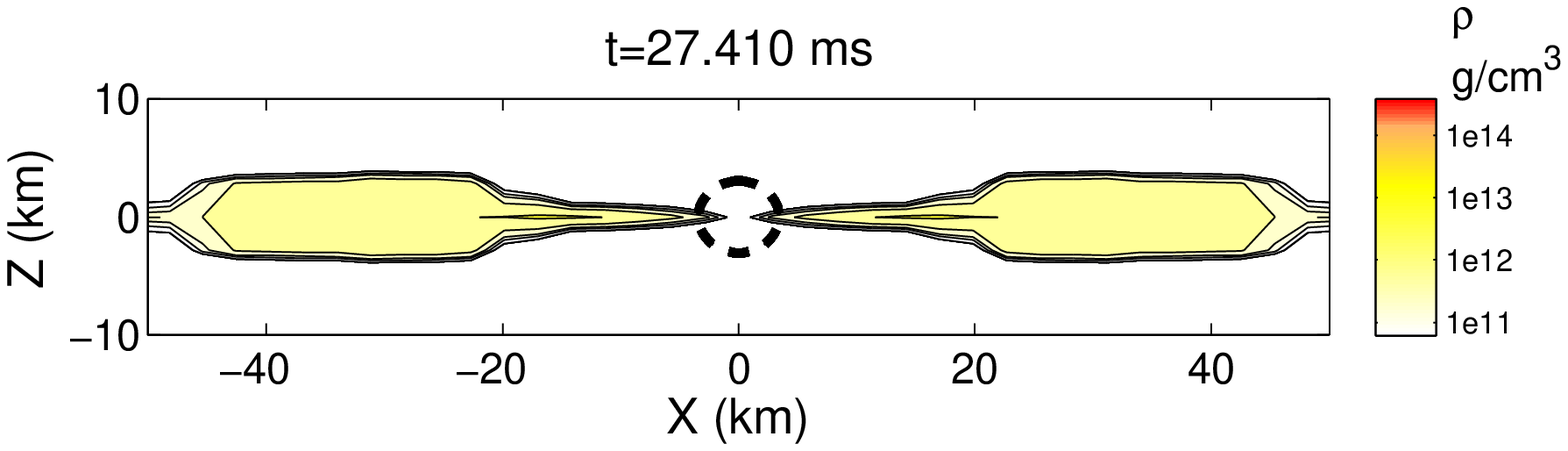}
\end{center}
\vskip -0.5cm
   \caption{Isodensity contours on the $(x,z)$ plane, highlighting the
     formation of a torus surrounding the central BH, whose AH
     is indicated with a thick dashed line. The data
     refers to the \textbf{low-mass} binary evolved with the
     \textbf{polytropic} EOS
     (\cf~Fig.~\ref{fig:BH_short_torus_poly_high}). A high-resolution version of this figure can be found at~\cite{weblink}.\label{fig:BH_short_torus_poly_low}}
\end{figure}

Also in this case, a large amount of matter with sufficient angular
momentum is found to be orbiting outside the BH in the form of an
accretion disc. Differently from the high-mass binary, however, the
torus here has a larger average rest-mass density (between $10^{12}$
and $10^{14}\ {\rm g/cm^3}$; see evolution of $\rho_{\rm max}$ in
Fig.~\ref{fig:poly-rho-low} after the AH), a larger extension in the
equatorial plane (between $20$ and $50\,{\rm km}$) but a comparable
vertical extension (below $10\,{\rm km}$). It also has a larger baryon
mass, which is initially $(M_{_{\rm T}})_0 = 0.1\,M_{\odot}$ and
becomes $(M_{_{\rm T}})_{3\,{\rm ms}} = 0.0787\, M_{\odot}$ after
$3\,\ms$ (see footnote on page~\pageref{footnote1} and
Table~\ref{table:FD}). The dynamics of the torus are summarized in
Fig.~\ref{fig:BH_short_torus_poly_low}, which shows the isodensity
contours on the $(x,z)$ plane; note that the panels refer to times
between $21.4\ {\rm ms}$ and $27.4\ {\rm ms}$ and thus to a stage in
the evolution which is between the last two panels of
Fig.~\ref{fig:rho2D_poly_low_xy}. A simple comparison between
Figs.~\ref{fig:BH_short_torus_poly_high}
and~\ref{fig:BH_short_torus_poly_low} is sufficient to capture the
differences between the tori in the two cases and also to highlight that
for a polytropic EOS the \textit{high-mass} binary produces a
\textit{lower-mass} torus (\cf~Table~\ref{table:FD} and see the
discussion in Sect.~\ref{sec:eam_losses}).

In analogy with what is seen for the high-mass binary, the torus has an
overall axisymmetric structure and is far from equilibrium. As a
result, it is subject to large oscillations, mostly in the radial
direction, at a frequency close to the epicyclic one. A more detailed
analysis of this will be presented in a companion paper~\cite{BGR08}.

Using again the isolated-horizon formalism we have estimated that the
final BH has in this case a mass $M_{_{\rm BH}}=2.60\,M_{\odot}$, spin
$J_{_{\rm BH}} = 5.24\,M^2_{\odot}=4.61\times10^{49}{\rm g}\,{\rm cm}^2{\rm s}^{-1}$ 
and thus a dimensionless spin $a\equiv J_{_{\rm BH}}/M^2_{_{\rm BH}} =0.76$
(\cf~Table~\ref{table:FD}). Interestingly, the dimensionless spin is 
lower in the low-mass binary.

It should also be remarked that the long time interval before which
the collapse takes place has prevented previous studies from the
complete calculation of the dynamics of NS binaries which would not
lead to the \textit{prompt} formation of a BH. The investigations of
refs.~\cite{Shibata06a,Anderson2007,Anderson2008}, for instance, are
limited to a few $\ms$ after the merger and should be contrasted with
the evolutions reported here that cover a timescale of $\sim 30\,\ms$,
also for the additional calculation of the gravitational waves. As a
result, our simulations represent, within an idealized treatment of
the matter, the first complete description of the inspiral and merger
of a NS binary leading to the \textit{delayed} formation of a BH.

\subsection{Ideal-fluid EOS: high-mass binary}
\label{IF_binaries_hm}

\begin{figure*}[ht]
\begin{center}
   \includegraphics[width=0.45\textwidth]{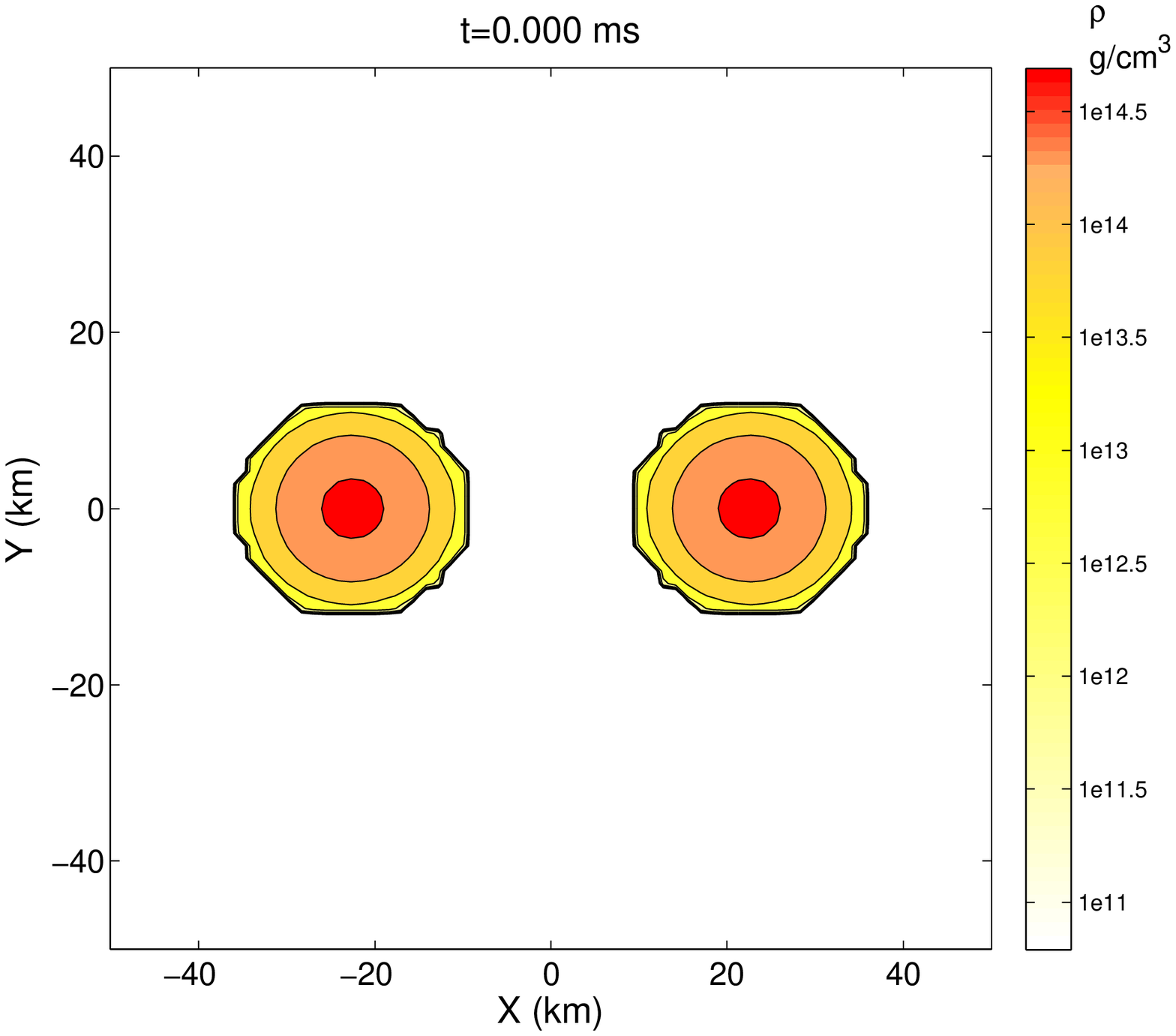}
   \hfill
   \includegraphics[width=0.45\textwidth]{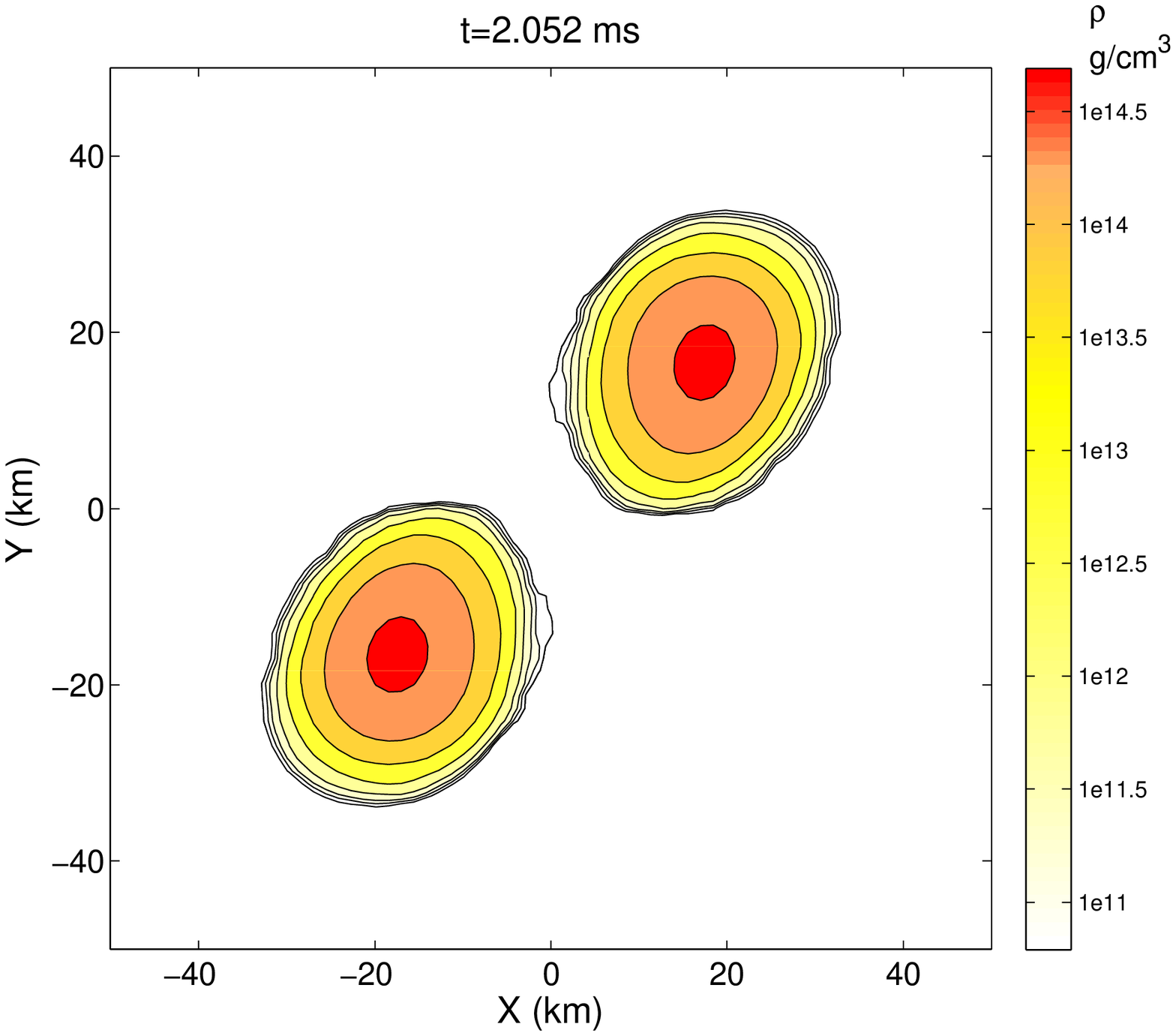}
   \includegraphics[width=0.45\textwidth]{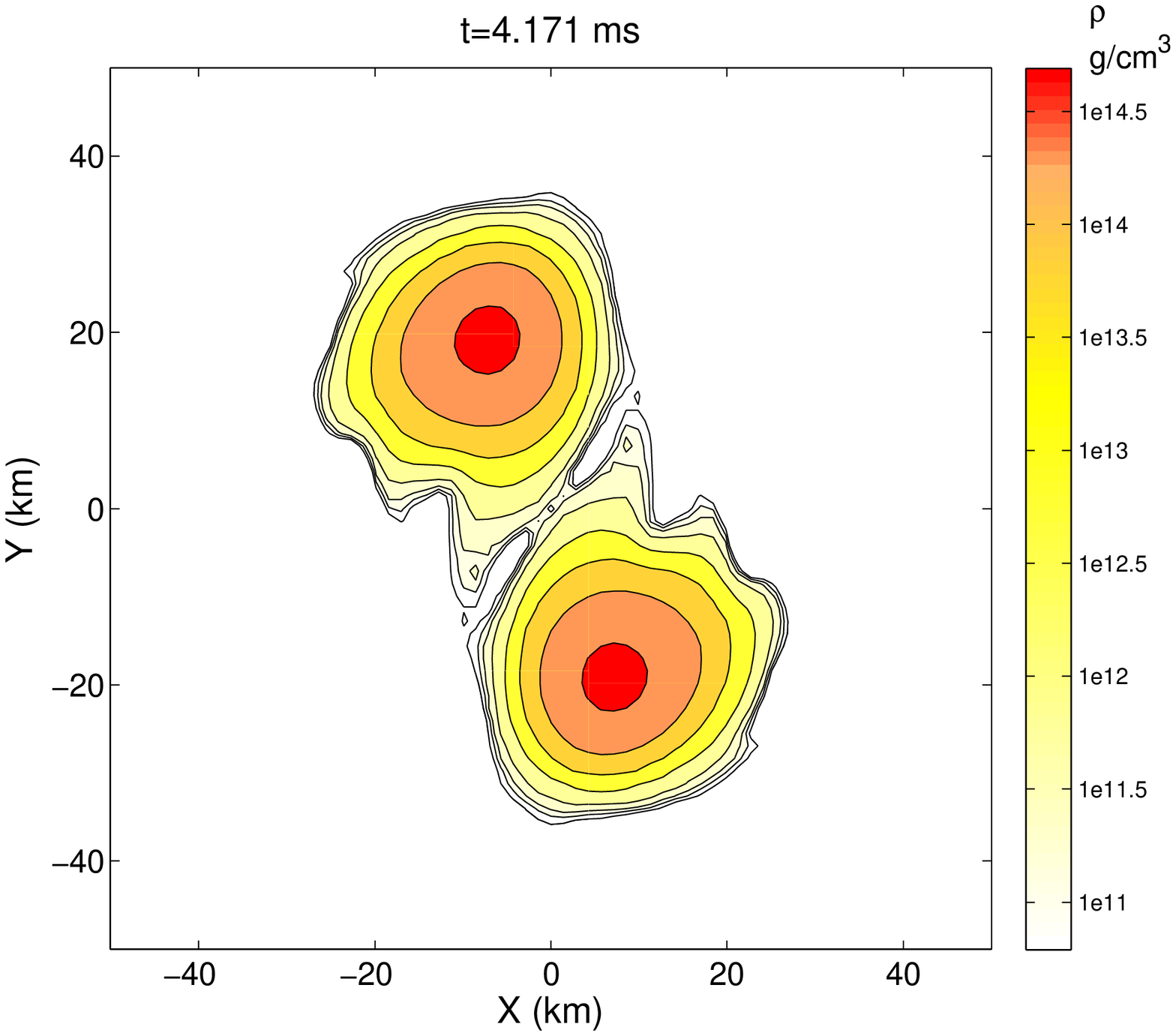}
   \hfill
   \includegraphics[width=0.45\textwidth]{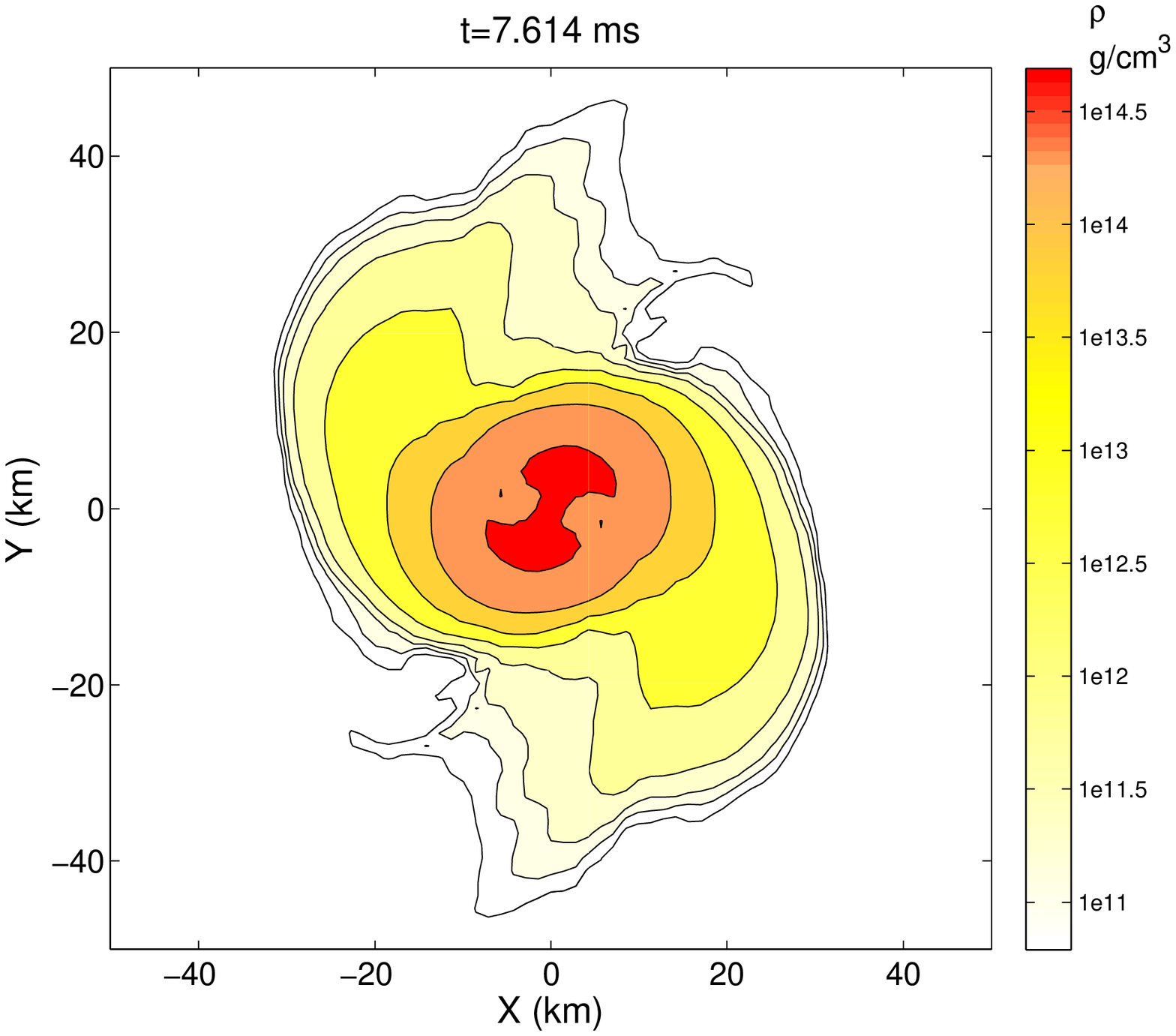}
   \includegraphics[width=0.45\textwidth]{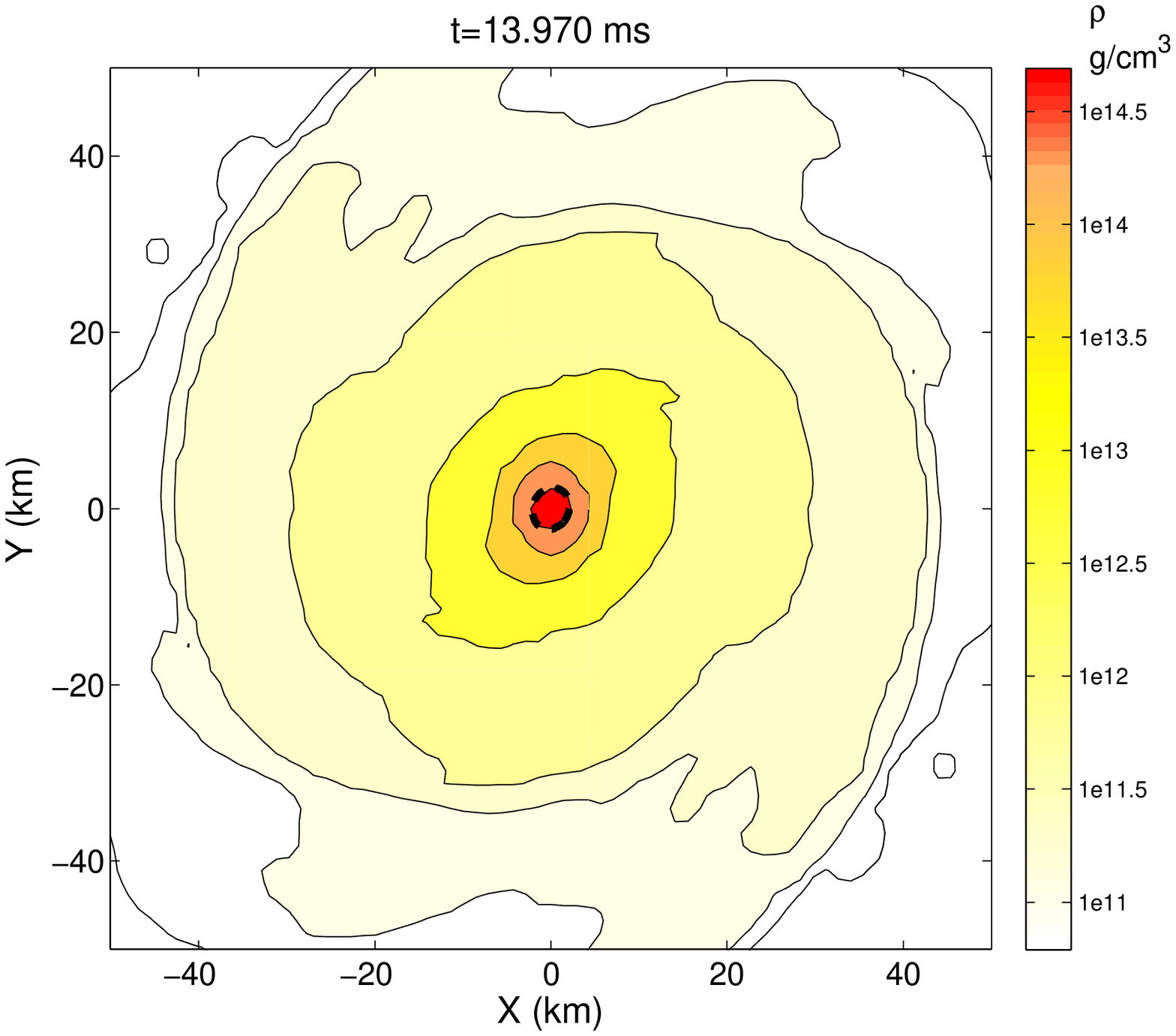}
   \hfill
   \includegraphics[width=0.45\textwidth]{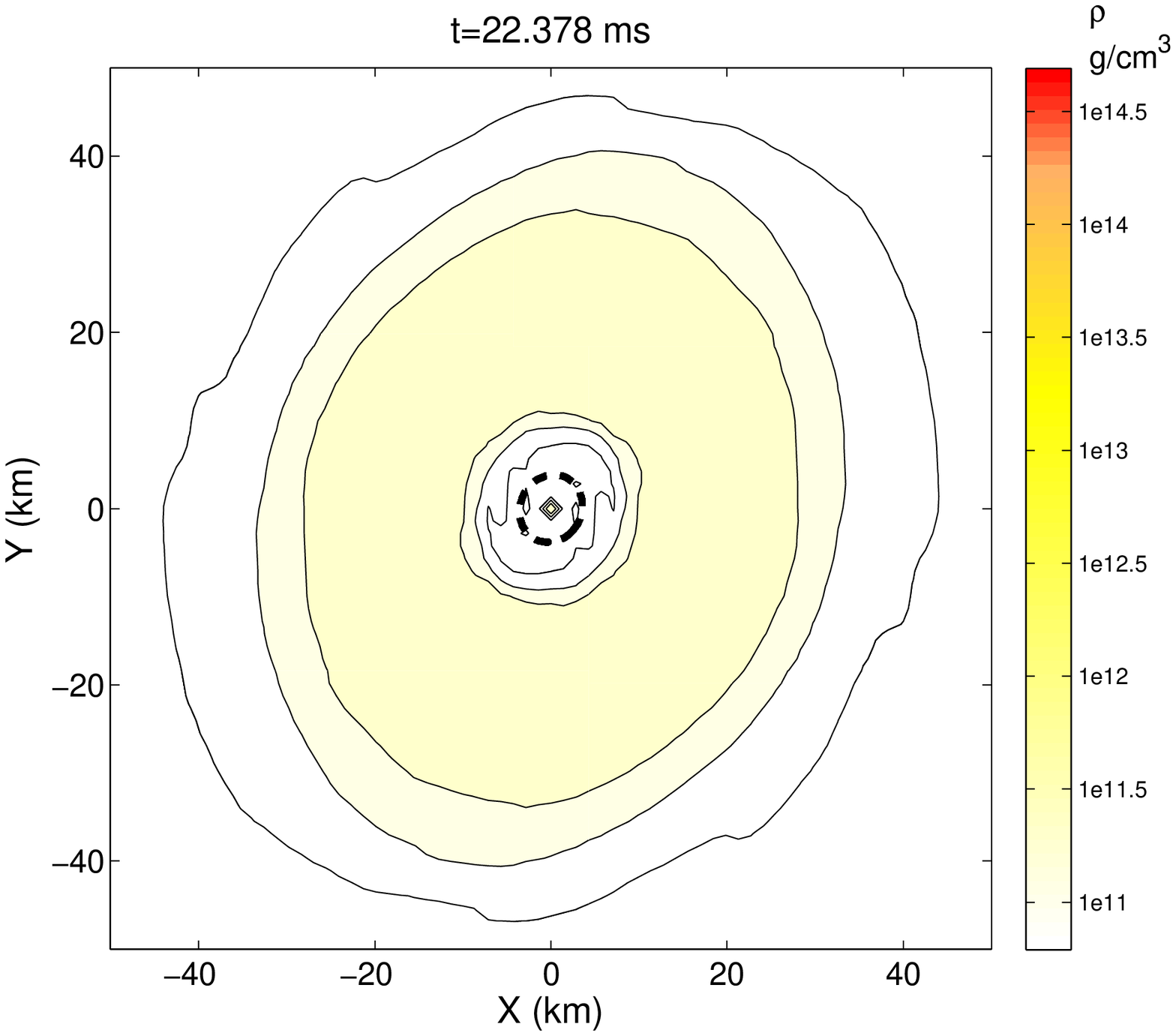}
\end{center}
\vskip -0.5cm
   \caption{Isodensity contours on the $(x,y)$ (equatorial)
     plane for the evolution of the \textbf{high-mass} binary with the
     \textbf{ideal-fluid} EOS (\ie model $1.62$-$45$-${\rm IF}$ in
     Table~\ref{table:ID}). The time stamp in $\ms$ is shown on the
     top of each panel, the color-coding bar is shown
     on the right in units of ${\rm g/cm}^3$ and the thick
     dashed line represents the AH. A high-resolution version of this figure can be found at~\cite{weblink}.\label{fig:BH_IF_short_xy}}
\end{figure*}

We now move on to discussing the dynamics of binary inspiral and
merger when the other EOS, the ideal-fluid one in eq.~\eqref{id
  fluid}, is used. As discussed in Sect.~\ref{hd_eqs}, while this is
an idealized and analytic EOS, it has the important property of being
non-isentropic and thus of allowing for the change of the thermal part
of the internal energy density (or, equivalently, of the
temperature). As we will show in the remainder of this Section, this
difference can lead to significant differences in the properties
and dynamics of the HMNS produced by the merger.  More specifically,
we concentrate on the evolution of model $1.62$-$45$-${\rm IF}$ in
Table~\ref{table:ID}, namely a binary in which each NS has a baryon
mass of $M_b=1.625\,M_{\odot}$ and an initial coordinate separation of
$45\,\km$. As for the previous binaries, we collect in
Fig.~\ref{fig:BH_IF_short_xy} some representative isodensity contours
on the equatorial plane.

\begin{figure}[ht]
\begin{center}
   \includegraphics[angle=-0,width=0.45\textwidth]{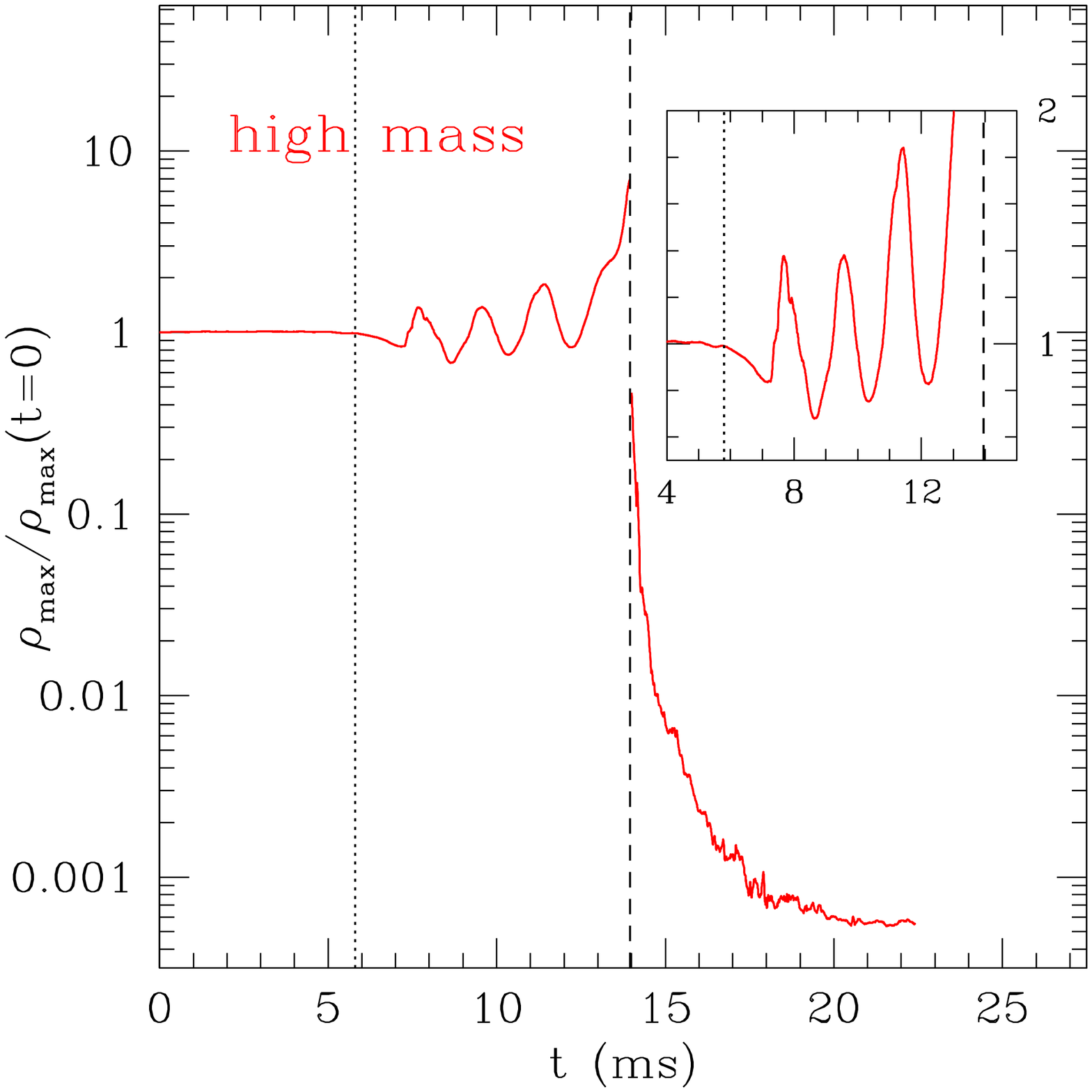}
\end{center}
\vskip -0.5cm
   \caption{Evolution of the maximum rest-mass density normalized to
     its initial value for the \textbf{high-mass} binary using the
     \textbf{ideal-fluid} EOS. Indicated with a dotted vertical line
     is the time at which the binary merges, while a vertical dashed
     line shows the time at which an AH is found. After this time, the
     maximum rest-mass density is computed in a region outside the
     AH. This figure should be compared with Fig.~\ref{fig:poly-rho-high}.}
  \label{fig:if-rho-high}
\end{figure}

As one would expect from PN considerations (which suggest that
finite-size effects are expected at orders equal or smaller than the
fifth~\cite{Blanchet02}), the bulk dynamics of the binary before the
merger are essentially identical to the one already discussed for
model $1.62$-$45$-${\rm P}$ and small differences are appreciable only
in the low-density layers of the stars, where the different tidal
fields cause comparatively larger amounts of matter to be stripped
from the surface; this can be appreciated by comparing the second and
third panels of Figs.~\ref{fig:rho2D_poly_high_xy}
and~\ref{fig:BH_IF_short_xy}. Indeed, this is a subtle point which is
worth remarking: when using the ideal-fluid EOS, the evolution
\textit{before} the merger (\ie during the inspiral) is not
isentropic. This is because small shocks are produced in the very
low-density layers of the stars as these orbit. These small shocks
channel some of the orbital kinetic energy into internal energy,
leading to small ejections of matter (\ie $\sim 10^{-6} M_{\odot}$),
and are thus responsible for the slight differences in the
inspiral. We also note that these shocks would appear quite
independently of the fact that the NSs are surrounded by an atmosphere
as they represent the evolution of small sound waves that, propagating
from the central regions of the stars, steepen as they move outwards;
we have checked that essentially identical results are obtained when
changing the threshold for the atmosphere of one or more orders of
magnitude (a discussion of this process for isolated stars evolved
within the Cowling approximation can be found in
ref.~\cite{Stergioulas04} and in ref.~\cite{Dimmelmeier06} for the
extension to a dynamical spacetime).

Besides this small difference, the merger takes places at almost the
same time as for model $1.62$-$45$-${\rm IF}$, namely after about
$2.5$ orbits, or equivalently after $5.8\ {\rm ms}$ from the beginning
of the simulation. However, the post-merger evolution of the HMNS is
considerably different. This is nicely summarized in
Fig.~\ref{fig:if-rho-high}, which reports the evolution of the maximum
rest-mass density normalized to its initial value and which, after the
AH is found, refers to the region outside the AH.
In this case shocks are allowed to form and the HMNS \textit{does not}
collapse promptly to a BH but, rather, undergoes very large
oscillations with variations of $100\%$ in the maximum of the
rest-mass density (\cf~Fig.~\ref{fig:if-rho-high}). These oscillations
are the result of what appears to be a dynamical bar-mode instability
which develops and is suppressed at least four times during the
post-merger phase. More specifically, after the first initial merger
at $t\sim 5\,\ms$, the two stellar cores break up again to produce a
bar-deformed structure, which rotates for more than a period before
disappearing as the cores merge again. This process takes
place four times and the merged object becomes increasingly more compact
as it loses angular momentum and thus spins progressively faster. This
behavior is clearly imprinted in the gravitational-wave signal as we
will illustrate in Sect.~\ref{gws_IF}.


%
\begin{figure}[t]
\begin{center}
   \includegraphics[angle=-0,width=0.45\textwidth]{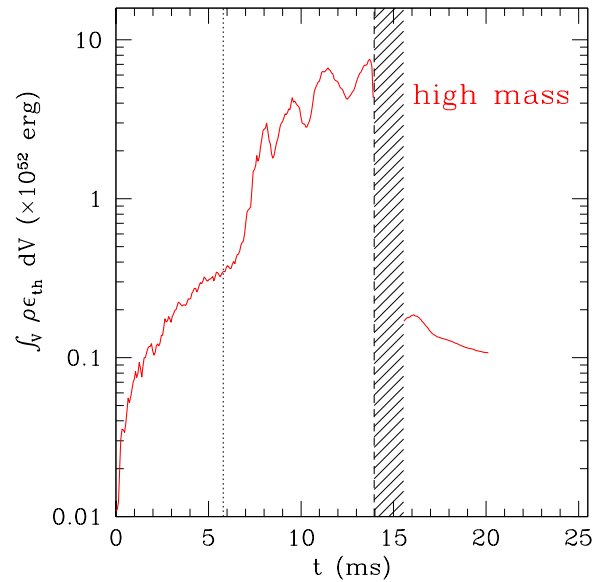}
\end{center}
\vskip -0.5cm
   \caption{Evolution of the coordinate volume-integral of the thermal
     part of the internal energy density. Note the secular increase in
     the thermal energy after the merger (vertical dotted line). The
     shaded area refers to a window in time in which the calculation
     of $\epsilon_{\rm th}$ via eq.~\eqref{eps_th} becomes inaccurate
     as a result of the steep gradients in the hydrodynamical
     variables developing inside the AH.}
  \label{fig:if-eps-high}
\end{figure}

Together with these large variations, the rest-mass density also
experiences a secular growth similar to the one already discussed for
the low-mass polytropic binary and, as discussed before, the increased
compactness eventually leads, at $t\sim 14\,\ms$, to the collapse to a
rotating BH. The use of the isolated-horizon formalism reveals that in
this case the final BH has a mass $M_{_{\rm BH}}=2.94\,M_{\odot}$,
spin $J_{_{\rm BH}} = 7.3\,M^2_{\odot}=6.4\times10^{49}{\rm g}\,{\rm
  cm}^2{\rm s}^{-1}$ and thus a dimensionless spin $a\equiv J_{_{\rm
    BH}}/M^2_{_{\rm BH}} =0.85$ (\cf~Table~\ref{table:FD}).

The explanation for this behavior in the post-merger phase and the
appearance, also at high masses, of a \textit{delayed} collapse to BH,
can be found by considering the component of the specific internal
energy which is produced by the shock heating. This can be done by
splitting the specific internal energy $\epsilon$ into a cold
component $\epsilon_{\rm cold} = {K \rho^{\Gamma-1}}/({\Gamma-1})$ and
into a thermal one $\epsilon_{\rm th}$, defined
as~\cite{ShibataTaniguchi2008}
\begin{equation}
\label{eps_th}
\epsilon_{\rm th} = \epsilon - \epsilon_{\rm cold} = 
\frac{p}{\rho(\Gamma-1)} - \frac{K \rho^{\Gamma-1}}{\Gamma-1}
\,.
\end{equation}
[Note that eq.~(13) of ref.~\cite{ShibataTaniguchi2008} contains a
  typo for the expression of $\epsilon_{\rm cold}$, which is corrected
  in expression~\eqref{eps_th}]. Fig.~\ref{fig:if-eps-high} then
reports the evolution of the coordinate volume-integral of the thermal
part of the internal energy density $\int_V \rho \epsilon_{\rm th}
dx^3$, which increases secularly soon after the merger (vertical
dotted line) [Indicated with the shaded area is the window in time in
  which the calculation of $\epsilon_{\rm th}$ via eq.~\eqref{eps_th}
  becomes inaccurate as a result of the steep gradients in the
  hydrodynamical variables developing inside the AH]. We recall that
in the case of an isentropic EOS (such as the polytropic EOS),
$\epsilon_{\rm th}=0$ and the quantity $\epsilon/\rho^{\Gamma-1}$ is
constant and proportional to the specific entropy of the
system. However, with a non-isentropic EOS (such as the ideal-fluid
EOS), entropy increases across shocks and the latter are clearly
present during the merger, thus leading to a local and global increase
of the specific internal energy, namely of $\epsilon_{\rm th}$. As a
result, soon after the merger (vertical dotted line in
Fig.~\ref{fig:if-eps-high}), the HMNS from an ideal-fluid high-mass
binary can rely on an additional pressure support, which allows it to
balance the gravitational forces at least for a few additional
$\ms$. Stated differently, the shocks produced at the merger are
responsible for a local and global increase of the temperature, which
will produce a global expansion of the HMNS and thus a reduction of
its compactness. The overall smaller compactness caused by the
increased internal energy can be appreciated by comparing the fourth
and fifth panels of Figs.~\ref{fig:rho2D_poly_high_xy}
and~\ref{fig:BH_IF_short_xy}.

A simple estimate for the temperature increase can be made by using
the thermal part of the specific internal energy $\epsilon_{\rm th}$
and by neglecting the thermal energy due to radiation, so that
$\epsilon_{\rm th} = 3 k T/(2 m_n)$, where $k$ is the Boltzmann
constant and $m_n$ the rest mass of a nucleon. In this way the
temperature is simply expressed as
\begin{eqnarray}
\label{t_estimate}
T &=& \frac{2 m_n}{3k(\Gamma-1)} 
\left[\frac{p}{\rho} - K \rho^{\Gamma-1}\right] \nonumber \\
 &\simeq& \frac{7.2174 \times 10^{12}}{\Gamma-1} \left[\frac{p}{\rho} - K
  \rho^{\Gamma-1}\right]\, {\rm K}\,.
\end{eqnarray}

\begin{figure}[ht]
\begin{center}
   \includegraphics[width=0.45\textwidth]{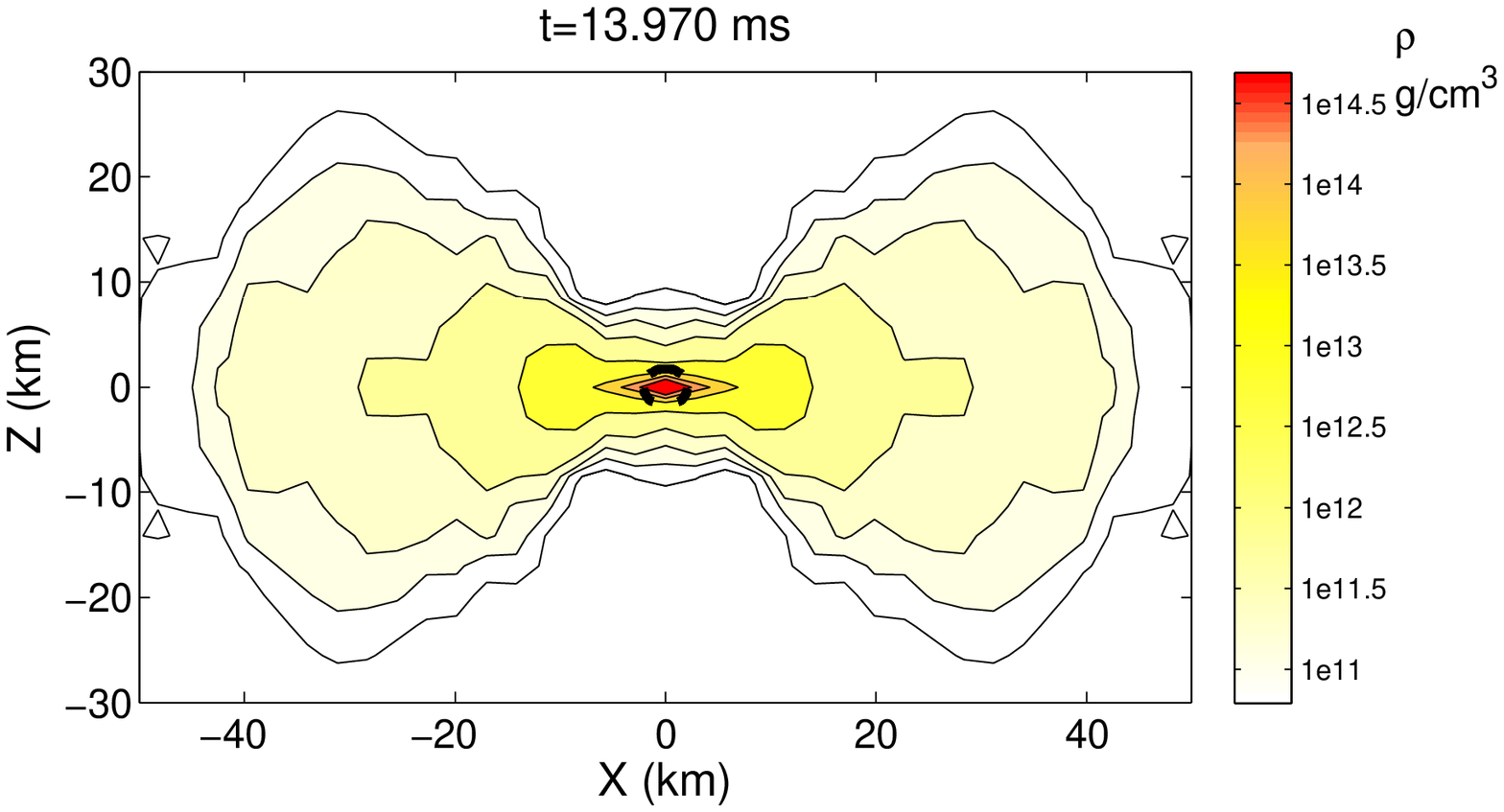}
   \includegraphics[width=0.45\textwidth]{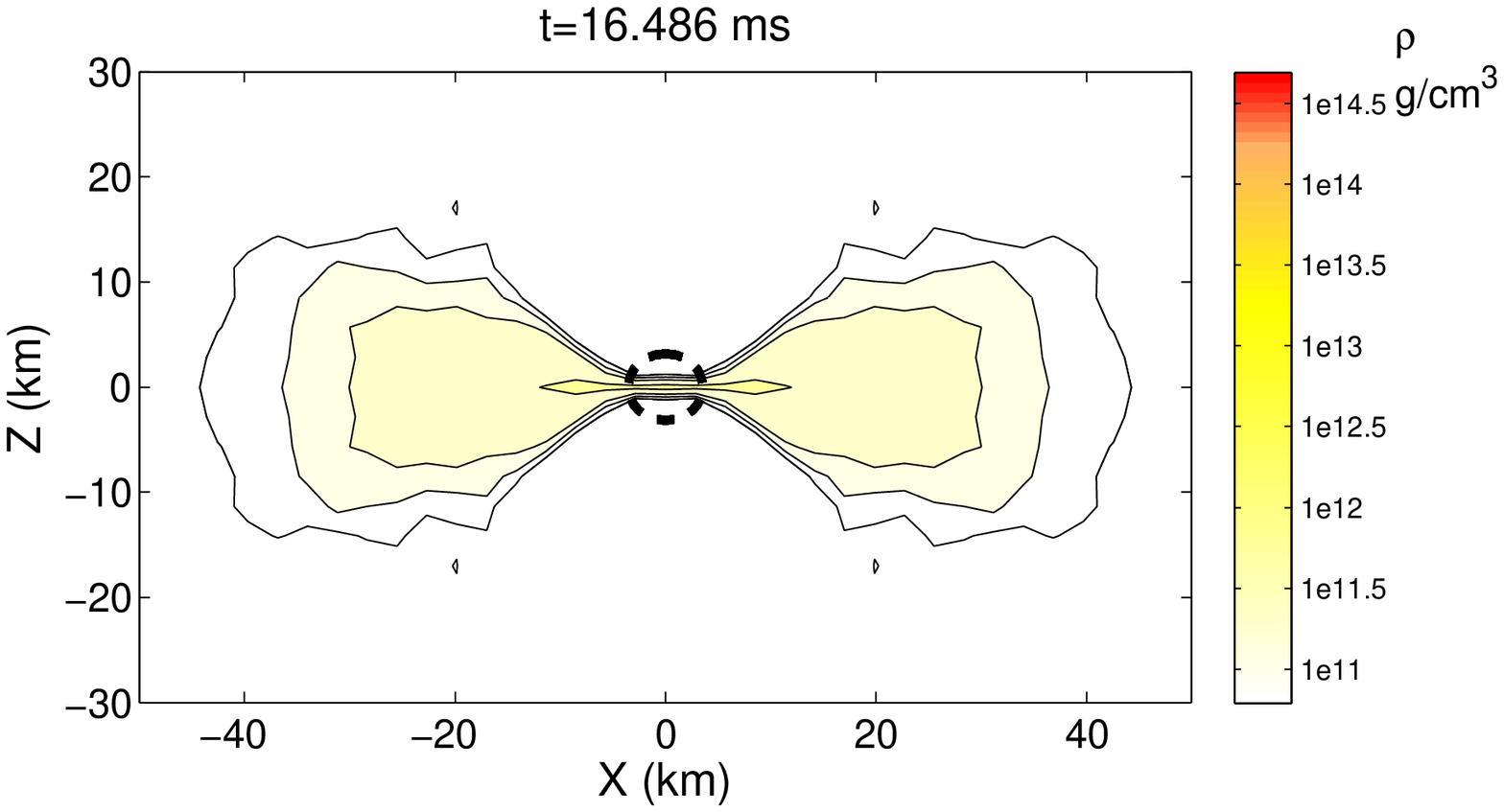}
   \includegraphics[width=0.45\textwidth]{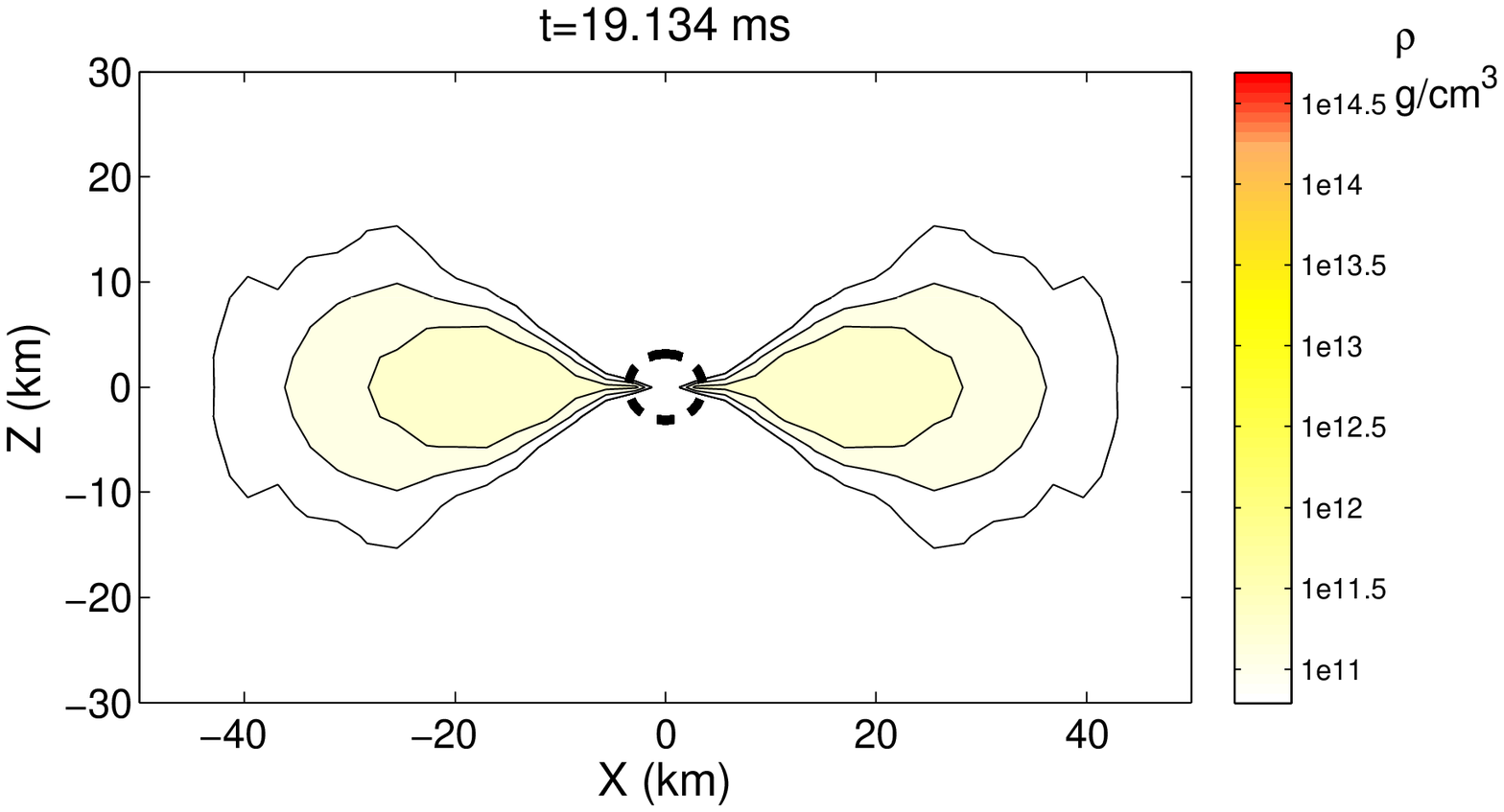}
   \includegraphics[width=0.45\textwidth]{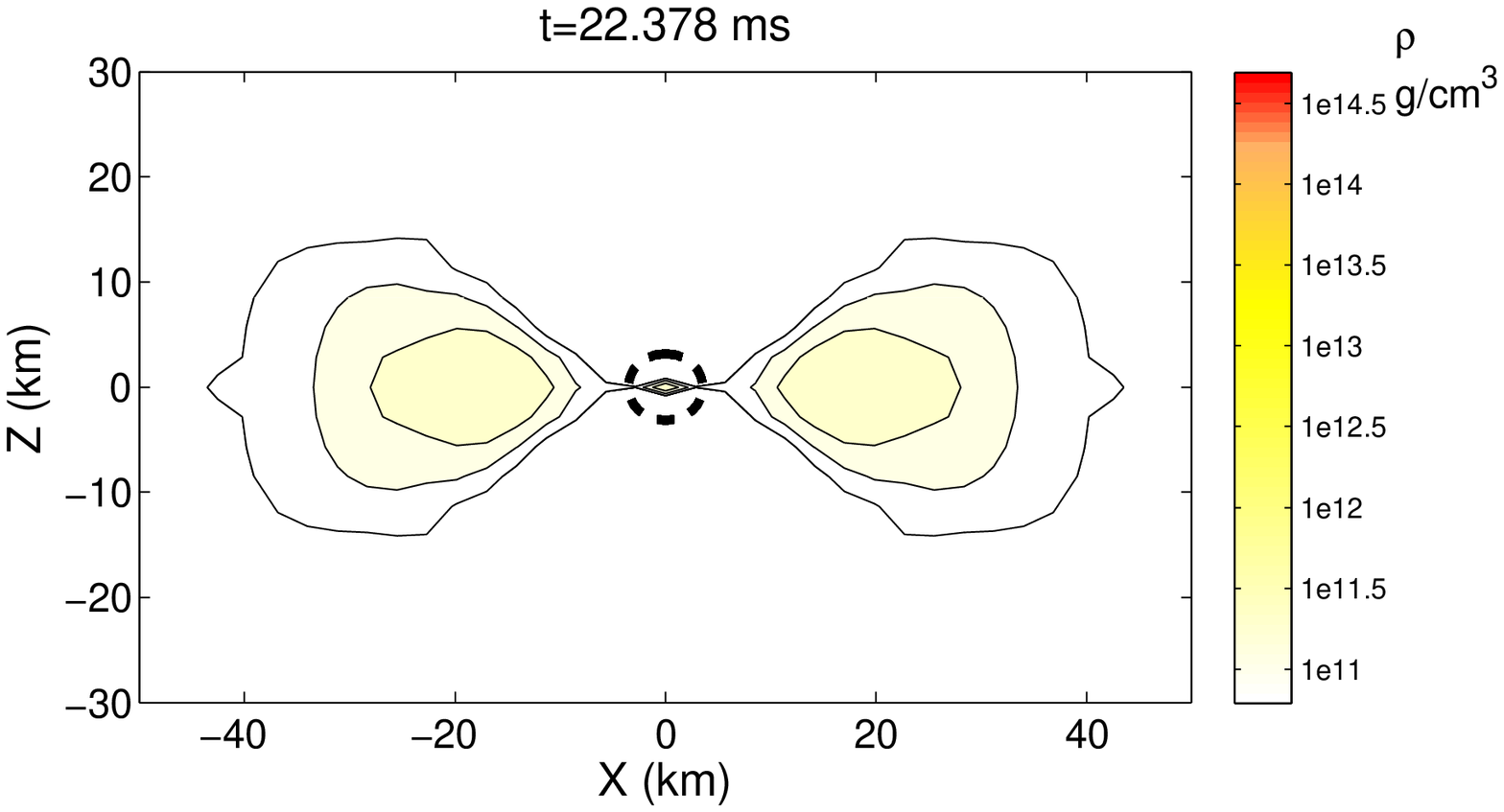}
\end{center}
\vskip -0.5cm
   \caption{Isodensity contours on the $(x,z)$ plane highlighting the
     formation of a torus surrounding the central BH, whose AH
     is indicated with a thick dashed line. The data
     refer to the \textbf{high-mass} binary evolved with the
     \textbf{ideal-fluid} EOS
     (\cf~Figs.~\ref{fig:BH_short_torus_poly_high},
     ~\ref{fig:BH_short_torus_poly_low} and the different vertical
     scales). A high-resolution version of this figure can be found at~\cite{weblink}.}
   \label{fig:BH_short_torus_IF_high}
\end{figure}

Using \eqref{t_estimate} it is then possible to estimate that the HMNS
has an initial temperature of $5\times 10^{10}\,{\rm K}$, which
rapidly increases to $5\times 10^{11}\,{\rm K}$ as the stellar cores
merge. The additional shocks produced by the large oscillations in the
post-merger phase can increase locally the temperature above these
values, with maximum values that can reach $2\times 10^{12}\,{\rm K}$.
Clearly, at such large temperatures the radiative losses, either via
photons or neutrinos, can become very important and lead to a
qualitative change from the evolution described here. While first
attempts of introducing the contribution of radiative losses in
general-relativistic calculations have recently been made
(see,~\eg refs.~\cite{ShibataSekiguchiTakahashi2006,Farris08}), we are
still far from a mathematically consistent and physically accurate
treatment of these processes, which we will include in future
works. For the time being it is sufficient to underline that, while it
is clear that the inclusion of radiative processes will lead, quite
generically, to a decrease in the survival time of the HMNS after the
merger, determining this time with any reasonable precision will
require not only the inclusion of radiative transport but also of a
more realistic treatment of the EOS and of the scattering properties
of the matter in the HMNS.

In the absence of a more detailed calculation of the radiative losses,
we can here resort to simpler back-of-the-envelope calculations to
assess the importance of radiative cooling in the post-merger phase
when taking into account neutrino emission and diffusion. Let us
therefore assume that the newly produced HMNS from a high-mass binary
is approximately spherical with an average radius of $R_{_{\rm
    HMNS}}\sim 20\,\km$, a mass of $M_{_{\rm HMNS}}\sim
3.2\,M_{\odot}$ and thus an average rest-mass density which is
essentially the nuclear rest-mass density, \ie $\rho_{_{\rm
    HMNS}}\sim\rho_{\rm nuc}\sim 3\times 10^{14}\,{\rm g/cm}^3$. We
can now consider two different cooling processes acting either via
modified-URCA emission (see Chap. 11 of ref.~\cite{Shapiro83}) or
through the more efficient direct-URCA
emission~\cite{Lattimer_etal_91}. Assuming an initial average
temperature of $T_{_{\rm HMNS}} \sim 10^{11}\,{\rm K}$, the HMNS would
cool down via modified-URCA processes to $T_{_{\rm HMNS}} \sim
10^{10}\ (10^{9})\,{\rm K}$ in about $20\,{\rm s}\ (1\ {\rm yr})$. On
the other hand, if the cooling takes place through the much more
efficient direct-URCA processes, the cooling time would be $\sim
3\,\ms\ (1\ {\rm min})$. Because the latter interval is smaller or
comparable with the $\sim 9\,\ms$ elapsing in the present calculations
between the formation of the HMNS and its collapse to a BH, we
conclude that radiative losses in the HMNS would accelerate its
collapse to a BH only if direct-URCA processes take place.

\begin{table*}[t]
  \caption{Summary of the results of the simulations: proper
    separation between the centers of the stars $d/M_{_{\rm ADM}}$;
    baryon mass $M_{b}$ of each star in solar masses; initial
    rest mass of the torus $(M_{_{\rm T}})_0$ (see footnote on
    page~\pageref{footnote1}); rest mass of the torus $3\,\ms$
    after the appearance of the AH $(M_{_{\rm T}})_{3\,{\rm ms}}$
    (actually $3\,\ms$ after the time when the AH mean radius has
    reached the value $2.1$, see footnote on
    page~\pageref{footnote1}); mass of the BH $M^{\rm IH}_{_{\rm
        BH}}$, as computed in the isolated-horizon formalism; angular
    momentum of the BH $J^{\rm IH}_{_{\rm BH}}$, as computed in the
    isolated-horizon formalism; BH spin parameter
    $a^{^{\rm IH}}\equiv (J_{_{\rm BH}}/M^2_{_{\rm BH}})^{\rm IH}$, as
    computed in the isolated-horizon formalism; ratio of the ADM
    mass carried by the waves to the initial ADM mass; ratio of the angular momentum
    carried by the gravitational waves to the initial angular momentum.}
\begin{ruledtabular}
\begin{tabular}{l|ccccccccc}
  Model &
\multicolumn{1}{c}{$d/M_{_{\rm ADM}}$}         &
\multicolumn{1}{c}{$M_{b}/M_{\odot}$}       &
\multicolumn{1}{c}{$(M_{_{\rm T}})_0/M_{\odot}$}   &
\multicolumn{1}{c}{$(M_{_{\rm T}})_{3\,{\rm ms}}/M_{\odot}$}   &
\multicolumn{1}{c}{$M^{^{\rm IH}}_{_{\rm BH}}/M_{\odot}$}  &
\multicolumn{1}{c}{$J^{^{\rm IH}}_{_{\rm BH}}({\rm g}\, {\rm cm}^2{\rm /s})$}  &
\multicolumn{1}{c}{$a^{^{\rm IH}}$} &
\multicolumn{1}{c}{$M_{_{\rm GW}}/M_{\odot}$}  &
\multicolumn{1}{c}{$J_{_{\rm GW}}/J(t=0)$}\\
%
\hline
$1.46$-$45$-${\rm P}$  & $14.3$ & $1.456$  & $0.1$  & $0.0787$ & $2.60$ & $4.61\times10^{49}$ & $0.76$ & $1.8\times10^{-2}$ & $0.21$ \\ 
$1.62$-$60$-${\rm P}$  & $16.8$ & $1.625$  & $0.04$ & $0.00115$& $3.11$ & $7.0\times10^{49}$ & $0.82$ & $9.6\times10^{-3}$ & $0.22$ \\ 
$1.62$-$45$-${\rm P}$  & $12.2$ & $1.625$  & $0.04$ & $0.0117$ & $2.99$ & $6.4\times10^{49}$ & $0.82$ & $9.3\times10^{-3}$ & $0.12$ \\ 
$1.46$-$45$-${\rm IF}$ & $14.3$ & $1.456$  & $--$   & $--$     & $--$  & $--$    & $--$ & $8.5\times10^{-3}$ & $0.15$\\ 
$1.62$-$45$-${\rm IF}$ & $12.2$ & $1.625$  & $0.2$  & $0.0726$ & $2.94$ & $6.4\times10^{49}$ & $0.84$ & $1.2\times10^{-2}$ & $0.17$ \\ 
%
\end{tabular}
\end{ruledtabular}
\vskip -0.25cm
\label{table:FD}
\end{table*}

Quite predictably, also the merger of a high-mass binary evolved with
the ideal-fluid EOS leads to the formation of a torus orbiting around
the BH. With respect to the high-mass polytropic binary, however, the
torus here has a different shape and a considerably larger vertical
extension. Indeed the ratio of the vertical and horizontal sizes is
$\sim 0.5$, while this was $\sim 0.1$ in the case of a polytropic EOS,
irrespective of the mass of the binary. Consequently, the measured
initial rest mass of the torus is of a factor $6$ larger than the one
of the corresponding high-mass polytropic binary, namely $(M_{_{\rm
    T}})_{3\,{\rm ms}} = 0.0726\,M_{\odot}$ instead of $(M_{_{\rm
    T}})_{3\,{\rm ms}} = 0.0117\, M_{\odot}$, $3\,\ms$ after the
first measure (\cf~Table~\ref{table:FD}). The average density, on the
other hand, is considerably smaller (between $10^{11}$ and
$10^{12}\ {\rm g/cm^3}$). 

The dynamics of the torus are summarized in
Fig.~\ref{fig:BH_short_torus_IF_high}, which shows the isodensity
contours on the $(x,z)$ plane; note again that the panels refer to
times between $14.0\ {\rm ms}$ and $22.4\ {\rm ms}$ and thus to a
stage in the evolution which is between the last two panels of
Fig.~\ref{fig:BH_short_torus_IF_high}. A simple comparison between
Figs.~\ref{fig:BH_short_torus_poly_high},
~\ref{fig:BH_short_torus_poly_low}
and~\ref{fig:BH_short_torus_IF_high} is sufficient to capture the
differences among the tori in the three different cases considered so
far.

In view of the discussion made above on the increased internal energy
content produced by the shocks in the case of the ideal-fluid EOS, the
formation of a vertically extended torus is not at all surprising, but
the obvious response of the matter of the torus to a larger (thermal)
pressure gradient in the vertical direction. Interestingly, the
maximum rest-mass density of the torus does not show the typical
harmonic behavior discussed so far in the case of the polytropic
binaries and produced by the quasi-periodic oscillations in the radial
direction. Rather, the maximum density shows a clear and monotonic
decrease with time as a result of the accretion of the torus onto the
BH (\cf~Fig.~\ref{fig:if-rho-high} for $t \gtrsim 14\,\ms$). At the
same time, the maximum of the internal energy in the torus is seen to
increase (\cf~Fig.~\ref{fig:if-eps-high} for $t \gtrsim
14\,\ms$). Both the higher temperature and the geometrically thick
shape of the torus produced in this case provide an important evidence
that the merger of a massive NS binary could lead to the physical
conditions behind the generation of a GRB. A more detailed analysis of
the energetics and properties of the torus (and in particular of its
variability in time) is needed to further support this possibility and
it will be presented in a future work~\cite{BGR08}.

\subsection{Ideal-fluid EOS: low-mass binary}
\label{IF_binaries_lm}

\begin{figure}[ht]
\begin{center}
   \includegraphics[angle=-0,width=0.45\textwidth]{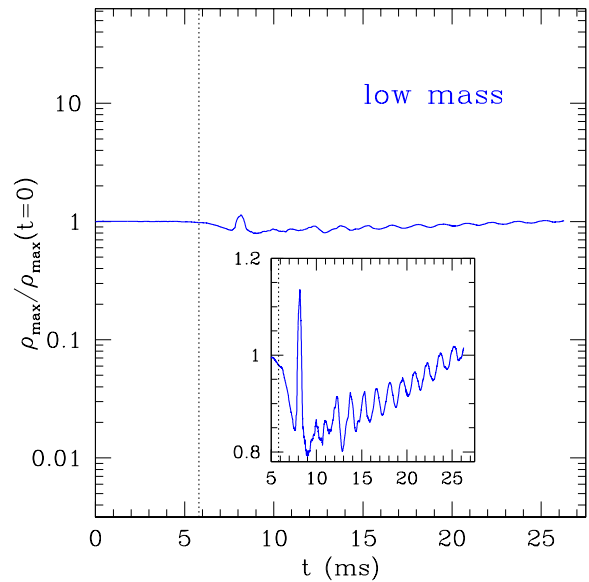}
\end{center}
\vskip -0.5cm
   \caption{Evolution of the maximum rest-mass density normalized to
     its initial value for the \textbf{low-mass} binary evolved using the
     \textbf{ideal-fluid} EOS. Indicated with a dotted vertical line
     is the time at which the binary merges. This figure should be
     compared with Fig.~~\ref{fig:poly-rho-high}, \ref{fig:poly-rho-low} 
     and \ref{fig:if-rho-high}, of which maintains the same scale.}
  \label{fig:if-rho-low}
\end{figure}

Despite it being significantly different from the evolution of both
the low-mass polytropic binary and of the high-mass ideal-fluid
binary, the dynamics of the low-mass ideal-fluid binary is rather
simple. In particular, the two NSs merge at essentially the same time
as the corresponding high-mass ideal-fluid binary (\ie $t\simeq
5.8\,\ms$) and produce a HMNS which is however not sufficiently
massive to collapse promptly to a BH. Rather, the HMNS undergoes a
bar-mode instability producing an $m=2$ deformation as the system tries
to reach a configuration which is energetically favourable. 

\begin{figure*}[t]
\begin{center}
   \includegraphics[width=0.48\textwidth]{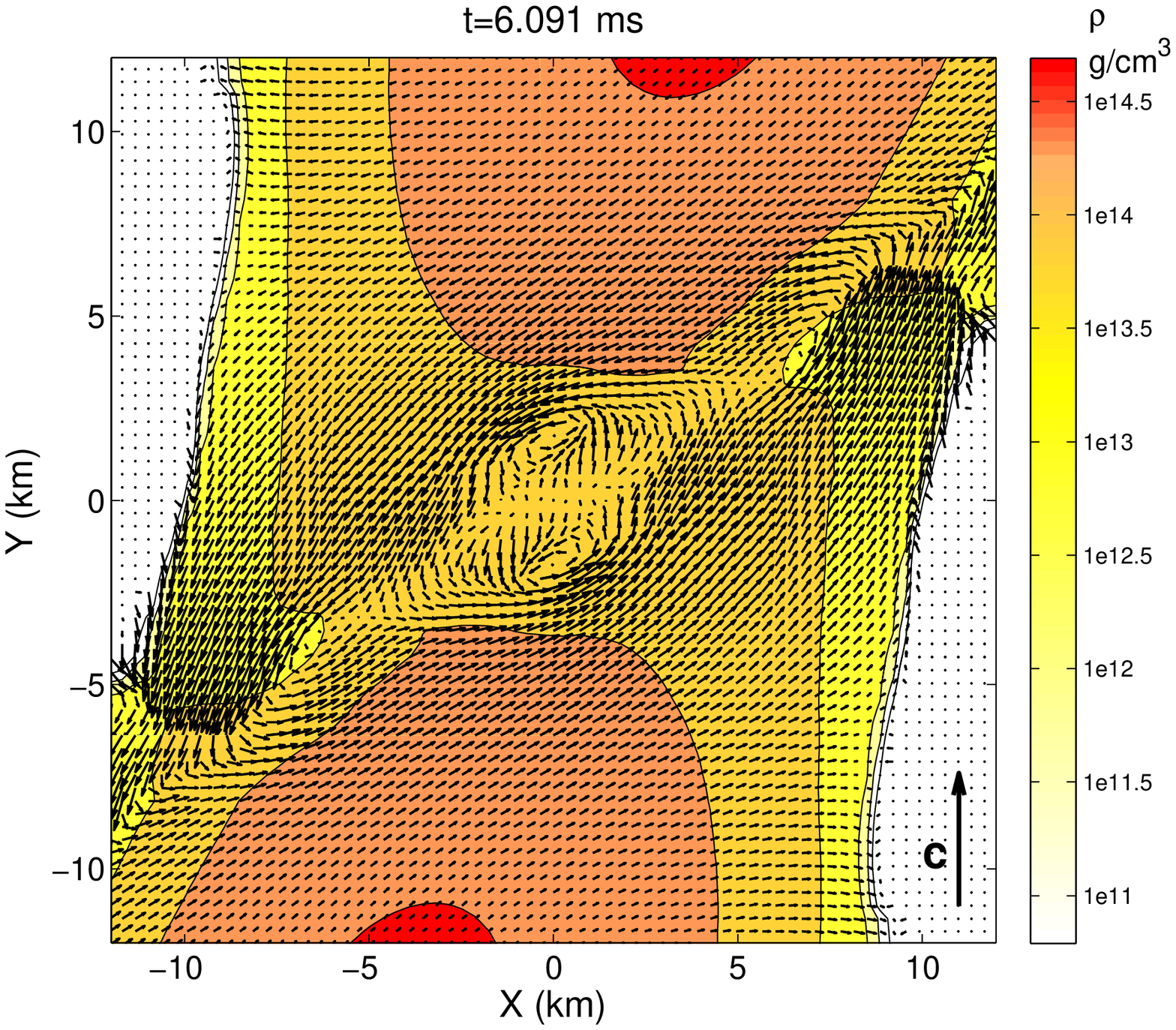}
   \includegraphics[width=0.5\textwidth]{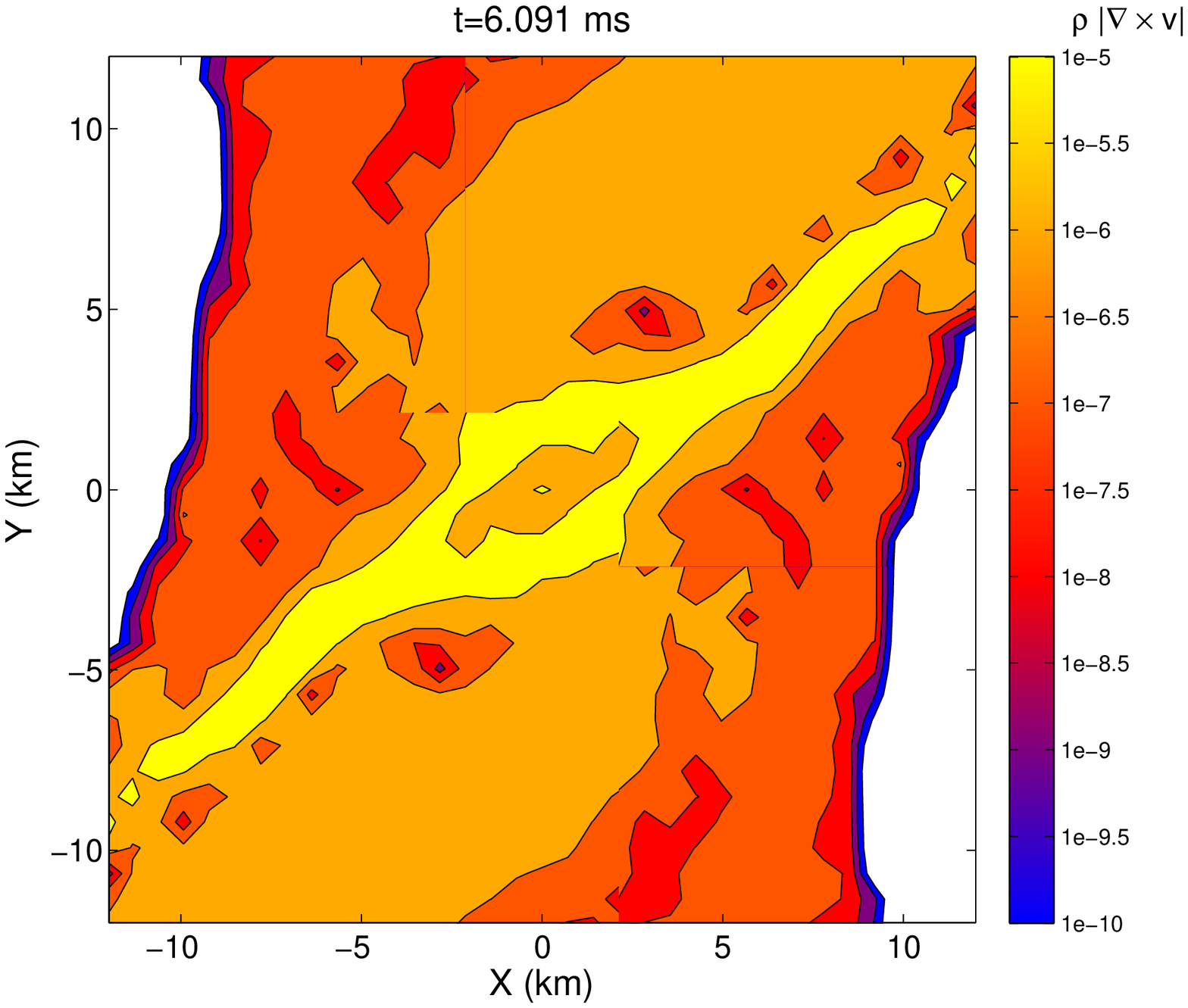}
\end{center}
\vskip -0.5cm
\caption{\textit{Left panel:} Isodensity contours and velocity vector
  field (with the orbital component removed) on the $(x,y)$
  (equatorial) plane at a selected time soon after the merger. Note
  the presence of localized vortices in the shear layer between the
  two stars. \textit{Right panel:} contours of the weighted vorticity
  $\rho |\nabla \times \boldsymbol{v}|^z$ (that is the rest-mass
  density multiplied by the module of the $z$ component of the
  vorticity) for the same time shown in the left panel. This rendering
  highlights that in the shear layer the vorticity can be up to three
  orders of magnitude larger than in the bulk of the stars. Both
  panels refer to a \textbf{high-mass} binary evolved with the
  \textbf{polytropic} EOS. A high-resolution version of this figure can be found at~\cite{weblink}.
  \label{fig:BH_short_vorticity}}
\end{figure*}

Either as a result of the $\pi$-symmetry imposed (and which prevents
the growth of the $m=1$ mode) or simply because the HMNS is very close
to the threshold of the bar-mode instability, the bar is seen to
persist for the whole time the calculations were carried out, \ie
$\sim 30\,\ms$ (see the discussion of ref.~\cite{Manca07} about under
what conditions a bar-mode deformation is expected to survive over a
longer timescale; recent additional work on this can also be found in
ref.~\cite{Saijo2008}). Note that the bar deformation remains only
approximately constant in time and that small oscillations in the
central rest-mass density can be measured. This is shown in
Fig.~\ref{fig:if-rho-low}, which reports the evolution of the maximum
rest-mass density normalized to its initial value. Indicated with a
dotted vertical line is the time at which the stars merge. This figure
should be compared with Figs.~\ref{fig:poly-rho-high},
\ref{fig:poly-rho-low} and~\ref{fig:if-rho-high}, of which it
maintains the same scale.

During this rather long period of time (corresponding to $\sim 16$
revolutions) the HMNS also loses large amounts of angular momentum
through gravitational radiation (see discussion in Sects.~\ref{gws_IF}
and~\ref{sec:eam_losses}). As a result, the compactness of the HMNS 
gradually increases and the central density shows the characteristic
secular increase already discussed for the previous binaries
(\cf~inset of Fig.~\ref{fig:if-rho-low}). The radiation-reaction
timescale is in this case much longer: the HMNS is not very massive
(its mass of $\approx 2.6\,M_\odot$ is only $\approx 10\%$ larger than
the supramassive limit) and is more extended as a result of the
increased internal temperature. As a result, the migration to the
unstable branch and the collapse to a BH will occur much later than
what was calculated and shown in Fig.~\ref{fig:if-rho-low}. Using the
latter to compute the growth rate of the maximum rest-mass density and
assuming that the collapse to a BH is triggered when $\rho_{\rm
  max}/\rho_{\rm max}(t=0) \simeq 2$
(\cf~Figs.~\ref{fig:poly-rho-high} and~\ref{fig:if-rho-high}), we
estimate that the collapse will take place at $t\sim 110\,\ms$. This
timescale should be compared with the corresponding one (\ie $\sim
21\,\ms$) obtained from the same initial data but evolved with a
polytropic EOS. Clearly, the increase in the internal energy via
shocks is responsible for this ``long-delay'' in the collapse to a BH.

As a final comment we note that a timescale of $\sim 110\,\ms$ is much
longer than what is computationally feasible at the moment. As a
result, the analysis of this binary will be limited to a time interval
of $\sim 30\,\ms$, which is, however, long enough to deduce its most
interesting properties (see discussion in Sects.~\ref{gws_IF}
and~\ref{sec:eam_losses}).

\subsection{Vortex sheet and Kelvin-Helmholtz instability}
\label{vs_and_khi}

\begin{figure*}[ht]
\begin{center}
   \includegraphics[width=0.45\textwidth]{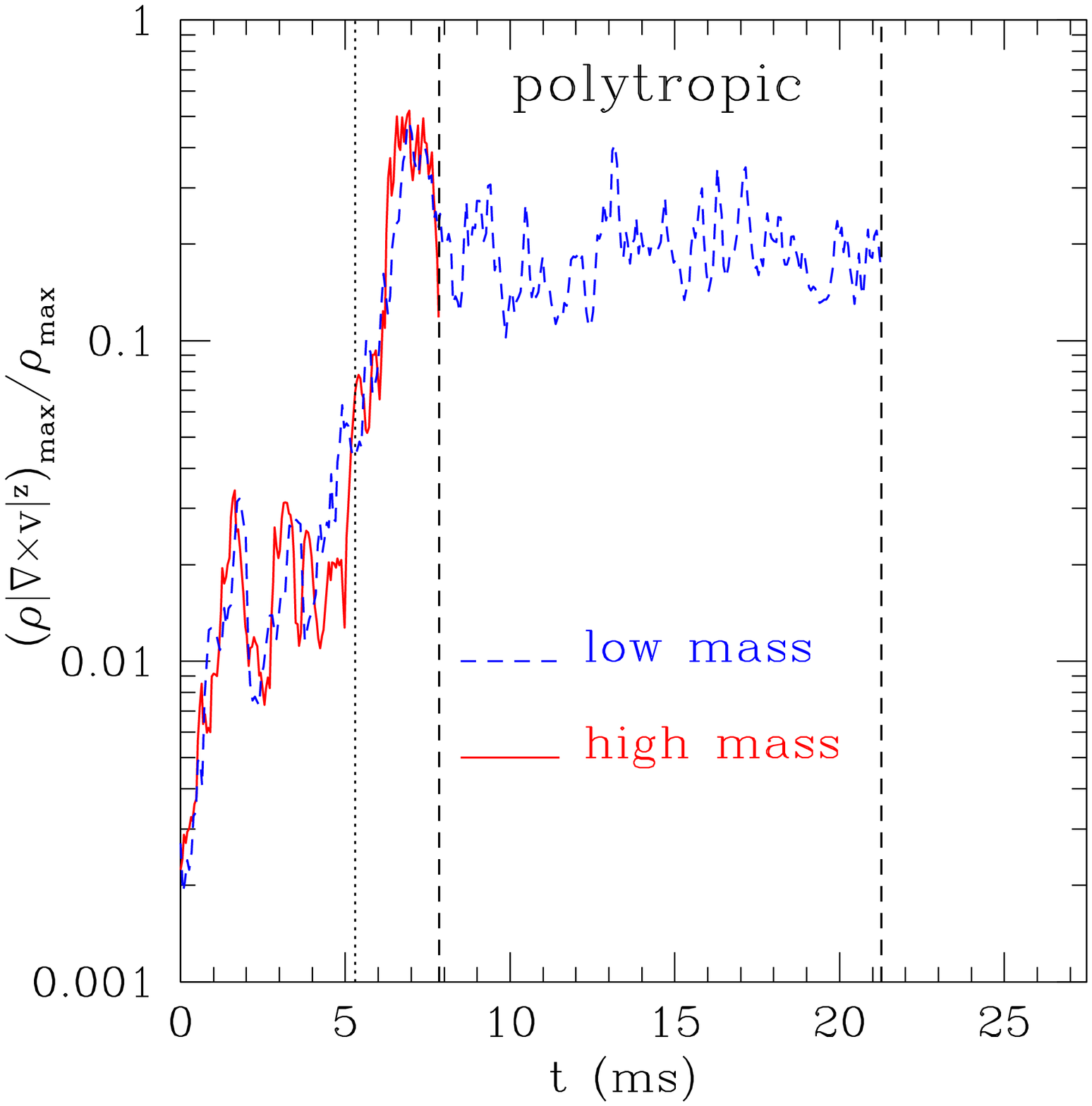}
  \hskip 1.0cm
   \includegraphics[width=0.45\textwidth]{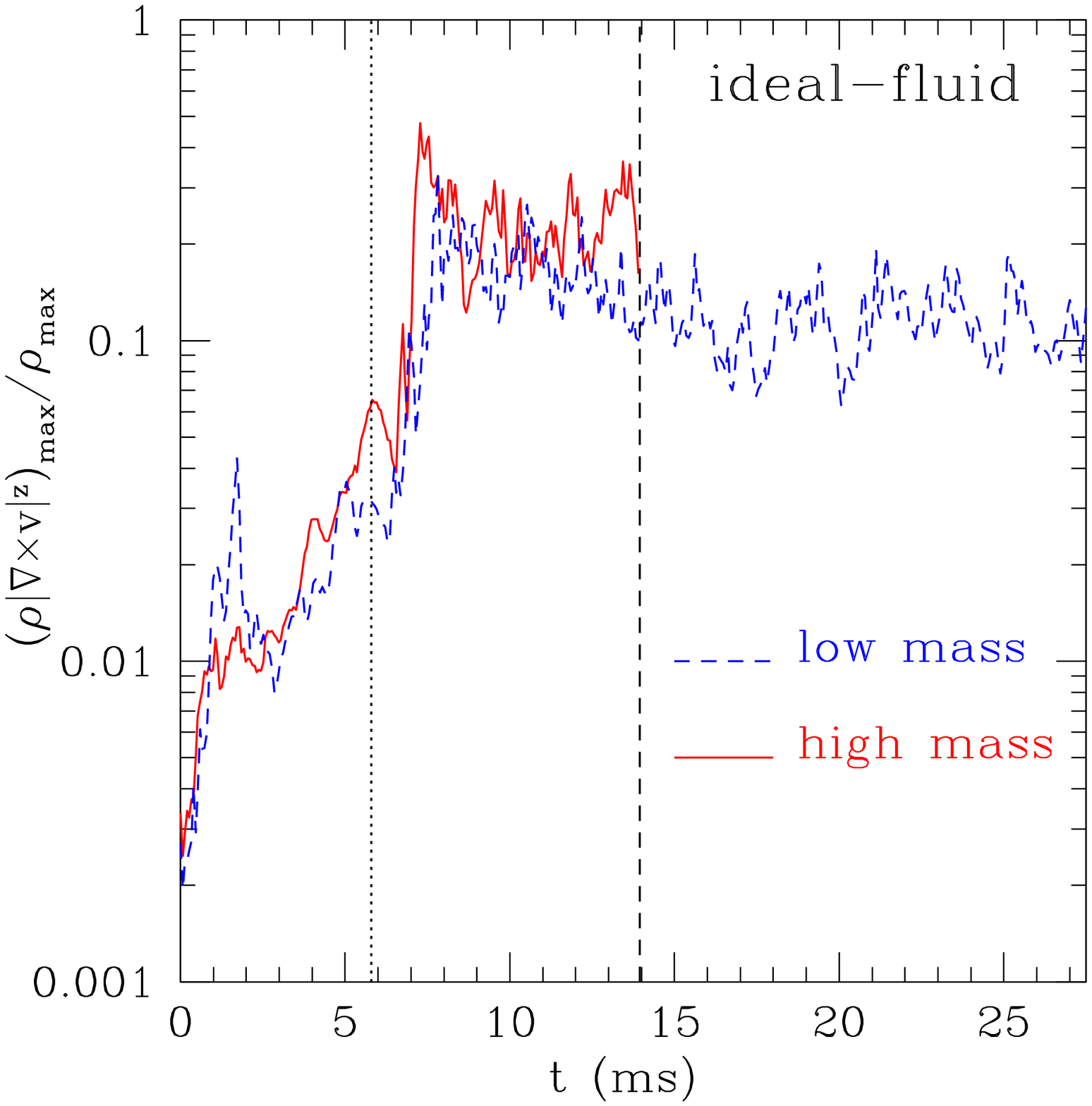}
\end{center}
\vskip -0.5cm
   \caption{\textit{Left panel:} Maximum of the weighted vorticity
     $\rho |\nabla \times \boldsymbol{v}|^z$ on the equatorial plane
     normalized by the maximum of the rest-mass density $\rho_{max}$
     during the evolution of the high (solid line) and low (dashed
     line) mass binaries evolved with the \textbf{polytropic}
     EOS. Indicated with a dotted vertical line is the time at which
     the binaries merge. Both the curves are plotted until the
     formation of an AH. \textit{Right panel:} The same as in the left
     panel but for the \textbf{ideal-fluid}
     EOS. \label{fig:BH_short_max_vort}}
\end{figure*}

As mentioned above, when the two stars come into contact a vortex
sheet (or shear interface) develops there where the tangential
components of the velocity exhibit a discontinuity (\ie the $x$ and
$y$ components of the three-velocity in our setup). This condition is
known to be unstable to very small perturbations and it can develop a
Kelvin-Helmholtz instability, which will curl the interface forming a
series of vortices~\cite{Chandrasekhar81}. This is indeed what we
observe in all our simulations, with features that are essentially not
dependent on the mass or on the EOS used.

In the left panel of Fig.~\ref{fig:BH_short_vorticity} we show the
isodensity contours and the velocity vector field on the equatorial
plane for the high-mass binary evolved with a polytropic EOS at a time
$t=6.091\,\ms$ when the presence of vortices is particularly
evident. The density is shown in units of ${\rm g/cm^3}$ and in the
bottom-right part of the plot an arrow is used as a reference for the
values of the velocity. Furthermore, in order to highlight the
formation of the shear interface, we have removed from the total
velocity field the orbital angular velocity defined as the angular
velocity of the stellar centers. The vector-field representation shows
rather clearly that the vortex sheet goes from the bottom-left corner
of the plot to the upper-right one.  Along this sheet one can
observe at least four main vortices, two of which are located at
$[x\approx \pm 7\,\km,~y\approx \pm 5\,\km]$, while the other two are
smaller and located at $[x\approx 0\,\km,~y\approx \pm 2\,\km]$. It is
worth remarking that, because these smaller vortices have a scale of
$\gtrsim 2\, \km \sim 1.3\,M_{\odot}$, they are well captured by our
resolution in the central regions which, we recall, is $h=
0.12\,M_{\odot}\approx 0.177\,\km$. Because the employed numerical methods are very weakly
dissipative on these scales, we believe that our description of the
Kelvin-Helmholtz instability is indeed accurate at the scales
presented. Of course, different resolutions will either remove some of
the vortices (as the resolution is decreased) or introduce new ones
(as the resolution is increased). In practice, we have found that a
vortex of scale $\lambda$ is lost when the resolution used is $h
\gtrsim 0.2\lambda$, probably because the intrinsic numerical
dissipation prevents their formation.

A different and novel way of showing the presence of a vortex sheet
and of the consequent development of a Kelvin-Helmholtz instability is
offered in the right panel of Fig.~\ref{fig:BH_short_vorticity}, which
shows the contours of the ``weighted vorticity'' on the equatorial
plane, \ie $\rho |\nabla \times \boldsymbol{v}|^z$. Although this
vector represents the Newtonian limit of the general-relativistic
vorticity tensor $\omega_{\mu \nu} = \partial_{[\nu}(h
  u_{\mu]})$~\cite{Teukolsky98}, it serves the purpose here of being
proportional to the latter and also of simpler calculation. Because
the color-coding is made in a logarithmic scale, the right panel of
Fig.~\ref{fig:BH_short_vorticity} clearly highlights that the
vorticity is not uniform in the merged object but that its value in
the vortex sheet is up to three orders of magnitude larger than in
the bulk of the stars. As stressed above, while both panels of
Fig.~\ref{fig:BH_short_vorticity} refer to the high-mass polytropic
binary, very similar results were obtained also for the low-mass
binary or with the ideal-fluid EOS.

To quantify the development of the Kelvin-Helmholtz instability and
measure its growth rate we have computed the maximum of the weighted
vorticity in the equatorial plane and plotted its time evolution in
Fig.~\ref{fig:BH_short_max_vort}, where it is also shown as divided by
the maximum of the rest-mass density to remove the contribution due to
the increase in $\rho$ after the merger.  Shown with different lines
are the weighted vorticities for the high-mass binary (solid line) and
for the low-mass binary (dashed line), evolved either with a
polytropic EOS (left panel) or with an ideal-fluid EOS (right
panel). Also indicated with a vertical dotted line is the time at
which the two NSs merge, while the two vertical dashed lines refer to
the times at which the AH is found in the two cases
(no evolution is shown past this time as the measure of the vorticity
becomes much more complex because of the turbulent motions in the
torus). It is evident that after an initial growth of a factor of a
few between $t=0$ and $t=2\,\ms$, probably produced by the transient
away from the initial data, the weighted vorticity remains
approximately small and constant. This stops at the time of the merger at
$t\approx 5\,\ms$ (\cf~dotted vertical line) when the weighted
vorticity grows exponentially of about two orders of magnitude. The
Newtonian perturbative expectation for the growth rate is $\sigma \sim
\pi v/\lambda$ where $v$ is the value of the velocity at the shear
interface and $\lambda$ is the wavelength of the smallest growing
mode; for $v \sim 10^{-2}$ and $\lambda \sim 2\,\km$, the measured
growth rate is $\sigma \simeq 10^3\,{\rm s}^{-1}$ and in reasonable
agreement with the Newtonian expectation.

The instability rapidly saturates when the two stellar cores merge; as a result, 
after $\sim 2\,\ms$ from its initial development it reaches a
quasi-stationary state. Note that the growth rate is essentially the
same for the high- and low-mass binary and for the two EOSs (\cf~the
two panels Fig.~\ref{fig:BH_short_max_vort}); however the evolution
after the saturation is different for the different masses. The
high-mass binaries collapse to a BH, while the HMNSs produced by the
low-mass binaries hang on for a longer time, during which the
instability persists at almost constant amplitude [for $\sim 13\,\ms$
(\cf~dashed line in Fig.~\ref{fig:BH_short_max_vort})].

 \begin{figure*}[t]
 \begin{center}
  \includegraphics[angle=-0,width=0.45\textwidth]{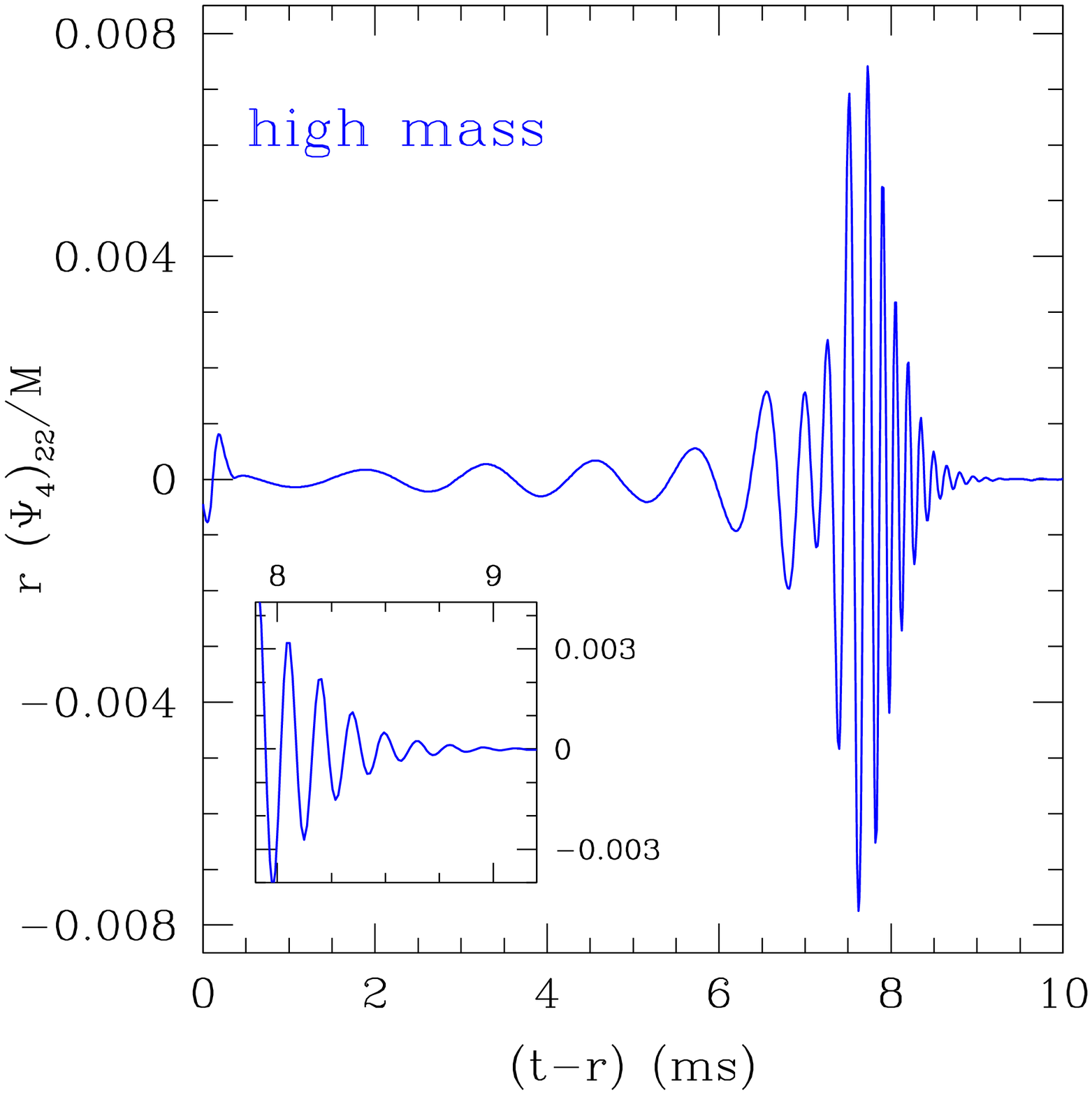}
  \hskip 1.0cm
  \includegraphics[angle=-0,width=0.45\textwidth]{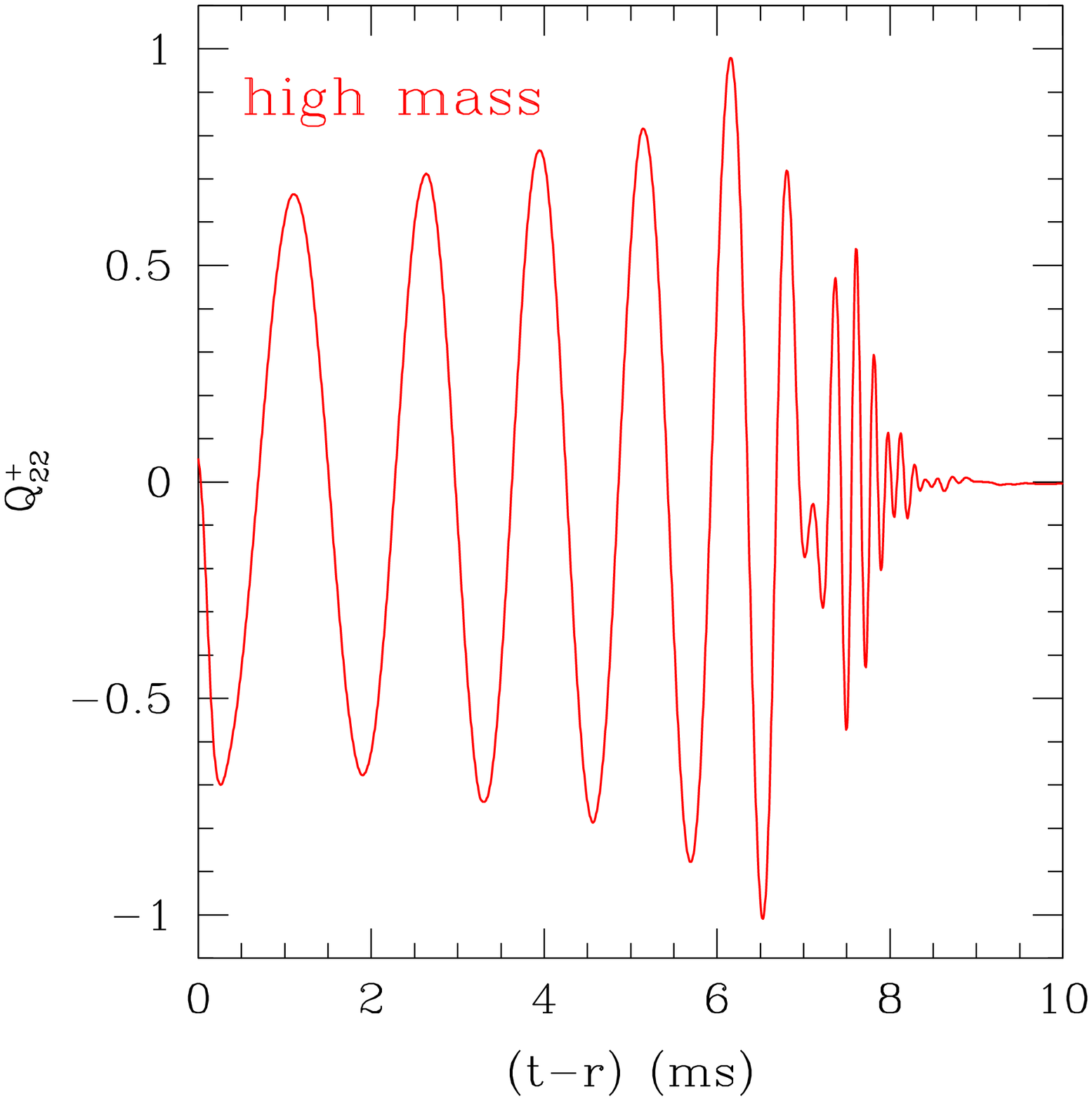}
 \end{center}
 \vskip -0.5cm
   \caption{\textit{Left panel}: Retarded-time evolution of the real
     part of the $\ell=m=2$ component of $r\Psi_4$ as extracted from a
     $2$-sphere at a coordinate radius $r=200\,M_{\odot}= 295\,\km$
     for the \textbf{high-mass} binary. Indicated in the inset is the
     final part of the signal corresponding to the BH quasi-normal
     ringing. The merger takes place at $(t-r)\sim 5.3\ {\rm
       ms}$. \textit{Right panel}: The same as in the left panel but
     shown in terms of the real part of the gauge-invariant quantity
     $Q^{+}_{22}$. In both cases the binaries have been evolved using
     the \textbf{polytropic} EOS.}
   \label{fig:psi4_pol_high}
 \end{figure*}

As a final remark we note that, even if this instability is purely
hydrodynamical, it can have strong consequences when studying the
dynamics of binary NSs in the presence of magnetic fields. Indeed, as 
first shown by ~\cite{Price06} in Newtonian simulations and
later briefly reported also by~\cite{Anderson2008} in
general-relativistic calculations, in the presence of a magnetic field
this instability leads to an exponential growth of the toroidal
component even if the initial magnetic field is a purely poloidal
one. In particular, it is reasonable to expect that even a moderate
initial poloidal magnetic field of $\approx 10^{12}\,G$ can be
increased up to values of order $10^{15}\,G$ through this
mechanism. Such high values of the magnetic fields are the ones
presumed to be behind the phenomenology in magnetars, but are also
thought to be the values necessary in order to extract sufficient
energy from a system composed by a torus orbiting around a BH and
power short hard GRBs. Work is now in progress for the investigation
of this mechanism in fully general-relativistic MHD using the code
presented in~\cite{Giacomazzo:2007ti}; results of this investigation
will soon be reported in a distinct article.

\section{Gravitational-wave emission}
\label{sec:gwe}

The accurate determination of the gravitational-radiation content of
the simulated spacetimes represents a delicate and yet fundamental
aspect of any modeling of sources of gravitational waves; in view of
this, we have implemented two different and equivalent methods to
compute the gravitational waves produced by the inspiralling
binaries. The possibility of a comparison between the two methods and
the cross-validation of the results provides us with additional
confidence that the extracted waveforms are not only numerically
accurate but also physically consistent.

The first method uses the Newman-Penrose formalism, which provides a
convenient representation for a number of radiation-related quantities
as spin-weighted scalars. In particular, the curvature scalar
\begin{equation}
  \Psi_4 \equiv -C_{\alpha\beta\gamma\delta}
    n^\alpha \bar{m}^\beta n^\gamma \bar{m}^\delta
  \label{eq:psi4def}
\end{equation}
is defined as a particular component of the Weyl curvature tensor,
$C_{\alpha\beta\gamma\delta}$, projected onto a given null frame
$\{\boldsymbol{l}, \boldsymbol{n}, \boldsymbol{m},
\bar{\boldsymbol{m}}\}$ and can be identified with the gravitational
radiation if a suitable frame is chosen at the extraction radius. In
practice, we define an orthonormal basis in the three-space
$(\hat{\boldsymbol{r}}, \hat{\boldsymbol{\theta}},
\hat{\boldsymbol{\phi}})$, centered on the Cartesian origin and
oriented with poles along $\hat{\boldsymbol{z}}$. The normal to the
slice defines a timelike vector $\hat{\boldsymbol{t}}$, from which we
construct the null frame
\begin{equation}
   \boldsymbol{l} = \frac{1}{\sqrt{2}}(\hat{\boldsymbol{t}} - \hat{\boldsymbol{r}}),\quad
   \boldsymbol{n} = \frac{1}{\sqrt{2}}(\hat{\boldsymbol{t}} + \hat{\boldsymbol{r}}),\quad
   \boldsymbol{m} = \frac{1}{\sqrt{2}}(\hat{\boldsymbol{\theta}} - 
     {\mathrm i}\hat{\boldsymbol{\phi}}) \ .
\end{equation}
We then calculate $\Psi_4$ via a reformulation of (\ref{eq:psi4def}) 
in terms of ADM variables on the slice~\cite{Shinkai94},
\begin{equation}
  \Psi_4 = C_{ij} \bar{m}^i \bar{m}^j,  \label{eq:psi4_adm}
\end{equation}
where
\begin{equation}
  C_{ij} \equiv R_{ij} - K K_{ij} + K_i{}^k K_{kj} 
    - {\rm i}\epsilon_i{}^{kl} \nabla_l K_{jk}\ .
\end{equation}

The gravitational-wave polarization amplitudes $h_+$ and $h_\times$
are then related to $\Psi_4$ by simple time
integrals~\cite{Teukolsky73}
\begin{equation}
\ddot{h}_+ - {\rm i}\ddot{h}_{\times}=\Psi_4 \ ,
\label{eq:psi4_h}
\end{equation}
where the double overdot stands for the second-order time derivative.

The second and independent method is instead based on the measurements
of the non-spherical gauge-invariant perturbations of a Schwarzschild
BH (see refs.~\cite{Abrahams97a,Rupright98,Rezzolla99a} for some
applications of this method to Cartesian-coordinate grids). In
practice, a set of ``observers'' is placed on $2$-spheres of fixed
Schwarzschild radius $r_{_{\rm S}}$, derived from the coordinate
(isotropic) radius via the standard formula
\begin{equation}
  r_{_{\rm S}} = r_{\rm iso} \left( 1 - \frac{M}{2r_{\rm iso}} \right)^2.
\end{equation}
where $M=M_{_{\mathrm{ADM}}}$ is assumed constant throughout the
simulation. On these $2$-spheres we extract the
gauge-invariant, odd-parity (or {\it axial}) current multipoles
$Q_{\ell m}^\times$ and even-parity (or {\it polar}) mass multipoles
$Q_{\ell m}^+$ of the metric
perturbation~\cite{Moncrief74,Abrahams95b}.  The $Q^+_{\ell m}$ and
$Q^\times_{\ell m}$ variables are related to $h_+$ and $h_\times$
as~\cite{Nagar05}
\begin{equation}
\label{eq:wave_gi}
h_+-{\rm i}h_{\times} =
  \dfrac{1}{\sqrt{2}r}\sum_{\ell,\,m}
  \Biggl( Q_{\ell m}^+ -{\rm i}\int_{-\infty}^t Q^\times_{\ell
          m}(t')dt' \Biggr)\,_{-2}Y^{\ell m}\ .
\end{equation}
Here $_{-2}Y^{\ell m}$ are the $s=-2$ spin-weighted spherical
harmonics and $(\ell, m)$ are the indices of the angular
decomposition.

\subsection{Waveforms from polytropic binaries}
\label{gws_pol}

In what follows we illustrate and discuss the gravitational-wave signal
produced by the inspiral and merger of the binaries discussed in
Sect.~\ref{sec:bd} and we start by discussing the waveforms produced
by the binaries evolved with the polytropic EOS.
 
 \begin{figure}[ht]
 \begin{center}
  \includegraphics[angle=-0,width=0.45\textwidth]{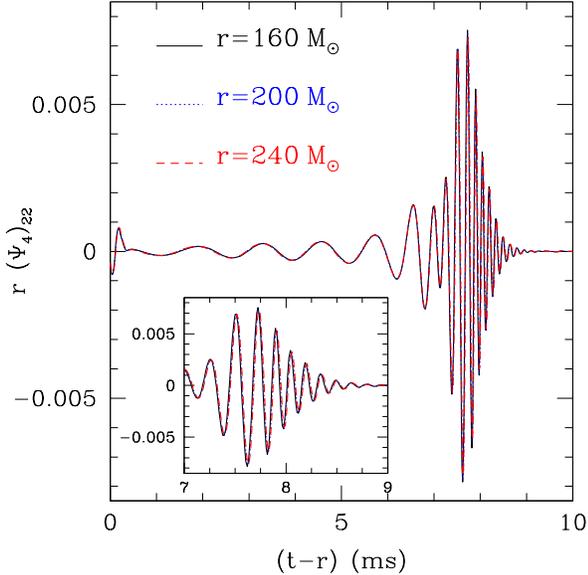}
 \end{center}
 \vskip -0.5cm
   \caption{Comparison of the real part of the $\ell=m=2$ component of
     $r\Psi_4$ for the \textbf{high-mass} binary evolved with the
     \textbf{polytropic} EOS when extracted at different radii: $r
         =160\,M_{\odot}= 236\,\km$ (solid line), $r
         =200\,M_{\odot}= 295\,\km$ (dashed line), and $r
         =240\,M_{\odot}= 354\,\km$ (dotted line).}
   \label{fig:detectors}
 \end{figure}

Figure~\ref{fig:psi4_pol_high}, in particular, shows in the left panel
the retarded-time evolution of the real part of the $\ell=m=2$
component of $r\Psi_4$ as extracted from a $2$-sphere at a coordinate
radius $r=200\,M_{\odot}= 295\,\km$ for the high-mass binary. Hereafter
$r=200\,M_{\odot}= 295\,\km$ will be the extraction radius for all the waveforms
presented, unless specified differently. Indicated in the inset is
the final part of the signal corresponding to the BH quasi-normal
ringing. We recall that the merger takes place at $(t-r)\sim 5.3\ {\rm
  ms}$ and that an AH is first found at $(t-r)=7.85\,\ms$. The
gravitational-wave signal during the inspiral is clearly very
well captured and remarkably reminiscent of the one observed in the
many binary BH simulations performed to date (see, for
instance,~\cite{Ajith:2007qp,Ajith:2007kx} and references therein) and
deviations from this type of waveforms are evident only at $(t-r)
\simeq 7\,\ms$, when the HMNS starts its collapse to a BH. The ability
of reproducing accurately the exponential decay of the quasi-normal
ringing is often a good indication of having reached a sufficient
level of accuracy as this involves the ability of measuring changes in
the fields on the smallest possible physical scales (\ie that of the
horizon). The clean quasi-normal ringing shown in the inset shows that
this is indeed the case for the simulations reported here.

 \begin{figure}[ht]
 \begin{center}
  \includegraphics[angle=-0,width=0.45\textwidth]{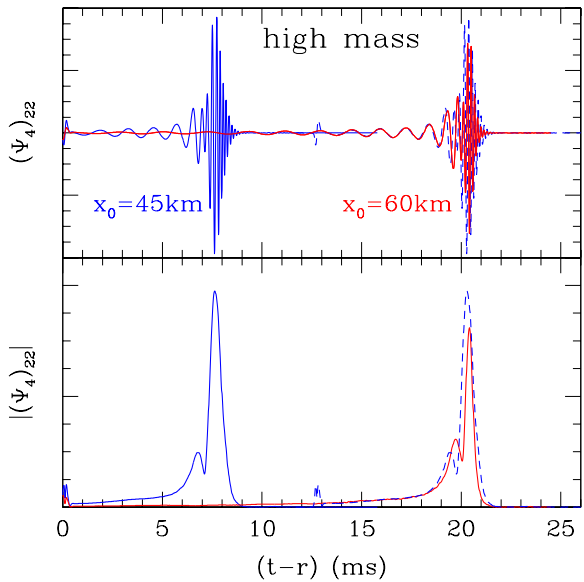}
 \end{center}
 \vskip -0.5cm
   \caption{Comparison of the real part of the $\ell=m=2$ component of
     $r\Psi_4$ (upper panel) and of its amplitude (lower panel) for
     the \textbf{high-mass} binaries evolved with the
     \textbf{polytropic} EOS starting from an initial separation of
     $45$ or $60\,{\rm km}$. Indicated with a dashed line are the
     values after a time-shift.}
   \label{fig:psi4_amp_pol_45_vs_60}
 \end{figure}

\begin{figure*}[t]
\begin{center}
 \includegraphics[angle=-0,width=0.45\textwidth]{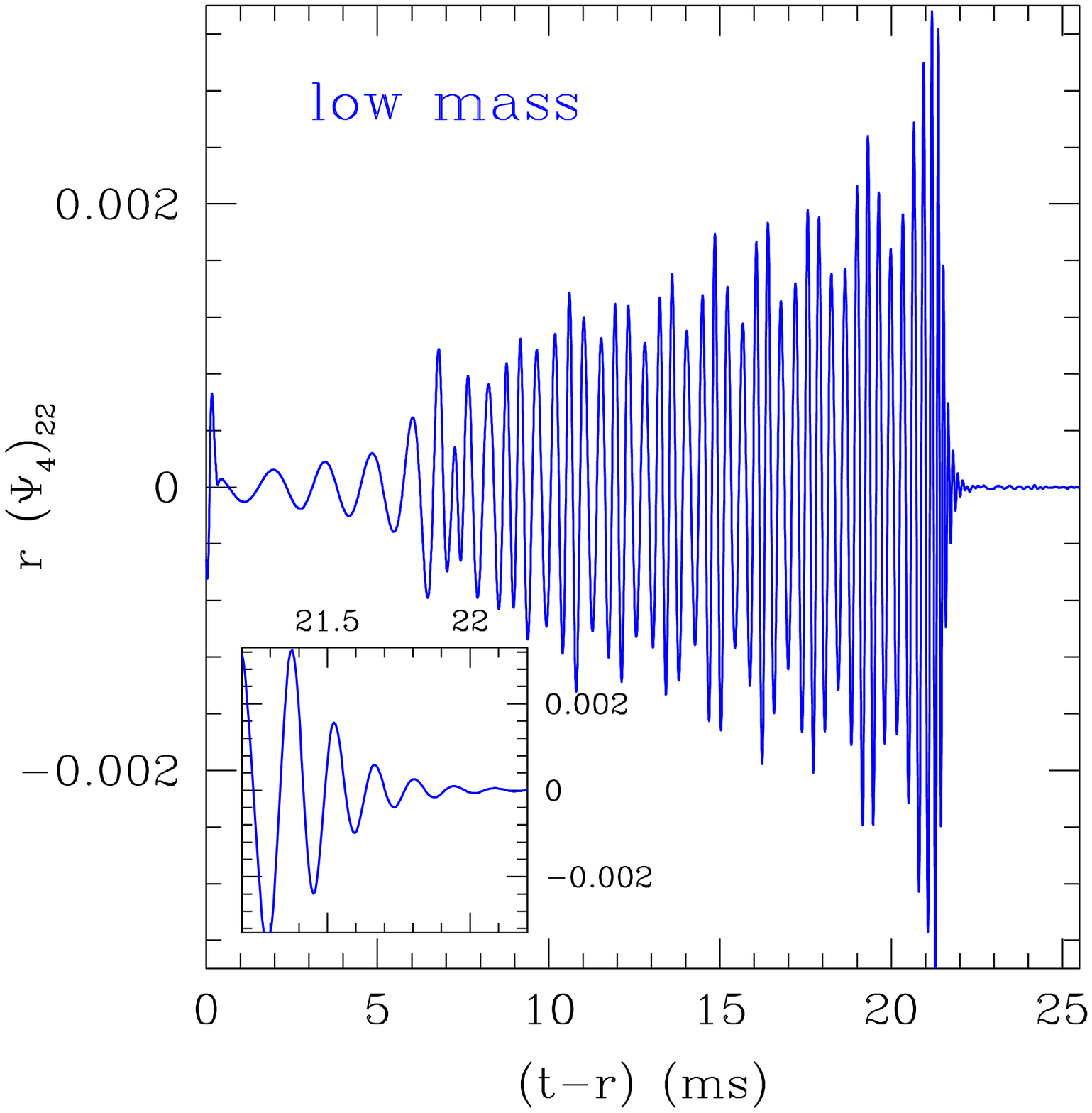}
 \hskip 1.0cm
 \includegraphics[angle=-0,width=0.45\textwidth]{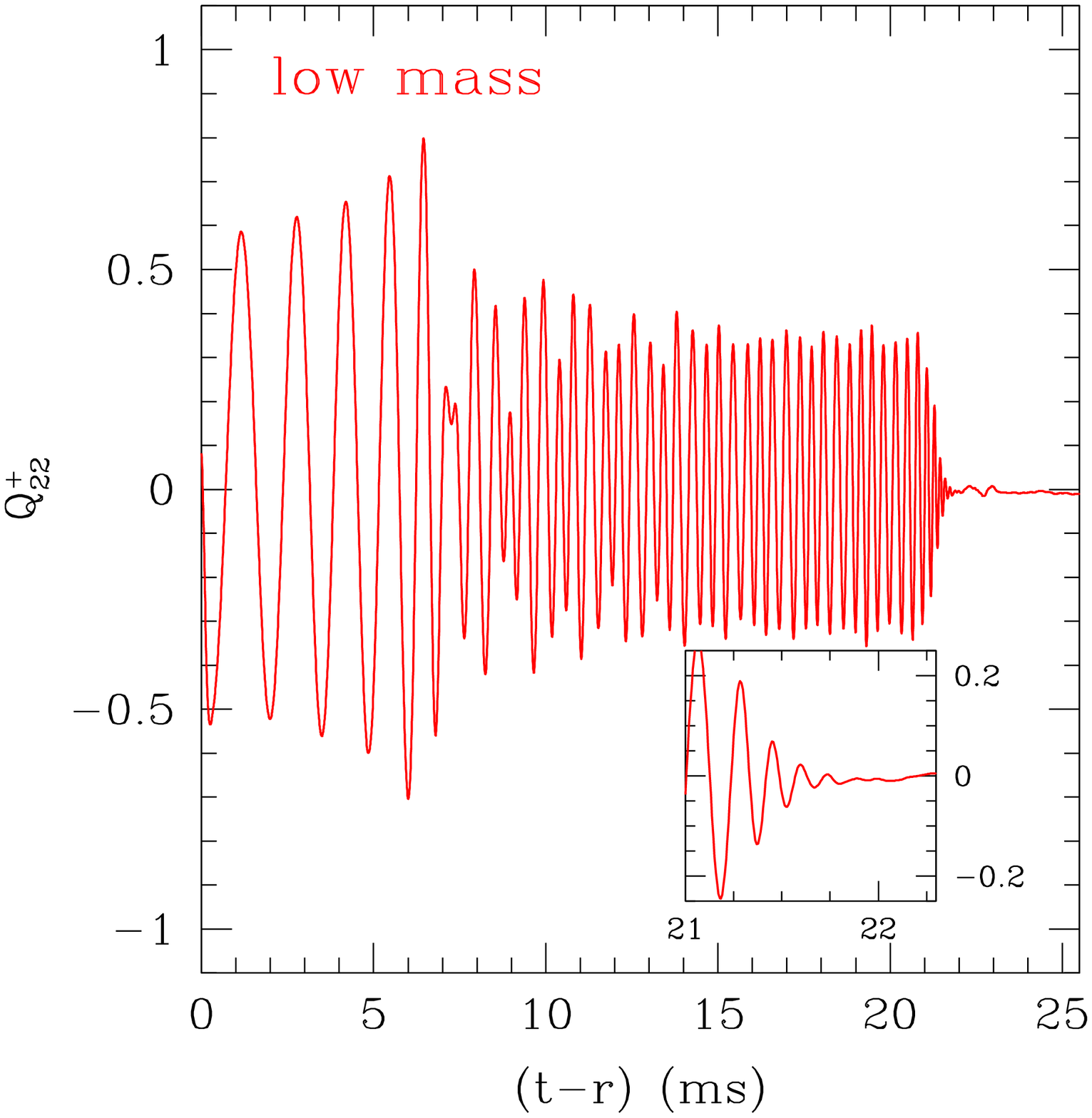}
\end{center}
\vskip -0.5cm
  \caption{\textit{Left panel}: Retarded-time evolution of the real
    part of the $\ell=m=2$ component of $r\Psi_4$ for the
    \textbf{low-mass} binary. Indicated in the inset is the final part
    of the signal corresponding to the BH quasi-normal ringing. The
    merger takes place at $(t-r)\sim 5.3\ {\rm ms}$. \textit{Right
      panel}: The same as in the left panel but shown in terms of the
    real part of the gauge-invariant quantity $Q^{+}_{22}$. In both
    cases the binaries have been evolved using the \textbf{polytropic}
    EOS.}
  \label{fig:psi4_pol_low}
 \end{figure*}

It should also be added that, because the newly formed BH is not in
vacuum but rather surrounded by a relativistic and accreting torus,
the gravitational-wave signal should not be expected to be
exponentially decaying to infinitesimal amplitudes during the
ringdown. This explains the tiny but nonzero oscillations which can be
seen after the ringdown and which are probably related to the
accretion of matter onto the BH. A comparison with the results of
ref.~\cite{Zanotti02,Zanotti05,Montero07} or with the perturbative
analysis of ref.~\cite{Nagar2007} could help to clarify the properties
of this signal.

The right panel of Fig.~\ref{fig:psi4_pol_high}, on the other hand,
shows the gravitational-wave signal in terms of the real part of the
gauge-invariant quantity $Q^{+}_{22}$. Because in this case the odd
perturbations have zero real and imaginary part, the time evolution of
the real (imaginary) part of $Q^{+}_{22}$ corresponds, modulo a
constant coefficient, to the time evolution of the $\ell=m=2$
component of $h_{+}\ (h_{\times})$. Note that the two waveforms are
clearly different, but this is simply because they differ by two time
derivatives [\cf~eqs.~\eqref{eq:psi4_h} and~\eqref{eq:wave_gi}]. In
fact, if a double time integral is made of the $\Psi_4$-waveforms, the
corresponding curve lies on top of the one for $Q^{+}_{22}$, after
a suitable normalization (see also Fig.~14 of
ref.~\cite{Pollney:2007ss} where this was shown in the case of binary
black holes).

The comparison offered by Fig.~\ref{fig:psi4_pol_high} is useful to
illustrate that, in contrast with what happens for binary BHs,
the amplitude of the $h_+$ and $h_{\times}$ polarizations does not
increase monotonically in time but, rather, is reduced as the two NSs
merge and as the HMNS collapses to a BH. Nevertheless, as we will
comment in Sect.~\ref{sec:eam_losses}, the energy loss rate is largest
during these stages (\cf~right panel of Fig.~\ref{fig:Erad}).

Another important validation that the signal extracted corresponds to
gravitational radiation can be obtained by verifying that $\Psi_4$
satisfies the expected ``peeling'' properties of the Weyl scalars,
\ie $r^{5-n} \Psi_n = {\rm const}$. This is illustrated in
Fig.~\ref{fig:detectors} which compares the real part of the
$\ell=m=2$ component of $\Psi_4$ when extracted at three considerably
different radii: $r=160\,M_{\odot}= 236\,\km$ (solid line), $r=200\,M_{\odot}= 295\,\km$
(dotted line), and $r =240\,M_{\odot}= 354\,\km$ (dashed line) (the last
radius is close to the outer boundary of our computational
domain). Clearly, the overlap among the different waveforms is very
good both in phase and in amplitude and indicates that already at $r
\sim 150\,M_{\odot}= 222\,\km$ gravitational waves can be extracted with
confidence. (A similar figure can be built using the Schwarzschild
perturbations and has not been shown here for compactness).
 
\begin{figure*}[t]
 \begin{center}
  \includegraphics[angle=-0,width=0.45\textwidth]{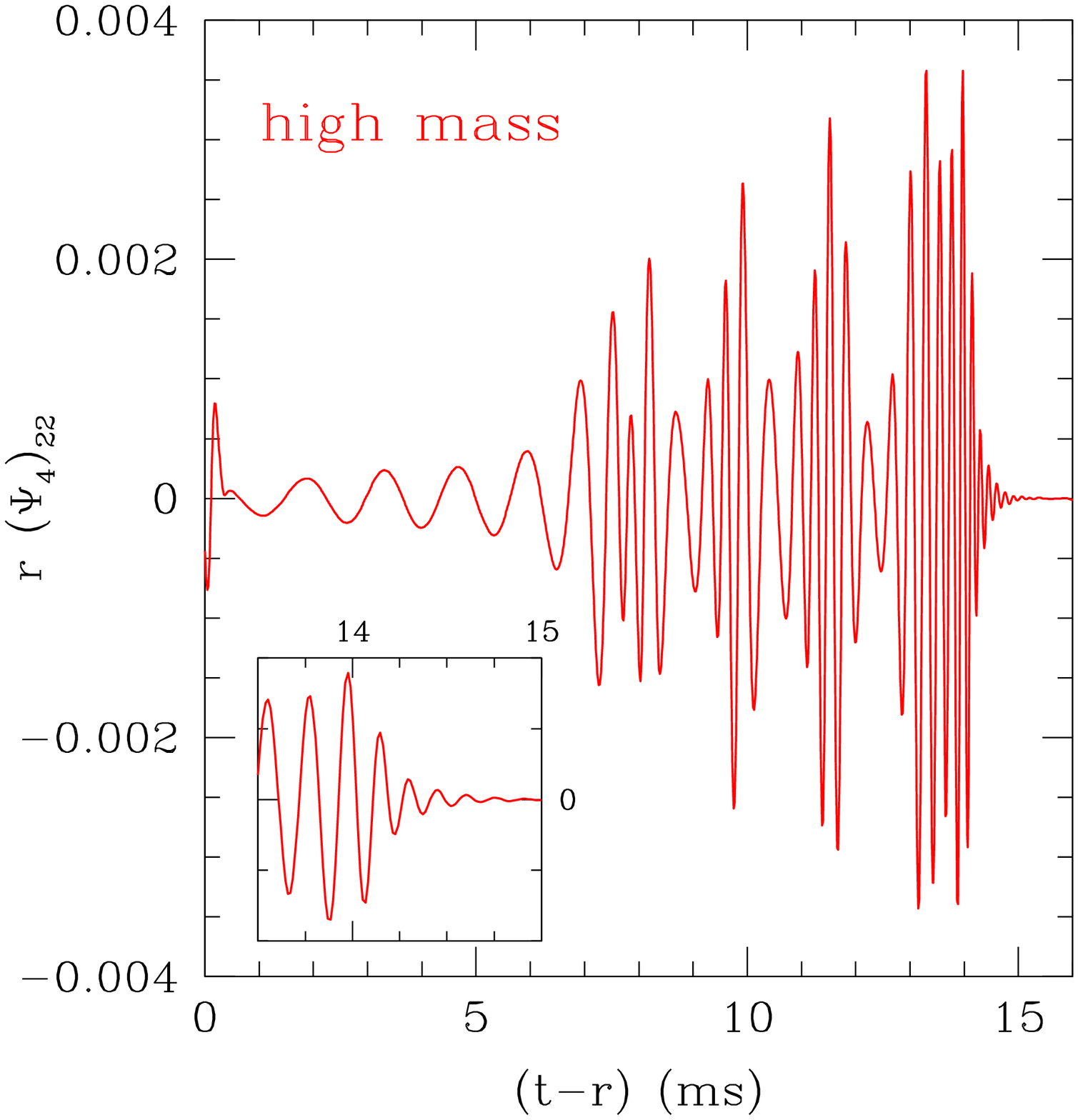}
  \hskip 1.0cm
  \includegraphics[angle=-0,width=0.45\textwidth]{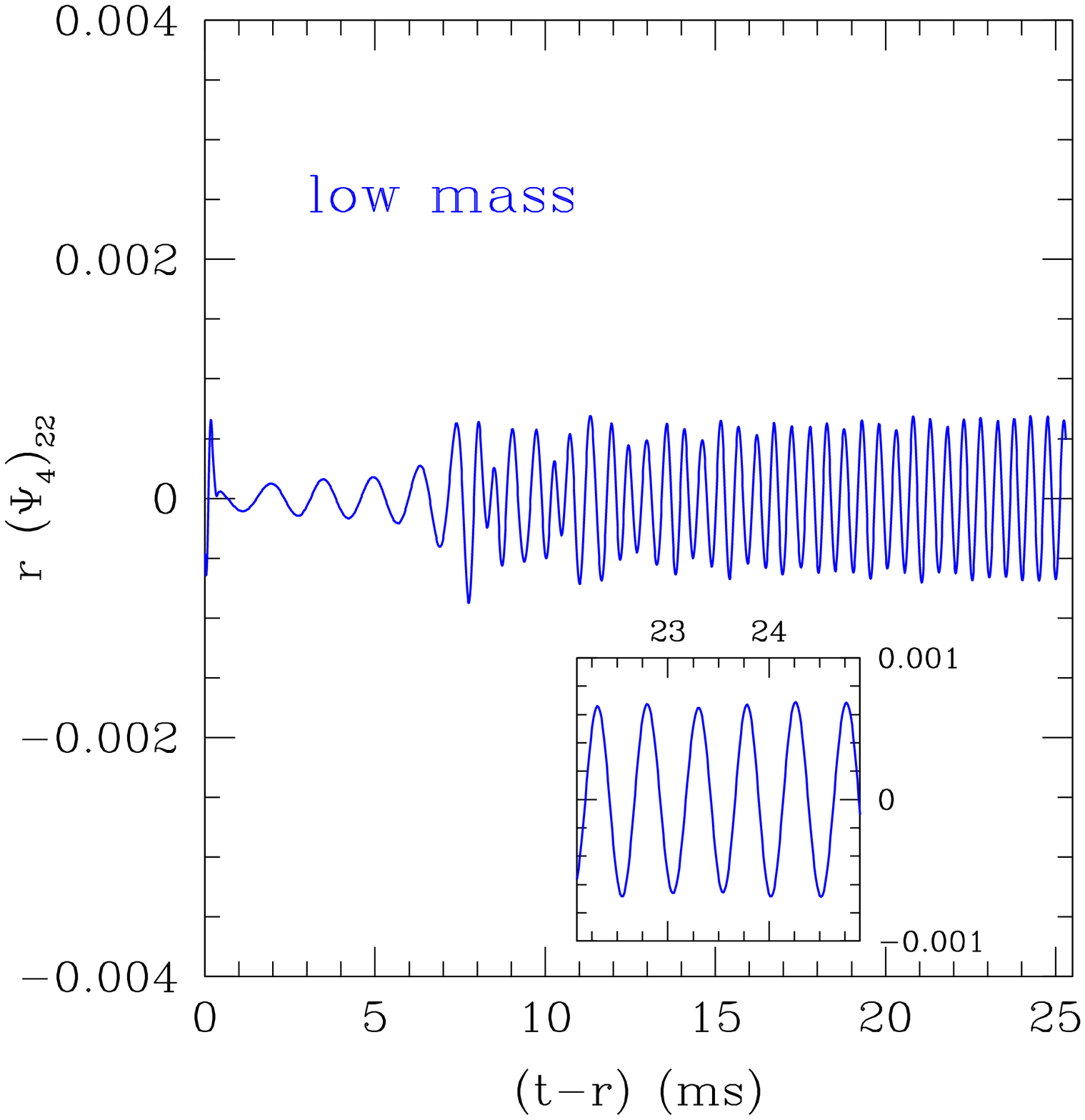}
 \end{center}
 \vskip -0.5cm
   \caption{\textit{Left panel:} Retarded-time evolution of the real
     part of the $\ell=m=2$ component of $r\Psi_4$ for the
     \textbf{high-mass} binary evolved with the \textbf{ideal-fluid}
     EOS. Indicated in the inset is the final part of the signal
     corresponding to the BH quasi-normal ringing. \textit{Right
       panel:} The same as in the left panel but for the \textbf{low-mass}
     binary. Here the inset does not refer to the BH quasinormal ringing;
     it highlights, instead, the periodic nature of the gravitational
     radiation after the merger. In both cases the merger takes place at $(t-r)\sim
     5.8\ {\rm ms}$.}
   \label{fig:psi4_IF_high_low}
 \end{figure*}

It is interesting now to reconsider the impact that different initial
separations of the same binary have on the emitted gravitational-wave
signal. This aspect was already discussed in Sect.~\ref{sec:poly_hm},
where the different dynamics were considered, and nicely summarized in
Figs.~\ref{fig:pol_high_45_vs_60}
and~\ref{fig:rho_pol_high_45_vs_60}. We recall that the conclusions
reached in Sect.~\ref{sec:poly_hm} were that the differences in the
evolution of the large-separation binary $1.62$-$60$-${\rm P}$ and of
its corresponding small-separation equivalent $1.62$-$45$-${\rm P}$
had to be found mostly in the internal structure and thus they were
absent in the pre-merger evolution of both the central rest-mass
density (\cf~Fig.~\ref{fig:rho_pol_high_45_vs_60}) and the proper
separation. A similar conclusion can be drawn also for the
waveforms and we show in Fig.~\ref{fig:psi4_amp_pol_45_vs_60} a
comparison in the real part of the $\ell=m=2$ component of $\Psi_4$
(upper panel) for the high-mass binaries evolved starting from an
initial separation of $45$ or $60\,{\rm km}$. Note that the waveform
for the $1.62$-$60$-${\rm P}$ binary contains more than $10$
gravitational-wave cycles and is, therefore, the longest
general-relativistic waveform computed to date.

An equivalent view of this comparison is shown in the lower panel of
Fig.~\ref{fig:psi4_amp_pol_45_vs_60} which reports instead the
amplitude of $\Psi_4$. Indicated with dashed lines in both panels are
the values after a suitable time shift. The good overlap in the
inspiral phase is what is expected on PN grounds; however, a closer
inspection also reveals that small differences do appear and these can
then be used as a measure of the high-order PN corrections coming from
compact binaries with finite size. More work and the use of long
waveforms are necessary to study this further.

\begin{figure*}[t]
 \begin{center}
  \vskip 1.0cm
  \includegraphics[angle=-0,width=0.45\textwidth]{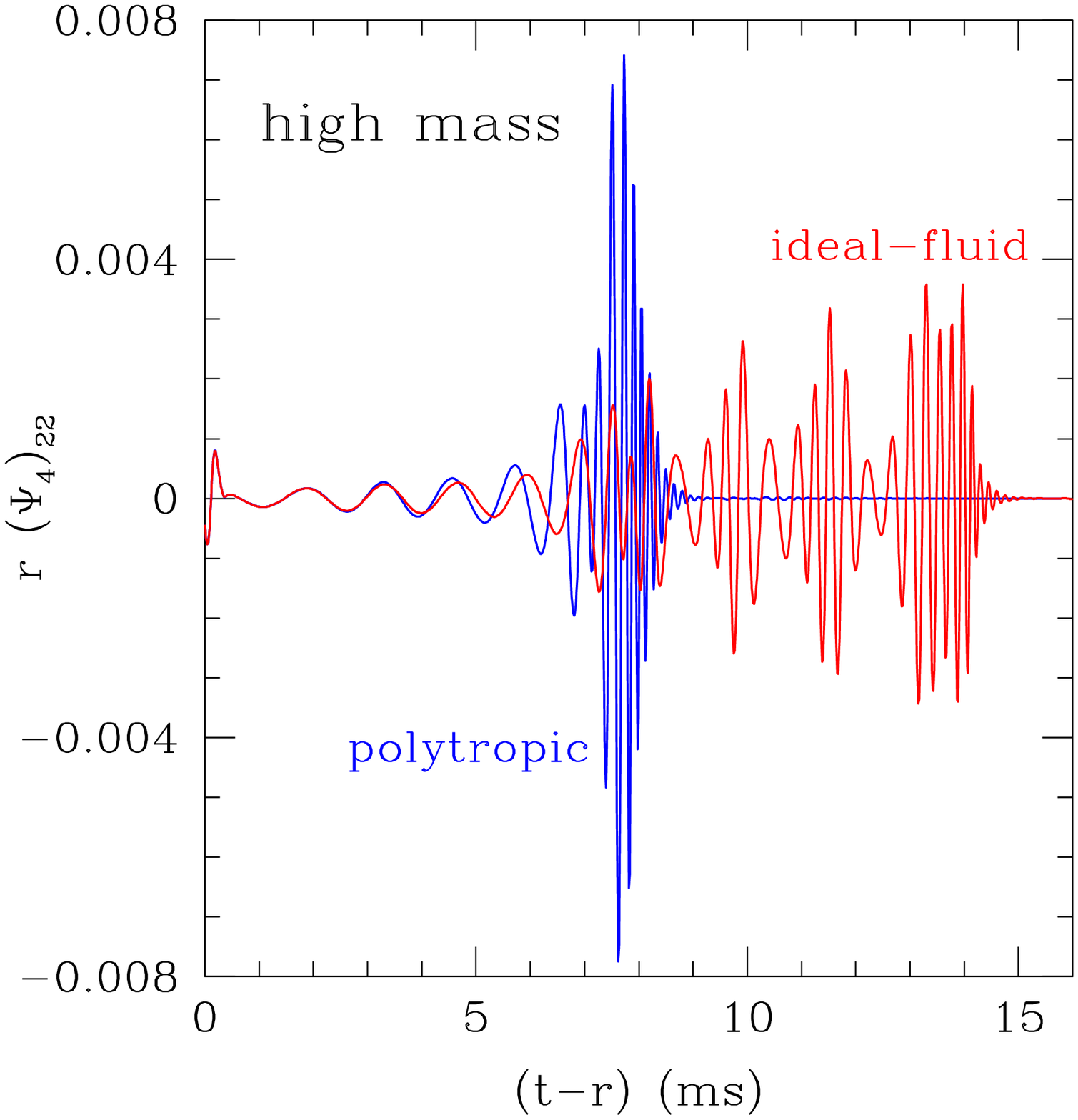}
  \hskip 1.0cm
  \includegraphics[angle=-0,width=0.45\textwidth]{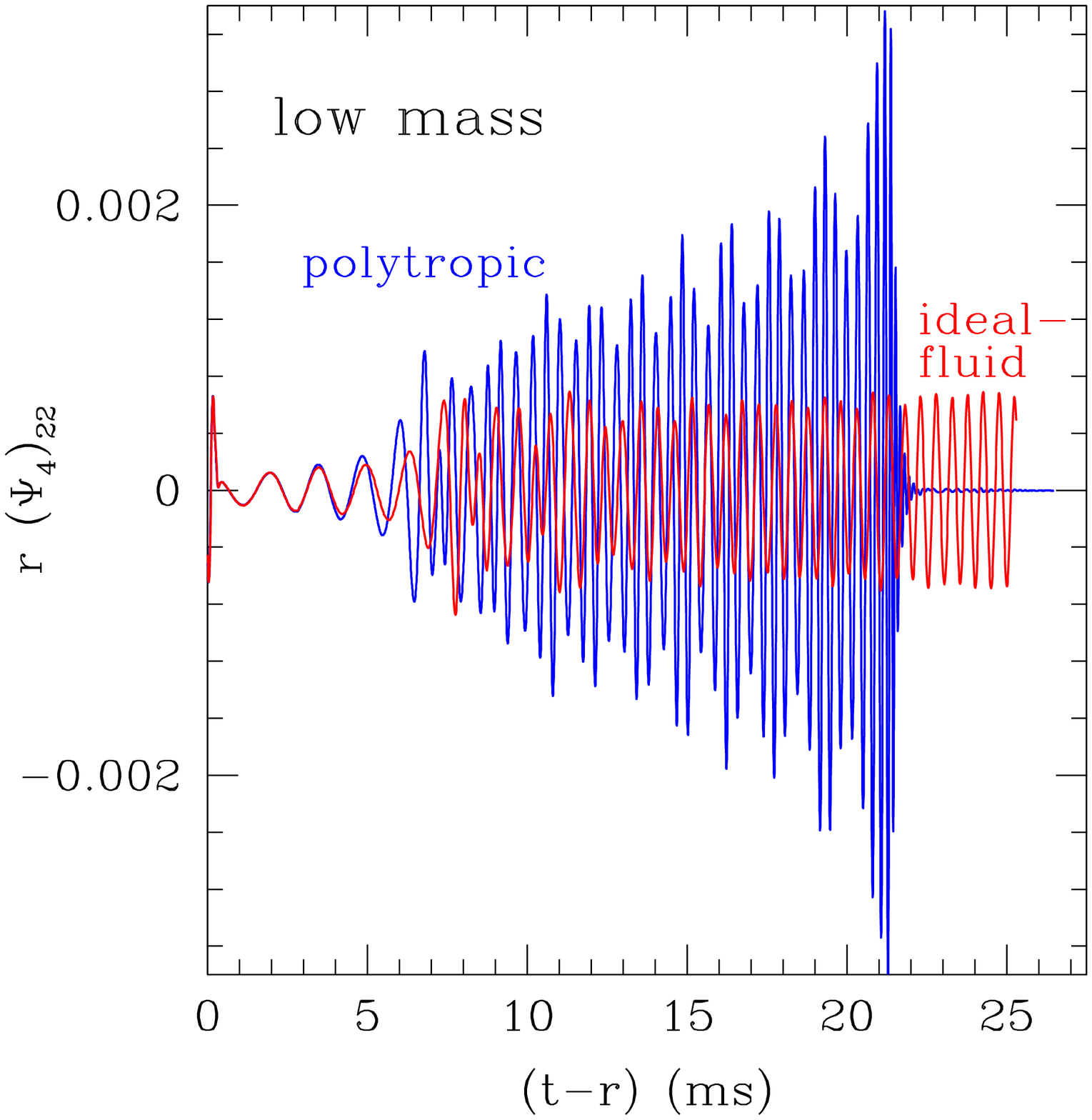}
 \end{center}
 \vskip -0.5cm
   \caption{\textit{Left panel:} Comparison in retarded-time evolution
     of the real part of the $\ell=m=2$ component of $r\Psi_4$ for the
     \textbf{high-mass} binary when evolved with the
     \textbf{polytropic} or with the \textbf{ideal-fluid}
     EOS. \textit{Right panel:} The same as in the left panel but for
     the \textbf{low-mass} binary.}
   \label{fig:psi4_pol_vs_IF}
 \end{figure*}

We conclude this Section by discussing the gravitational-wave signal
emitted by the low-mass binary and reported in
Fig.~\ref{fig:psi4_pol_low}. Also in this case we show in the left
panel the retarded-time evolution of the real part of the $\ell=m=2$
component of $r\Psi_4$, while in the right panel the real part of the
gauge-invariant quantity $Q^{+}_{22}$. As mentioned in the previous
Section, the HMNS has a prominent $m=2$ bar deformation and gradually
evolves towards a configuration which is unstable to gravitational
collapse through the emission of gravitational waves. The loss of
energy and angular momentum progressively reduces the centrifugal
support and increases the compactness of the HMNS which, as a result,
spins more rapidly. This is particularly clear in the evolution of
$\Psi_4$, which is shown in the left panel of Fig.~\ref{fig:psi4_pol_low} and
which exhibits the typical increase in amplitude and frequency of the
gravitational-wave signal. This runaway behavior ends at the time of the 
formation of the BH, which then rings down exponentially as shown in the two
insets. A rapid comparison of Figs.~\ref{fig:psi4_pol_high} and
Fig.~\ref{fig:psi4_pol_low} is sufficient to appreciate the marked
differences introduced in the evolution of the binary by a different
initial mass. In the following Section this comparison will be carried
out also across different EOSs (\cf~Fig.~\ref{fig:psi4_pol_vs_IF}).

\subsection{Waveforms from ideal-fluid binaries}
\label{gws_IF}

As mentioned when discussing the dynamics of ideal-fluid binaries, the
significant differences that emerged both for the evolution of high-
and low-mass binaries are reflected in their gravitational-wave
emission. We recall that ideal-fluid binaries will experience a
considerable increase of their internal energy (temperature) as a
result of the shocks produced at the merger. As a result, a high-mass
binary exhibits a delay in the collapse to BH of $\sim 8\,\ms$, which
should be contrasted with the corresponding $\sim 3\,\ms$ obtained for
the same binary when evolved with a polytropic EOS. Similarly, a
low-mass binary will show a much longer delay, which we estimated to be
$\sim 105\,\ms$ and which is to be contrasted with the corresponding
$\sim 16\,\ms$ obtained for the same binary when evolved with a
polytropic EOS.

This is nicely summarized in Fig.~\ref{fig:psi4_IF_high_low}, whose
left panel shows the retarded-time evolution of the real part of the
$\ell=m=2$ component of $r\Psi_4$ for the high-mass binary. As
commented in Sect.~\ref{IF_binaries_hm}, the HMNS undergoes repeatedly
a dynamical bar-mode instability which develops and is suppressed at
least four times during the post-merger phase, as the two stellar cores
merge. The HMNS becomes increasingly more compact as it loses angular
momentum and thus spins progressively faster. This behavior is clearly
imprinted in the gravitational-wave signal and it is easy to
distinguish the four stages of the bar development at times $t\sim
8,\,10,\,12,$ and $14\,\ms$, respectively. The last one is accompanied
also by the gravitational collapse to BH and exhibits a well-captured
quasi-normal ringing.

The right panel of Fig.~\ref{fig:psi4_IF_high_low}, on the other hand,
refers to the low-mass binary and has a straightforward
interpretation: the HMNS produced has a small $m=2$ deformation and is
still too far from the instability threshold to the collapse to a
BH. Rather, the bar rapidly reaches an equilibrium configuration which
persists over the $16$ revolutions over which the calculations were
carried out. The resulting waveforms are produced at twice the
frequency of the revolution of the bar, \ie at $\sim 2\,\khz$, and show a
remarkably constant amplitude (\cf~inset in the right panel of
Fig.~\ref{fig:psi4_IF_high_low}). It is still unclear whether the
stability of the deformation is the result of the bar being very close
to the dynamical instability threshold or the result of the imposed
$\pi$-symmetry, which prevents the growth and coupling of the
$m=1$ and $m=2$ modes~\cite{Manca07,Saijo2008}. Clarifying this point
will require calculations which are at least twice as expensive but it
will be essential to determine whether the corresponding
gravitational-wave spectrum will be characterized by a large and
predominant peak at $\sim 2\,\khz$ (\cf~right panel of
Fig.~\ref{fig:hs_psd_pol_vs_IF}).

Figure~\ref{fig:psi4_pol_vs_IF} offers in its left panel a comparison
in retarded-time evolution of the real part of the $\ell=m=2$
component of $r\Psi_4$ for the high-mass binaries when evolved with
the polytropic or with the ideal-fluid EOS (\cf~left panels of
Figs.~\ref{fig:psi4_pol_high} and~\ref{fig:psi4_IF_high_low}). When
shown in the same graph, it becomes much easier to appreciate the
impact that the non-isentropic nature of the ideal-fluid EOS has on
the dynamics of the merger and, most importantly, on the
gravitational-wave emission. Clearly, when the waveforms from merging
binary NSs will be detected, they will effectively provide the Rosetta
stone for the deciphering of the stellar structure and EOS. In
addition, the comparison in Fig.~\ref{fig:psi4_pol_vs_IF} can also be
used to gauge the possible range of behaviors that a more realistic
treatment of the matter may yield. Both a polytropic and an
ideal-fluid EOS, in fact, can be considered as the extremes of such a
behavior, with either a perfectly adiabatic evolution in which shocks
(and hence shock heating) cannot occur, or with an evolution in which
local increases of the temperature through shocks are allowed but
cannot lead to radiative processes. Furthermore, because neutrinos or
photons are expected to be trapped and would be able to leave the
system only on diffusion timescales, any realistic EOS will lead to
evolutions similar to those observed with the non-isentropic
ideal-fluid EOS.

Finally, the right panel of Fig.~\ref{fig:psi4_pol_vs_IF}, is the same
as in the left panel but for the low-mass binaries (\cf~left panel of
Figs.~\ref{fig:psi4_pol_low} and right panel of
Fig.~\ref{fig:psi4_IF_high_low}). Also in this case the analogies and
differences have a straightforward interpretation and underline the
importance of considering the time between the merger and the collapse
to BH as an important indicator of the properties of the binary.

\subsection{Energy and Angular-Momentum Losses}
\label{sec:eam_losses}

\begin{figure*}[t]
 \begin{center}
  \includegraphics[angle=-0,width=0.45\textwidth]{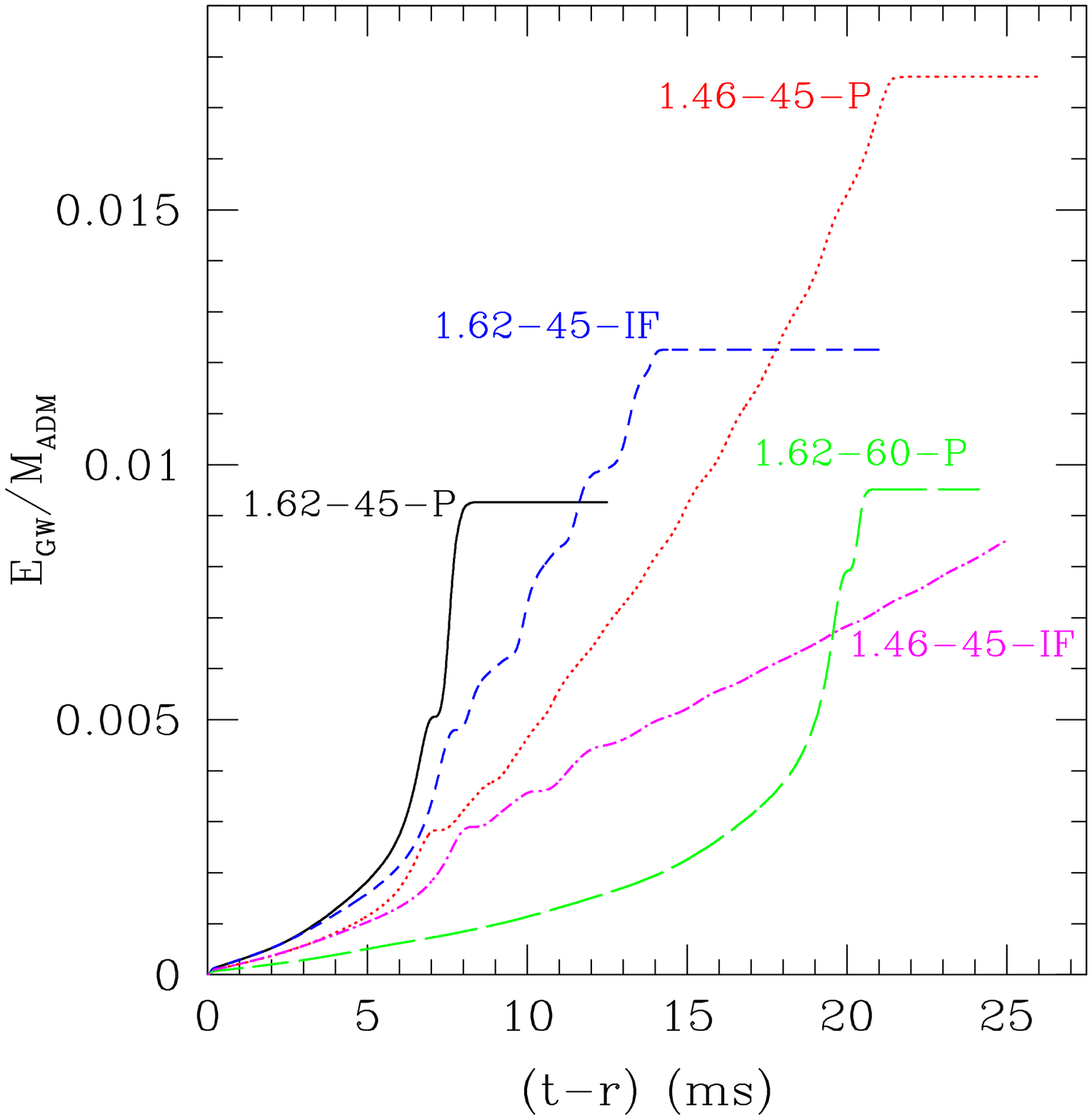}
  \hskip 1.0cm
  \includegraphics[angle=-0,width=0.45\textwidth]{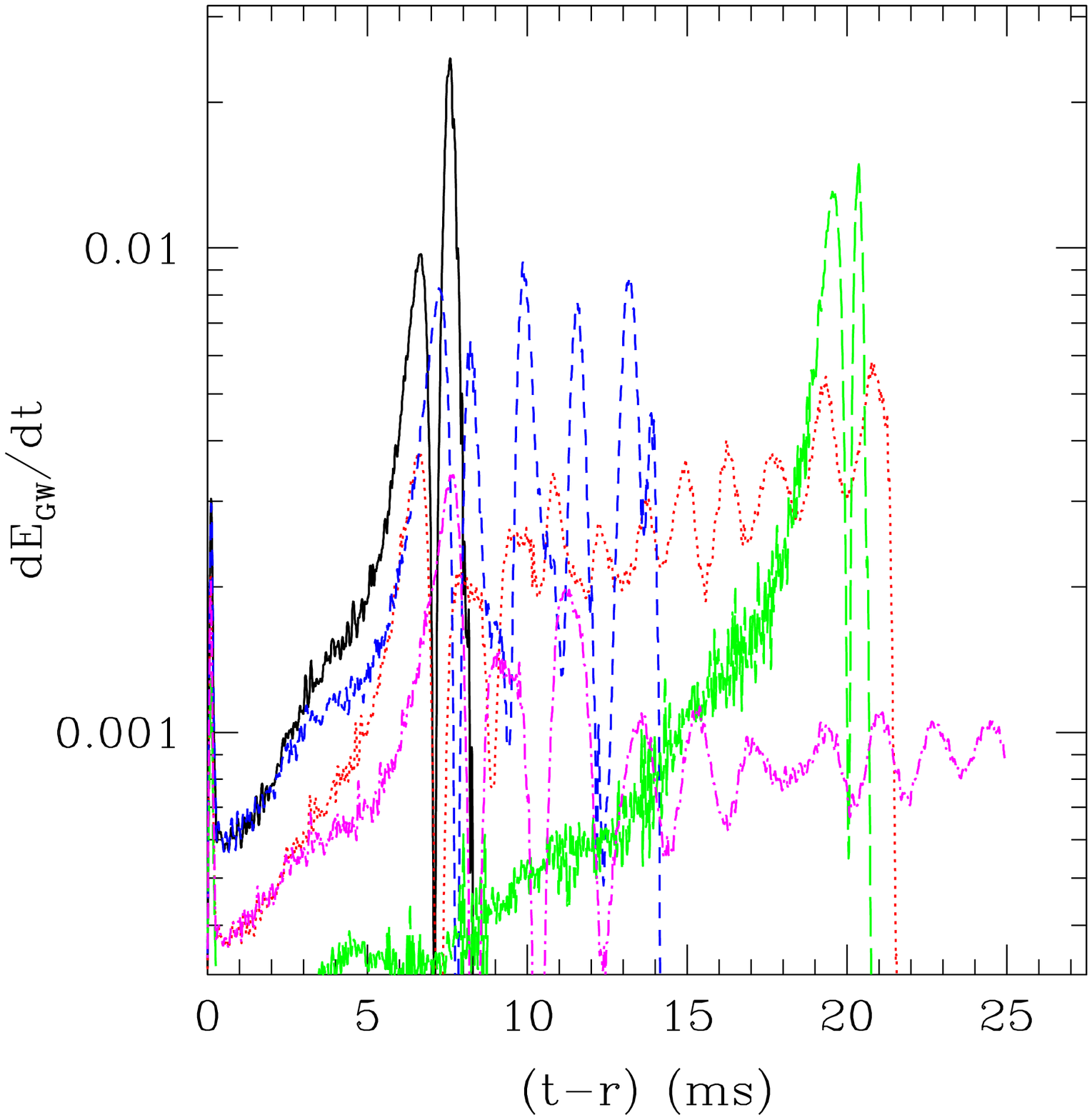}
 \end{center}
 \vskip -0.5cm
   \caption{\textit{Left panel}: Energy emitted in gravitational waves
     for the high-mass binary evolved with a polytropic EOS (solid
     line), for the low-mass binary evolved with a polytropic EOS
     (dotted line), for the high-mass binary evolved with the
     ideal-fluid EOS (dashed line) and for the low-mass binary evolved
     with the ideal-fluid EOS (dot-dashed line). Note that the largest
     amount of radiation comes from the low-mass binary whose emission
     has not been computed before. Indicated with a long-dashed line
     is the high-mass polytropic binary starting at $60\,{\rm
     km}$. \textit{Right panel}: The same as in the left panel but for
     the rate of the energy loss. Note that the largest burst of
     radiation is produced by the high-mass polytropic binary at the
     time of the prompt collapse to a BH.}  \label{fig:Erad}
 \end{figure*}

We have computed the energy and the angular momentum carried away by
gravitational waves using the even and odd-parity perturbations,
$Q_{\ell m}^{+}$ and $Q_{\ell m}^\times$, respectively. The rate of
energy loss, simply given by~\cite{Nagar05}
\begin{equation}
\frac{\mathrm{d}E_{_{\rm GW}}}{\mathrm{d}t}=\frac{1}{32\pi}\sum_{\ell,\,m}\left(\left|\frac{\mathrm{d}Q_{\ell m}^{+}}{\mathrm{d}t}\right|^2+\left|Q_{\ell m}^\times\right|^2\right)\, ,
\label{eq:en_loss}
\end{equation}
is shown in the right panel of Fig.~\ref{fig:Erad} for all the
low-mass and high-mass binaries considered here. In the left panel of
the same figure we show the value of $E_{_{\rm GW}}$ normalized to the
initial ADM mass of the system $M_{_{\rm ADM}}$ as a function of the
retarded time $t-r$ where $r=200\,M_{\odot}= 295\,\km$ is the radius at which
the waveforms were extracted. In both panels the solid line refers to
the high-mass polytropic model $1.62$-$45$-${\rm P}$, the dashed line
to the high-mass ideal-fluid case $1.62$-$45$-${\rm IF}$, the dotted
line to the low-mass polytropic binary $1.46$-$45$-${\rm P}$, the
dotted-dashed line to the low-mass ideal-fluid one $1.46$-$45$-${\rm
  IF}$ and finally the long-dashed line to the high-mass polytropic
model with an initial separation of $60\, \km$, namely
$1.62$-$60$-${\rm P}$.

From the right panel of Fig.~\ref{fig:Erad} it is evident that all the
models have a first maximum in the energy emission rate soon after the
merger. This initial increase in the emission rate is related to the
last part of the inspiral phase, when the amplitude and the frequency
of the gravitational-wave signal increase. After this first peak,
however, the emission rate has a substantial drop, which is common to
all the models and it is due to a very short (\ie $\ll 1\, \ms$)
transition phase in which the deviations from axisymmetric are
smaller. We now concentrate on describing the different dynamics of
the different models after this initial common part, {\it i.e.} on the
emission rate related to the evolution of the system after the merger.

In the case of the two high-mass polytropic binaries,
\ie $1.62$-$45$-${\rm P}$ and $1.62$-$60$-${\rm P}$, there is also a
second peak in the energy emission at the time of the collapse and,
except for the different times at which the merger and the subsequent
collapse to BH take place, their profiles are very similar, with a
total energy emitted which is $\sim 0.01\, M_{_{\rm ADM}}$. This
second peak, which has an amplitude comparable to or higher than the
first one, is simply related to the increase in amplitude and
frequency of the gravitational waves emitted during the collapse (see
also Fig.~\ref{fig:psi4_amp_pol_45_vs_60}). In the case of the
high-mass binary evolved with an ideal-fluid EOS, however, the
emission rate exhibits four peaks after the merger and this is due
to the different post-merger dynamics. As already discussed in
Sect.~\ref{IF_binaries_hm}, instead of collapsing promptly to a BH as
the polytropic one, this system forms a bar-shaped HMNS with the
high-density cores of the two NSs periodically merging and bouncing
until sufficient angular momentum is carried away and the collapse
starts. These periodic bounces and mergers of the two cores determine
the several peaks seen in the emission rates. At the end, the total
energy radiated through gravitational-waves is larger than the one
emitted in the polytropic case and is $\simeq 0.012\, M_{_{\rm ADM}}$.

For the two low-mass binaries, $1.46$-$45$-${\rm P}$ and
$1.46$-$45$-${\rm IF}$, on the other hand, the emission rate is always
smaller than for the high-mass binaries, but it shows several peaks
and for a longer time. This is related to the dynamics of the
bar-deformed HMNSs that rotate for several stellar periods before
collapsing to BHs. As a result, even if the emission rate is smaller,
the total energy emitted in gravitational waves is much larger and in
the case of the low-mass polytropic binary is $\simeq 0.018\,M_{_{\rm
    ADM}}$ at the time of the collapse, while for the low-mass
ideal-fluid binary it can be estimated to be $\approx 0.04\,M_{_{\rm
    ADM}}$ when extrapolating the time of the collapse to $t\approx
110\, \ms$ (see discussion in Sect.~\ref{IF_binaries_lm}).

The two panels in Fig.~\ref{fig:Erad} are particularly useful to
appreciate and quantify the differences that emerge among different
binaries in the inspiral phase and, later on, in the post-merger
phase. It is particularly instructive to consider the similarity in
the evolutions of binaries having the same initial separation and
mass, but different EOS, \ie $1.62$-$45$-${\rm P}$ and
$1.62$-$45$-${\rm IF}$ or $1.46$-$45$-${\rm P}$ and $1.46$-$45$-${\rm
  IF}$. We recall that these sets of binaries have exactly the same
initial data and hence the differences during the inspiral are due
uniquely to the role played by the EOS. As clearly shown in the left
panel of Fig.~\ref{fig:Erad}, these differences are very small, so
that $1.62$-$45$-${\rm P}$ and $1.62$-$45$-${\rm IF}$ have lost to
gravitational waves essentially the same amount of mass at the time of
the merger, although the latter actually takes place at slightly
different times (\ie $t\sim 5.3\,\ms$ for $1.62$-$45$-${\rm P}$ and
~$t\sim 5.8\,\ms$ for $1.62$-$45$-${\rm IF}$; see the discussion in
Sec.~\ref{IF_binaries_hm}). Because an identical comment also applies
for $1.46$-$45$-${\rm P}$ and $1.46$-$45$-${\rm IF}$, we conclude that
the EOS introduces major differences in the binary evolutions only
\textit{after} the merger.

On the contrary, for binaries having the same EOS but different masses
(\eg binaries $1.62$-$45$-${\rm P}$ and $1.46$-$45$-${\rm P}$), also
the evolution \textit{before} the merger is different and can
contribute to different post-merger evolutions (see the comment below
on the angular-momentum losses).

\begin{figure*}[t]
 \begin{center}
  \vskip 0.75cm
  \includegraphics[angle=-0,width=0.45\textwidth]{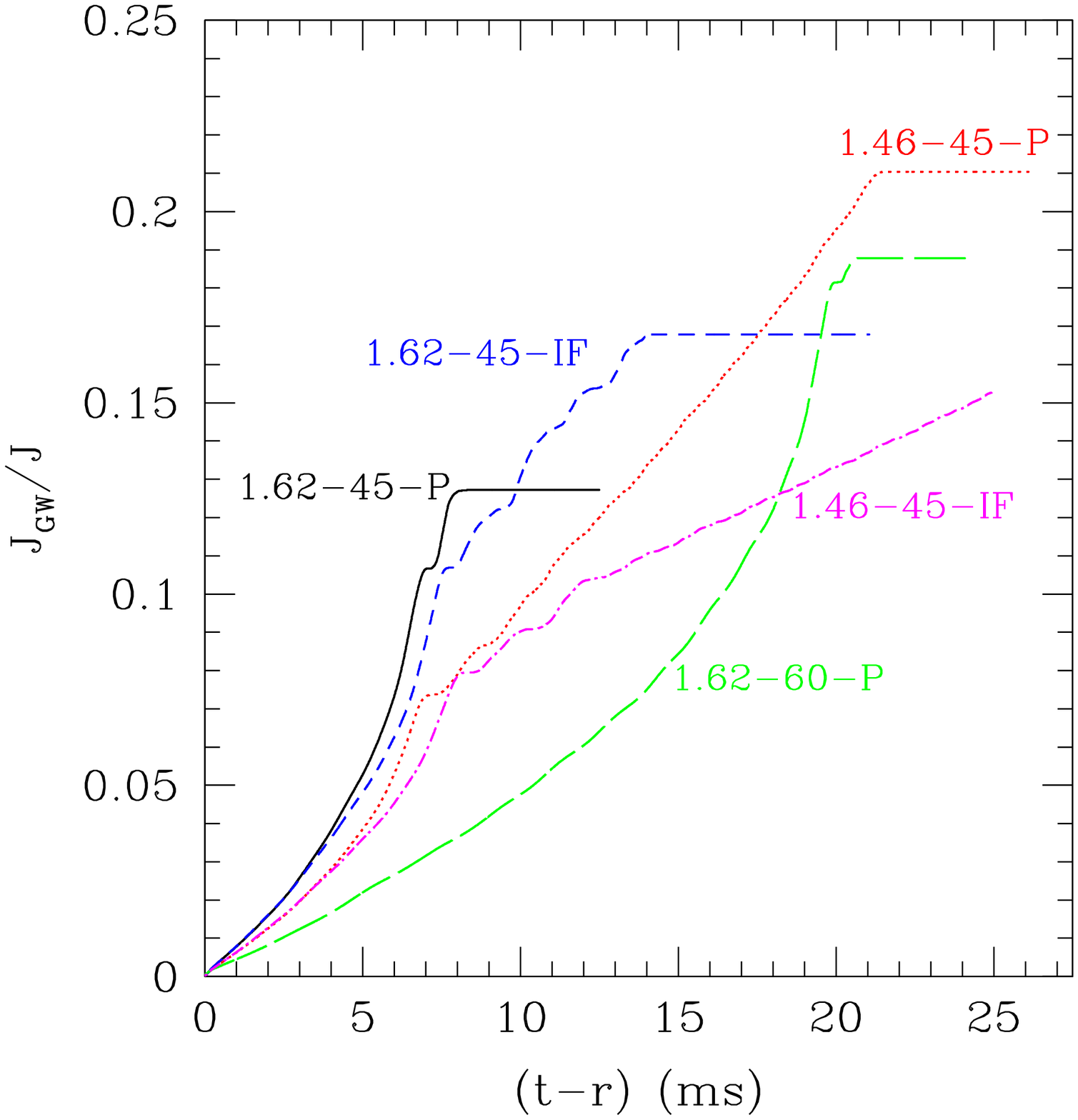}
  \hskip 1.0cm
  \includegraphics[angle=-0,width=0.45\textwidth]{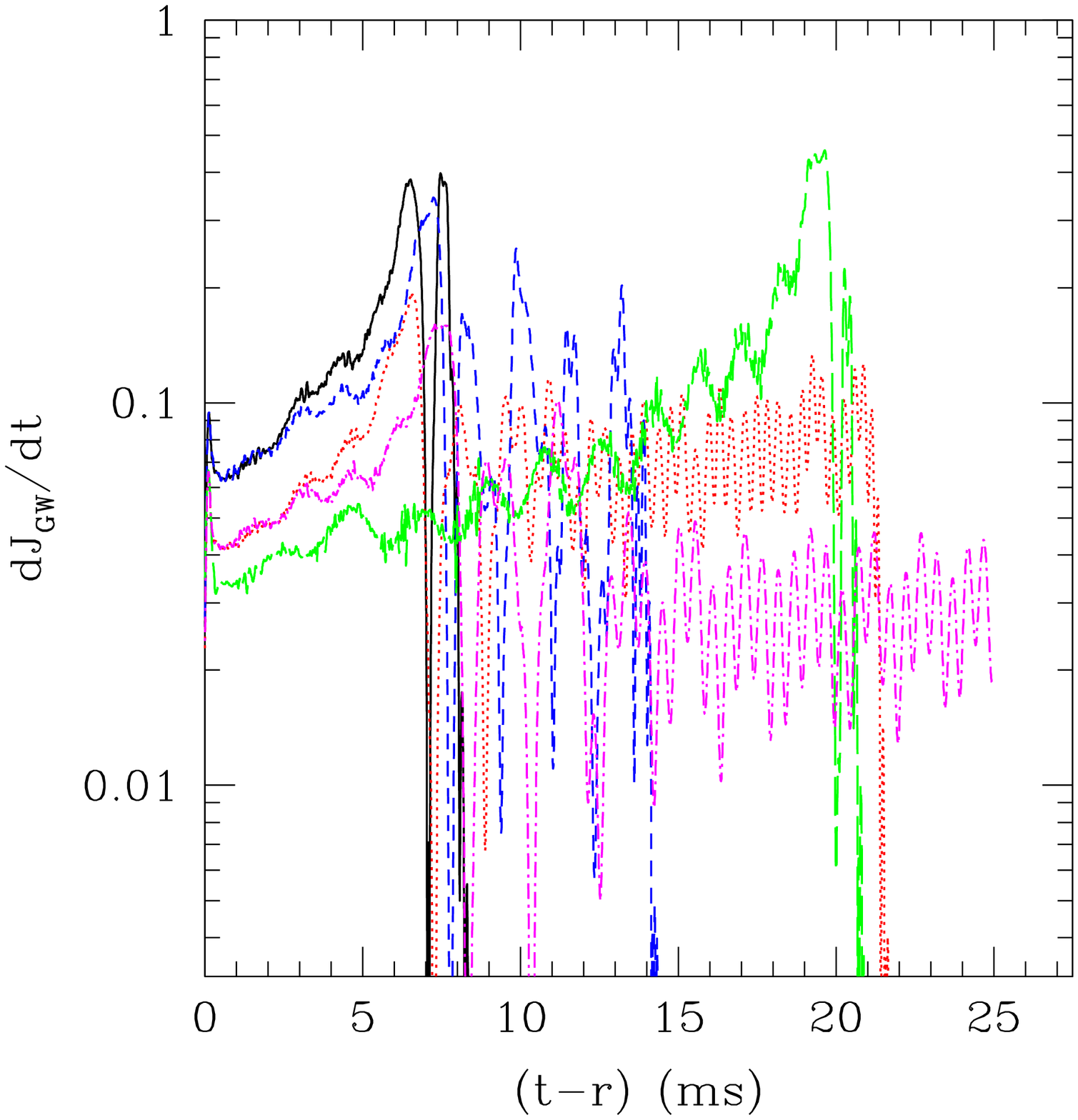}
 \end{center}
 \vskip -0.5cm
   \caption{\textit{Left panel}: The same as in Fig.~\ref{fig:Erad}
     but for the orbital angular momentum normalized to its initial
     value (\cf Table~\ref{table:ID}). \textit{Right panel}: The same
     as in Fig.~\ref{fig:Erad} but for the rate of loss of orbital
     angular momentum.}  \label{fig:Jrad}
 \end{figure*}

In a similar way, we have computed the angular-momentum loss
as~\cite{Nagar05}
\begin{eqnarray}
\label{dJdt_RWM}
&&\frac{dJ_{_{\rm GW}}(t)}{dt}=\frac{1}{32\pi}\sum_{\ell, m}{\rm i} m
        \left[\frac{dQ^{+}_{\ell m}}{dt}
        \left(Q^{+}_{\ell m}\right)^*+\right. \nonumber \\
&&\hskip 3.0cm
        \left. Q_{\ell m}^{\times}\int_{-\infty}^{t}
        \left(Q_{\ell m}^{\times}\right)^*(t')dt'\right]
\end{eqnarray}
and, in analogy with Fig.~\ref{fig:Erad}, of which we use the same
line-type convention, we show in the left panel of Fig.~\ref{fig:Jrad}
the loss of angular momentum normalized to the initial angular
momentum of the system and in the right panel the loss rate.


Overall, the angular-momentum losses and loss rates follow rather
closely the behaviour already discussed for the energy, namely there
is very little difference during the inspiral for binaries having the
same mass. When looking more carefully, however, it is possible to
note that the evolution of the angular momentum shows small
differences after about $3\ms$ even for binaries with the same mass,
\eg $1.62$-$45$-${\rm P}$ and $1.62$-$45$-${\rm IF}$. This
time corresponds roughly to that of the first orbit, after which the
non-isentropic evolution of model $1.62$-$45$-${\rm IF}$
will have changed the stellar structure only slightly but in a manner
sufficient to produce a different emission of gravitational waves and
hence a different loss of angular momentum. Indeed it is clear from
the left panel of Fig.~\ref{fig:Jrad} that the merger begins when the
two binaries have lost $\sim 1\%$ of their initial angular momentum
but that this takes place at different times for the two binaries,
happening earlier for model $1.62$-$45$-${\rm P}$ which is isentropic,
more compact and with a larger quadrupole moment.

More marked are the differences seen when comparing binaries differing
only in the mass (\eg binaries $1.62$-$45$-${\rm P}$ and
$1.46$-$45$-${\rm P}$ or binaries $1.62$-$45$-${\rm IF}$ and
$1.46$-$45$-${\rm IF}$). We recall that these two sets of binaries
essentially merge at the same time and it is then apparent from
Fig.~\ref{fig:Jrad} that at the time of the merger the high-mass
binary will have lost a larger relative amount of the initial orbital
angular momentum. As a result, the matter orbiting outside the AH when
this forms will also have a smaller amount of angular momentum and is
therefore more likely to be more rapidly accreted. This explains why
the high-mass polytropic binary $1.62$-$45$-${\rm P}$ produces a torus
with a smaller rest mass than the low-mass polytropic binary
$1.46$-$45$-${\rm P}$, both at the AH formation and after $3\,\ms$
(\cf~Table~\ref{table:FD})\footnote{Since we cannot follow the
  low-mass ideal-fluid binary till BH formation we cannot verify that
  this conclusion holds also for the ideal-fluid binaries, although we
  expect so.}.

This behaviour indicates that, at least for binaries having the same
EOS, the rate of loss of angular momentum during the inspiral phase
plays an important role in determining the final mass of the torus and
that the models that lose less angular momentum during the inspiral,
hence comparatively \textit{low-mass} binaries, are expected to have
comparatively \textit{high-mass} tori. This confirms what is already
observed in ref.~\cite{Shibata06a}.

Note, however, that such a simple conclusion is strictly true for
binaries having the same EOS and when no radiative losses are taken
into account. Under more generic conditions, however, the EOS is also
expected to play an important role and a representative example comes
from comparing the high-mass binaries $1.62$-$45$-${\rm P}$ and
$1.62$-$45$-${\rm IF}$. In this case, in fact, the loss of angular
momentum during the inspiral is essentially the same (\cf~left panel
of Fig.~\ref{fig:Jrad}), but it is substantially different after the
merger, with a loss of angular momentum which is at least $50\%$
larger for the ideal-fluid binary. Yet, because of the increased
pressure support the latter produces a torus with a mass which is
$\sim 7$ times larger than the corresponding one for the polytropic
binary.

\subsection{Gravitational-Wave Spectra and Signal-to-Noise Ratios}
\label{sec:SNR}

\begin{figure*}[t]
\begin{center}
 \includegraphics[angle=-0,width=0.45\textwidth]{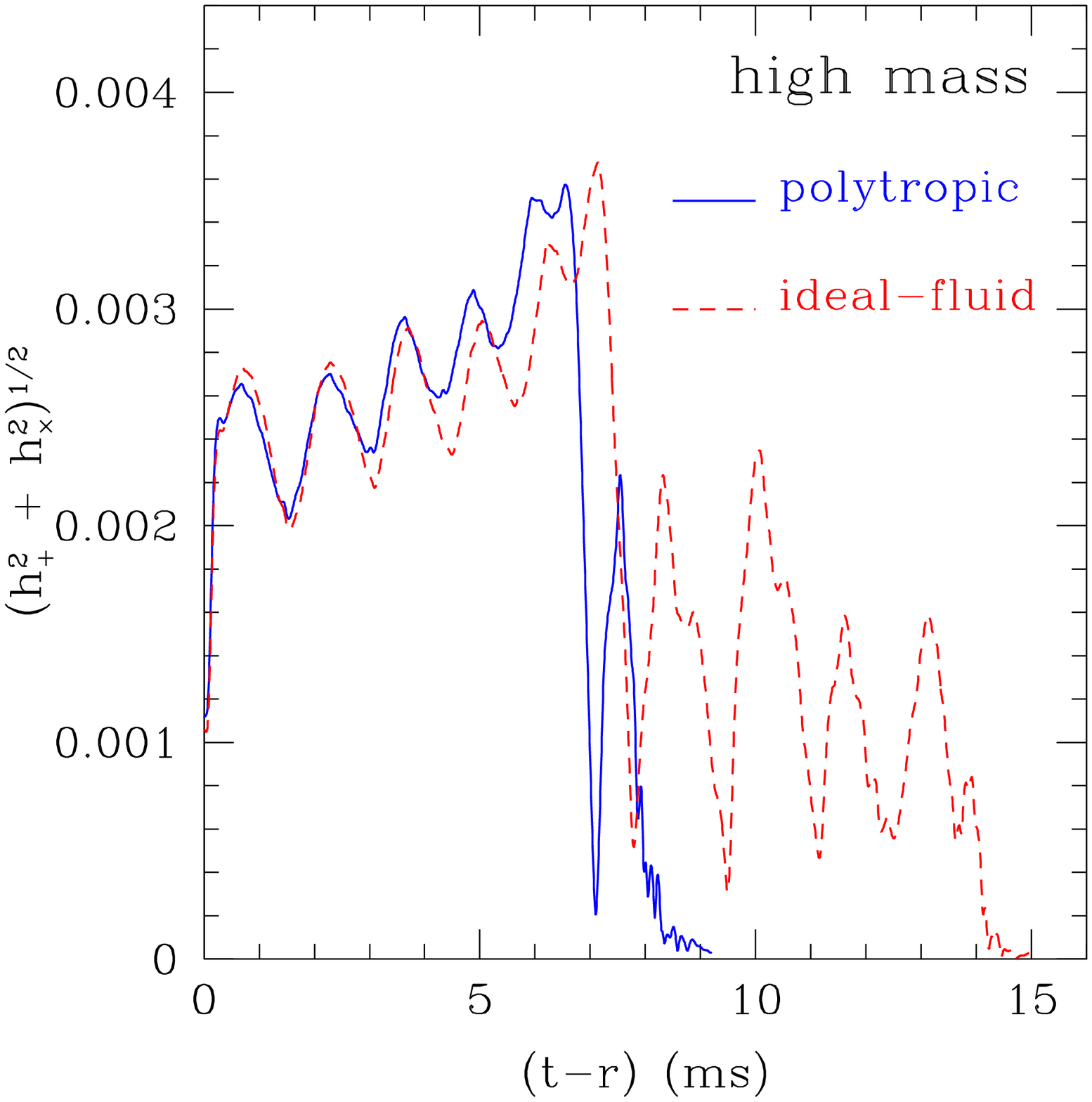}
 \hskip 1.0cm
 \includegraphics[angle=-0,width=0.45\textwidth]{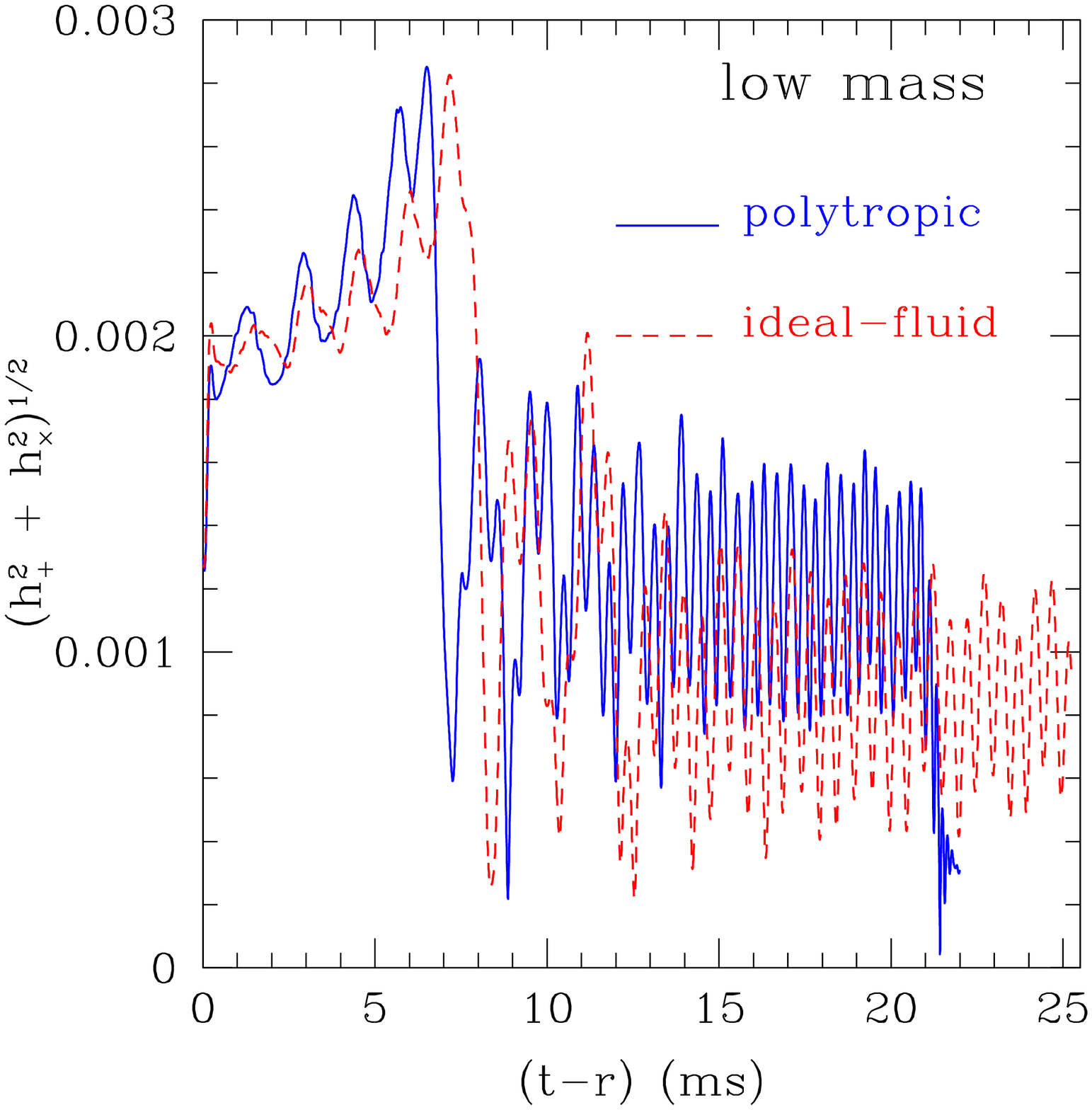}
\end{center}
\vskip -0.5cm
  \caption{\textit{Left panel}: Comparison of retarded-time
    evolution of the amplitude the $\ell=m=2$ component of
    $h=(h^2_{+}+h^2_{\times})$ for the \textbf{high-mass} binary
    when evolved with the \textbf{polytropic} (solid line) or with
    the \textbf{ideal-fluid} (dashed line) EOS; \textit{cf.},
    Fig.~\ref{fig:psi4_pol_vs_IF}, left panel. \textit{Right panel}:
    The same as in the left panel but for the \textbf{low-mass}
    binary; \textit{cf.}, Fig.~\ref{fig:psi4_pol_vs_IF}, right
    panel.}  \label{fig:hs_ampl_pol_vs_IF}
\end{figure*}

We have also studied and compared the amplitudes and frequencies of
the gravitational-wave signal produced by the different models. In
particular in Fig.~\ref{fig:hs_ampl_pol_vs_IF} we plot the amplitude
of the $\ell=m=2$ component of the total gravitational-wave amplitude
$\sqrt{h^2_{+}(t)+h^2_{\times}(t)}$ [where we neglect the
  contribution of the spin-weighted spherical harmonic $_{-2}Y^{22}$
  in equation~\eqref{eq:wave_gi}], for four different binaries, all
starting from an initial separation of $45\,\km$, as a function of the
retarded time $t-r$, where $r=200\,M_{\odot}= 295\,\km$ is the radius at which
the signal was extracted. In particular, in the right panel we show
the evolution of the gravitational-wave amplitude for the low-mass
binaries $1.46$-$45$-$*$ evolved using a polytropic EOS (solid line)
and an ideal-fluid EOS (dashed line) while in the left panel we show
the same but for the high-mass binaries $1.62$-$45$-$*$. We recall
that for all of these models the merger takes place after $\approx
5\,\ms$, which corresponds to the time when the amplitude reaches
its maximum. The slight difference in the position of these maxima
between the polytropic and the ideal-fluid binaries is related to the
difference in the time of the merger and is $\lesssim 1\,\ms$.

Since the dynamics in the inspiral are very similar, the two high-mass
binaries have a very similar and increasing amplitude, up to the
merger. Note, however, that the increase is not monotonic and this is
due mostly to the presence of a nonzero eccentricity. As commented in
Sect.~\ref{sec:poly_hm}, a good part of the eccentricity is due to
gauge effects (and is significantly reduced when the shift vector is
initially set to zero), but a small portion of it is also genuinely
present in the initial data. Fortunately this spurious eccentricity
has only a small impact in the power-spectral density (PSD) of the
gravitational-wave signal and it is easy to isolate being it at $\sim
4$ times the orbital frequency. The evolution of the amplitude in the
post-merger phase is rather different and, while it drops
significantly for the polytropic binary, it remains at large values
for the ideal-fluid binary as a result of the delayed collapse to BH;
as we will comment later on, this will have an impact also on the
detectability of this signal.

The two low-mass binaries in the right panel of
Fig.~\ref{fig:hs_ampl_pol_vs_IF} also show a similar evolution up to
the merger with an increase of the amplitude which is modulated by
eccentricity and reaches its maximum at the merger. Of course, the
maximum value reached in this case is lower than the one obtained in
the high-mass cases. After the merger the amplitude is reduced by a
factor of $\sim 2$ and remains to that level for the $\approx 15 \ms$ 
which separate the merger and the collapse to a BH. In the
case of the ideal-fluid binary, on the other hand, the post-merger
amplitude is smaller and essentially constant for the whole
time the simulation was carried out. As mentioned already, this binary
is expected to collapse to a BH on a timescale of $\sim 110\,\ms$.

We next consider the gravitational-wave emission in the frequency
domain and for this we have computed the power-spectral density
of the effective amplitude
\begin{equation}
{\tilde h}(f)\equiv \sqrt{\frac{{\tilde h}^2_{+}(f)+
	{\tilde h}^2_{\times}(f)}{2}}\,,
\end{equation}
where $f$ is the gravitational-wave frequency and where
\begin{equation}
{\tilde h}_{+,\times}(f) \equiv \int_0^{\infty}
	e^{2\pi {\rm i} f t} h_{+,\times}(t) dt
\end{equation}
are the Fourier transforms of the gravitational-wave amplitudes
$h_{+,\times}(t)$, built using only the largest $\ell=m=2$ multipole.

\begin{figure*}[t]
 \begin{center}
  \vskip 1.0cm
  \includegraphics[angle=-0,width=0.45\textwidth]{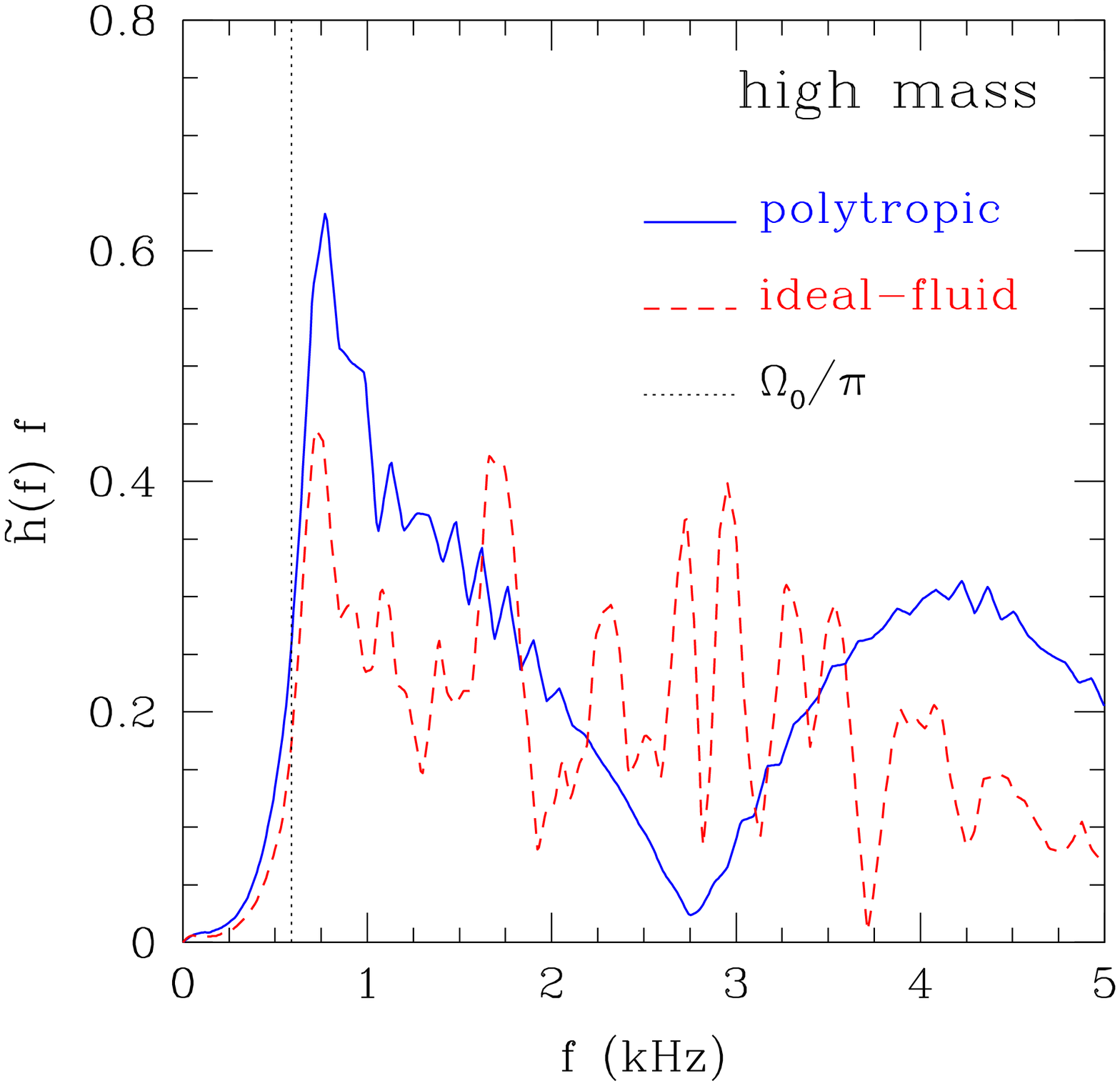}
   \hskip 1.0cm
  \includegraphics[angle=-0,width=0.45\textwidth]{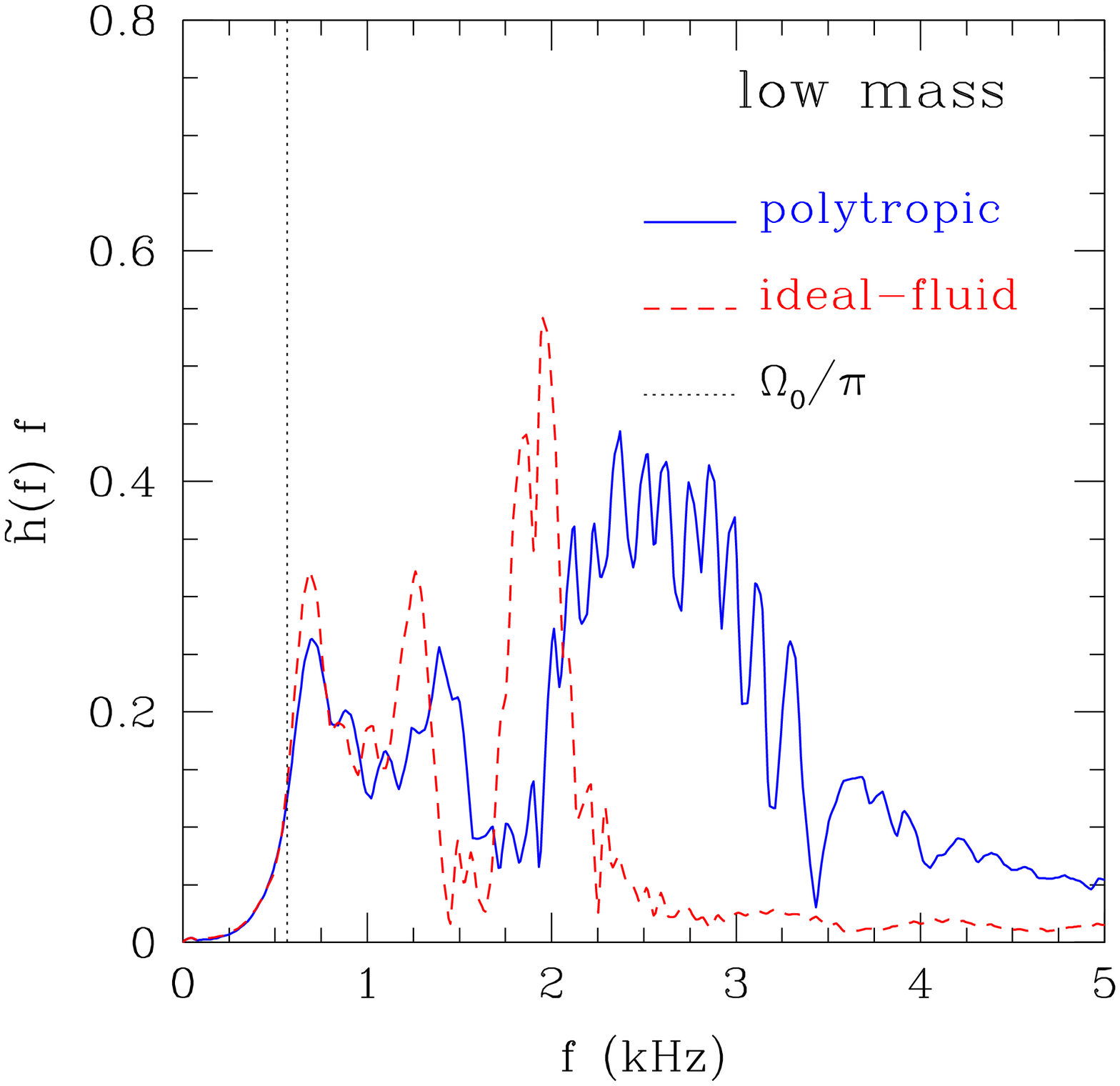}
 \end{center}
 \vskip -0.5cm
   \caption{\textit{Left panel}: Comparison of the PSD of the
     $\ell=m=2$ component of $h(f)f$ for the \textbf{high-mass} binary
     when evolved with the \textbf{polytropic} (solid line) or with
     the \textbf{ideal-fluid} (dashed line) EOS; \textit{cf.},
     Fig.~\ref{fig:psi4_pol_vs_IF}, left panel. \textit{Right panel}:
     The same as in the left panel but for the \textbf{low-mass}
     binary; \textit{cf.}, Fig.~\ref{fig:psi4_pol_vs_IF}, right
     panel. Indicated with a vertical long-dashed line is twice the
     initial orbital frequency.}
   \label{fig:hs_psd_pol_vs_IF}
\end{figure*}

In Fig.~\ref{fig:hs_psd_pol_vs_IF} we compare the spectral
distribution of the quantity $\tilde{h}(f)f$ for the high-mass
binaries (left panel) and the low-mass binaries (right panel) when
evolved with the two EOSs. In both cases we use a solid line for the
polytropic binaries and a dashed line for the binaries evolved with
the ideal-fluid EOS.  Also indicated in both panels with a vertical
long-dashed line is the frequency corresponding to twice the initial
orbital frequency $f_0\equiv \Omega_0/(2\pi)$ where $f_0=283\,\hz$ for
the low-mass binaries and $f_0=295\,\hz$ for the high-mass ones. These
frequencies are representative of the signal at the beginning of the
simulated inspiral and thus represent lower cut-off frequencies, below
which the PSD is not meaningful. On the other hand, the peaks in the
PSDs observed at frequencies slightly larger than the orbital ones are
very important as they refer to the power emitted during the inspiral.

\begin{table*}[t]
  \caption{Signal-to-noise ratio (SNR) computed for different
    detectors assuming a source at $10\,\mpc$. The different columns
    refer to: the proper separation between the centers of the stars
    $d/M_{_{\rm ADM}}$; the baryon mass $M_{b}$ of each star in solar
    masses; the total ADM mass $M_{_{\rm ADM}}$ in solar masses, as
    measured on the finite-difference grid; the approximate
    quasi-normal mode frequency of the fundamental mode $f_{_{\rm
        QNM}}$ in $\khz$; the SNR for Virgo, LIGO, Advanced LIGO and
    GEO.}
\begin{ruledtabular}
\begin{tabular}{l|ccccccccc}
Model &
\multicolumn{1}{c}{$d/M_{_{\rm ADM}}$} &
\multicolumn{1}{c}{$M_{b}\ (M_{\odot})$} &
\multicolumn{1}{c}{$M_{_{\rm ADM}}\ (M_{\odot})$} &
\multicolumn{1}{c}{$f_{_{\rm QNM}}(\khz)$} &
\multicolumn{1}{c}{SNR~(Virgo)}&
\multicolumn{1}{c}{SNR~(LIGO)}&
\multicolumn{1}{c}{SNR~(Adv.~LIGO)}&
\multicolumn{1}{c}{SNR~(GEO)}\\
\hline
$1.46$-$45$-${\rm P}$  & $14.3$ & $1.456$ & $2.681$ & $7.3$ & $1.92$ & $1.33$ & $12.54$ & $0.57$\\
$1.46$-$45$-${\rm IF}$ & $14.3$ & $1.456$ & $2.681$ & --    & $2.08$ & $1.45$ & $13.52$ & $0.62$\\
$1.62$-$45$-${\rm P}$  & $12.2$ & $1.625$ & $2.982$ & $6.7$ & $2.15$ & $1.48$ & $13.29$ & $0.63$\\
$1.62$-$45$-${\rm IF}$ & $12.2$ & $1.625$ & $2.982$ & $7.0$ & $2.29$ & $1.57$ & $14.42$ & $0.67$\\
$1.62$-$60$-${\rm P}$  & $16.8$ & $1.625$ & $2.987$ & $6.5$ & $3.97$ & $3.15$ & $35.52$ & $1.48$
\end{tabular}
\end{ruledtabular}
\vskip -0.25cm
\label{table:SNR}
\end{table*}

The PSD for the high-mass polytropic binary (left panel of
Fig.~\ref{fig:hs_psd_pol_vs_IF}) is quite simple, as it shows, besides
the inspiral peak, also a peak at $f\approx 4\,\khz$, corresponding to
the collapse of the HMNS (\cf~left panel of
Fig.~\ref{fig:hs_ampl_pol_vs_IF}). Note that the PSD shown does not
include the frequency of the fundamental QNM of the newly produced
BH. Using the approximate expression~\cite{Leaver85,Echeverria89}
\begin{equation}
f_{_{\rm QNM}} \approx 3.23 \left(\frac{M_{_{\rm BH}}}{10\,M_{\odot}}\right)^{-1} 
	[1 - 0.63(1-a)^{0.3}]\quad \khz\,,
\end{equation}
this frequency is $f_{_{\rm QNM}} \simeq 6.7\,\khz$ for the BH
produced by this binary (\cf~Table~\ref{table:SNR}).

The PSD for the high-mass ideal-fluid binary, on the other hand, is
more complex, with the inspiral peak at $f\approx 0.75\,\khz$ being
accompanied by a number of other peaks, the most prominent having a
similar amplitude at $f\approx 1.75\,\khz$ and $f\approx
3\,\khz$. These additional peaks (and also the smaller ones between
the two) are obviously related to the post-merger phase at $t \gtrsim 5
\,\ms$ and, in particular, to the dynamics of the HMNS
formed after the merger and especially to the dynamics of the cores of
the two NSs, which merge and bounce several times before the HMNS
collapses to a BH, producing a small peak at $f\approx
4\,\khz$. Also in this case even the fundamental QNM has a frequency
$f_{_{\rm QNM}} \simeq 7.0\,\khz$ (\cf~Table~\ref{table:SNR}) and is
therefore outside the range shown in Fig.~\ref{fig:hs_psd_pol_vs_IF}.

In a similar way it is possible to interpret the PSDs of the low-mass
binaries. The polytropic one, in particular, shows an excess power at
$f\approx 0.75\,\khz$ due to the inspiral but also a very broad peak
between $f\approx 2\,\khz$ and $f\approx 3.5\,\khz$, that is related
to the dynamics of the bar-deformed HMNS formed after the merger and
persisting for several milliseconds. Also in this case a small excess
power is seen at $f\gtrsim 4\,\khz$ and is associated with the
collapse to BH, whose fundamental QNM has a frequency $f_{_{\rm QNM}}
\simeq 7.3\,\khz$. Interestingly, the low-mass ideal-fluid PSD does
not show the broad peak but a very narrow and high-amplitude one at
$f\approx 2\khz$. This is obviously related to the long-lived bar
deformation of the HMNS, which we have followed for $\sim 16$
revolutions (as computed from the cycles of the quadrupolar
gravitational radiation). At this stage it is unclear whether this
prominent peak will survive when the simulations are repeated without
the use of a $\pi$-symmetry and more conclusive results on this will
be presented elsewhere~\cite{BGR08}. Note that the high-frequency part
of the PSD for the low-mass ideal-fluid binary (\ie for $f\gtrsim
2\,\khz$) is essentially zero, because of the absence of a collapse to
BH, which for this binary takes place in an excessively long time.

A fundamental piece of information necessary to assess the relevance
of binary NSs as sources of gravitational waves comes from the
calculation of the SNR which we have computed for interferometric
detectors such as Virgo, LIGO, Advanced LIGO and GEO. For all the
models discussed above, including the high-mass polytropic binary with
a larger initial separation of $60\,\km$, the SNR has been computed as
\begin{equation}
\label{eq:SNR}
\left(\frac{S}{N}\right)^2=4\int_0^\infty
	\frac{|\tilde{h}_+(f)|^2}{S_h(f)} df \; ,
\end{equation}
where $S_h(f)$ is the noise power-spectral density for a given
detector. The results computed assuming a source at a distance of
$10\, {\rm Mpc}$ are reported in Table~\ref{table:SNR} and show that,
while a detection is ideally possible with the current interferometers
[the SNR is ${\cal O}(1)$], it is unlikely in practice, given the small
event rate at such distances, \ie $\approx 0.01 {\rm yr}^{-1}$. On the
other hand, larger SNRs ${\cal O}(10)$ can be obtained with advanced
detectors. This also means that a detection of these sources up
to a distance of $100\, {\rm Mpc}$ will be possible and so there will a higher event
rate. Interestingly, binaries of the same mass, but described by a 
non-isentropic EOS have a slightly higher SNR and this is simply due
to the increase in the delay for the collapse to BH.

Both the small range in which the masses of NSs fall and the low
sensitivity of present detectors in the high-frequency region, where a
lot of the power is emitted, underline the importance of the inspiral
phase for the detection. This is particularly evident when comparing
the large SNR of signals in which the inspiral is a significantly long
part. The signal for the high-mass polytropic binary $1.62$-$60$-${\rm
  P}$ starts from an initial separation of $60\,\km$ and
spans over more than $5$ orbits, resulting in a SNR which is a factor
of $3$ larger than the one of the other binaries, which have an initial
separation of $45\,\km$ and merge in little more than $2$ orbits. This
result strongly motivates the investigation, both through simulations
and PN approximations, of binaries inspiralling over timescales longer
than the already long ones presented here. Stated differently, the
study of longer simulations can be used to assess when the lower-order
PN expressions are very accurate, while the study of the final part of
the inspiral (say the last two orbits) can be used to determine those
higher-order PN effects that have not been worked out analytically
yet.

\section{Conclusions}

We have discussed accurate general-relativistic simulations of binary
systems of equal-mass NSs which inspiral starting from irrotational
configurations in quasi-circular orbit. Spanning over $\sim 30\,\ms$,
our simulations are the longest of their kind and provide the first
complete (within an idealized treatment of the matter) description of
the inspiral and merger of a NS binary leading to the \textit{prompt}
or \textit{delayed} formation of a BH and to its ringdown.

More specifically, we have considered binary NSs with two different
initial masses: low-mass binaries with $M_{_{\rm ADM}}=2.681
M_{\odot}$ and high-mass binaries with $M_{_{\rm ADM}}=2.982
M_{\odot}$. Such binaries have then been evolved using two different
EOSs: namely an isentropic (\ie polytropic) EOS and a non-isentropic
(\ie ideal-fluid) EOS. Despite the use of only simple, analytical
EOSs, we were able to reproduce some of the aspects that a more
realistic EOS would yield. In particular, we have shown that the
polytropic EOS leads either to the \textit{prompt} formation of a
rapidly rotating BH surrounded by a dense torus in the high-mass case,
or, in the low-mass case, to a HMNS which develops a bar, emits large
amounts of gravitational radiation and eventually experiences a
\textit{delayed} collapse to BH. Conversely, we have shown that the
ideal-fluid EOS inevitably leads to a further delay in the collapse to
BH as a result of the larger pressure support provided by the
temperature increase via shocks. In this case the temperature in the
formed HMNS can reach values as high as $10^{11}-10^{12}{\rm K}$, so
that the subsequent dynamics and especially the time of the collapse
can be reduced if cooling mechanisms, such as the direct-URCA process,
take place.

With the exception of the low-mass ideal-fluid binary, whose HMNS is
expected to collapse to BH on a timescale which is computationally
prohibitive (\ie $\sim 110\,\ms$), all the binaries considered lead
to the formation of a BH surrounded by a rapidly rotating torus. The
masses and dimensions of the tori depend on the EOS, but are
generically larger than those reported in previous independent
studies, with masses up to $\approx 0.07M_{\odot}$. Confirming what was
reported in ref.~\cite{Shibata06a}, we have found that the amount of
angular momentum lost during the inspiral phase can influence the mass
of the torus for binaries that have the same EOS. In particular,
the models that lose less angular momentum during the inspiral,
the comparatively \textit{low-mass} binaries, are expected to have
comparatively \textit{high-mass} tori. A more detailed study of the
dynamics of the torus (especially when produced from non-equal-mass
binaries) and of its implication for short hard GRBs will be the
subject of a following paper~\cite{BGR08}.

Most of the binaries considered have an initial coordinate separation
of $45\,\km$ and merge after $\sim 2$ orbits or, equivalently, after
$\sim 6\,\ms$. However, we have also considered a high-mass polytropic
binary with an initial coordinate separation of $60\,\km$, which
merges after $\sim 5$ orbits or, equivalently, after $\sim
20\,\ms$. As a stringent test of the accuracy of our results we have
carried out a systematic comparison between identical binaries
starting at different initial separations. Such a comparison, which
has never been performed before, has shown that there is an excellent
agreement in the inspiral phase (as expected from the lowest-order PN
approximations), but also small differences at the merger and in the
subsequent evolution. These results provide us with confidence on our
ability to perform long-term accurate simulations of the inspiral
phase, and also open the prospect of investigating higher-order PN
corrections.

Besides the study of the bulk dynamics of the two NSs, we have also
investigated the small-scale hydrodynamics of the merger and the
possibility that dynamical instabilities develop. In this way we have
provided the first quantitative description of the onset and
development of the Kelvin-Helmholtz instability, which takes place
during the first stages of the merger phase, when the outer layers of
the stars come into contact and a shear interface forms. The instability
curls the interface forming a series of vortices which we were able to
resolve accurately using the higher resolutions provided by the AMR
techniques. Since the development of this instability is essentially
independent of the EOS used or of the masses of the NSs, it could have
important consequences in the generation of large magnetic
fields. Also this aspect will be further investigated in a subsequent
work~\cite{BGR08}.

Given the importance of binary NSs as sources of gravitational waves,
special attention in this work has been dedicated to the analysis of
the waveforms produced and to their properties for the different
configurations. In particular, we have found that the largest loss rates
of energy and angular momentum via gravitational radiation develop at
the time of the collapse to BH and during the first stages of the
subsequent ringdown. Nevertheless, the configurations which emit the
highest amount of energy and angular momentum are those with lower
masses, since they do not collapse promptly to a BH. Instead they
produce a violently oscillating HMNS, which emits copious
gravitational radiation, while rearranging its angular-momentum
distribution, until the advent of the collapse to BH. We have also
found that although the gravitational-wave emission from NS binaries
has spectral distributions with large powers at high frequencies (\ie
$f \gtrsim 1\,\khz$), a signal-to-noise ratio as large as $3$
can be estimated for a source at $10\,\mpc$ using the sensitivity of
currently operating gravitational-wave interferometric detectors.

Several aspects of the simulations reported here could be improved and
probably the most urgent among them are the use of more realistic EOSs
and the inclusion of magnetic fields via the solution of the MHD
equations. Recent calculations~\cite{Price06,Anderson2008} have indeed
shown that the corrections produced by strong magnetic fields could be
large and are probably very likely to be present. Work is in progress
towards these improvements using the code developed in
ref.~\cite{Giacomazzo:2007ti}. The results of these investigations
will be presented in forthcoming works.

\begin{appendix}

\section{Characterizing the truncation error}

\subsection{The influence of  numerical methods}
\label{sec:influence num methods}

\begin{figure}[t]
 \begin{center}
    \includegraphics[angle=-0,width=0.45\textwidth]{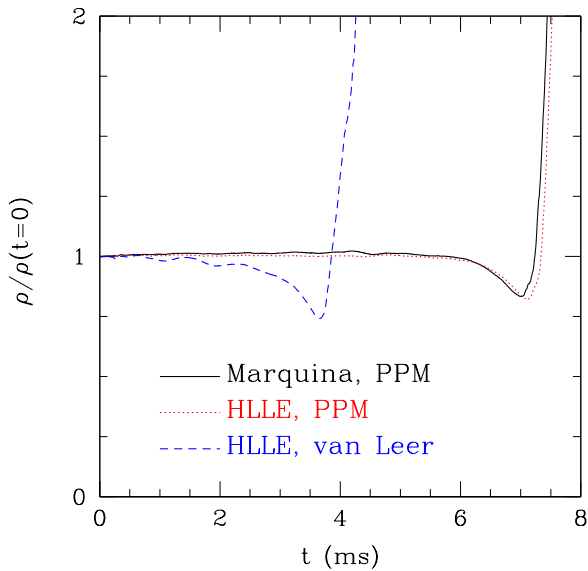}
 \end{center}
 \vskip -0.5cm
 \caption{Comparison of the rest-mass density normalised to its value
   at $t=0$ for evolutions performed with different numerical methods;
   the solid line refers to an evolution performed
   using the Marquina flux formula and a PPM reconstruction, the dotted 
   line to HLLE-PPM and the dashed line
   to HLLE-TVD (van Leer slope limiter). These data refer to an initial 
   configuration not present in Table~\ref{table:ID} (see text for details)
   and to an evolution with the \textbf{ideal-fluid}
   EOS.}
   \label{fig:num-visc}
 \end{figure}

The inherent numerical viscosity of the numerical method used for the
reconstruction of the variables on cell interfaces is crucial to
determine the time of the merger. As one might expect, lower-order
reconstruction schemes result in an anticipated merger due to their
higher numerical viscosity, as Fig.~\ref{fig:num-visc} shows (for a
review of the numerical methods implemented in {\tt Whisky},
see~\ref{hd_eqs}). The results in the test simulations presented in
the figure were produced through the evolution of initial data that
are not listed in Table~\ref{table:ID}, {\it i.e.} proper separation
between the centers of the stars $d/M_{_{\rm ADM}}=12.6$; baryon mass
of each star $M_{b}=1.78\,M_{\odot}$; total ADM mass $M_{_{\rm
    ADM}}=3.24\,M_{\odot}$; angular momentum $J=9.93\,M^2_{\odot}=
8.75\times 10^{49}\,{\rm g cm^2/s}$; initial orbital angular velocity
$\Omega_0=9.39\times10^{-3}= 1.9\, {\rm rad/ms}$; approximate mean
radius of each star $R=8.4\,M_{\odot}= 12\,\km$; ratio of the polar to
the equatorial coordinate radius of each star $r_p/r_e=0.945$.

In particular, in Fig.~\ref{fig:num-visc} we show the differences in
the evolution of the rest-mass density normalized to its initial value
when different numerical methods are used for the evolution: the solid
line refers to an evolution performed using the Marquina flux formula
and a PPM reconstruction (which is our usual choice), the dotted line
to the HLLE approximate Riemann solver with PPM reconstruction and the
dashed line to the HLLE solver with TVD reconstruction (in particular,
the van Leer slope limiter was used). Smaller changes in the merger
time and in the evolution of the HMNS are observed also by changing
some parameters of the PPM reconstruction method, in particular those
related to the shock detection, that is the parameters that define how
big a jump in the evolved variable has to be, in order to be
considered a discontinuity and treated as such.

We have found instead that the choice of approximate Riemann solver
does not influence significantly the evolution of the coalescence. As
one can see from Fig.~\ref{fig:num-visc}, when coupled with the PPM
reconstruction, both the Marquina and the HLLE solvers produce very
similar dynamics and the time of the merger is almost the same. The
situation changes when a lower-order reconstruction method, such as the van
Leer one, is used. In this case the numerical viscosity is large
and the time of the merger is very different, \ie $\approx 4\, \ms$
instead of $\approx 6.5\, \ms$.

From these tests one can then learn that the numerical viscosity of
the evolution method is very important in this scenario, being
responsible for changes in the dynamics and also in the estimate of
the gravitational-wave emission. Of course, one should always employ
the least viscous method available.

\begin{figure}[t]
 \begin{center}
    \includegraphics[angle=-0,width=0.45\textwidth]{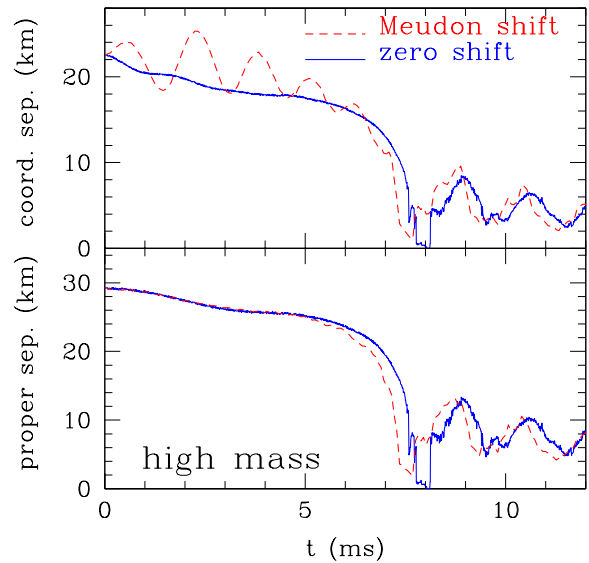}
 \end{center}
 \vskip -0.5cm
  \caption{Comparison of the time evolution of the coordinate
    separation (upper panel) and the proper separation (lower panel)
    between the stellar centres in case the initial Meudon shift is
    used (dashed line) and in case the initial shift is set to zero
    (continuous line). }
   \label{fig:zeroshift}
 \end{figure}

\subsection{The influence of the initial gauge conditions}
\label{sec:0shift}

We have found that using the shift profile given in the Meudon data
introduces a considerable amount of gauge dynamics, which can be
avoided by setting the initial shift to zero. We recall that the
Meudon shift condition is determined through the Killing equation
which is implicit in the quasi-equilibrium assumption for binary
systems~\cite{Gourgoulhon01}.  A clear way to highlight this feature
is a comparison of the time evolution of the coordinate separation
between the stellar centres. This is shown in
Fig.~\ref{fig:zeroshift}, which offers a comparison of the time
evolution of the coordinate separation (upper panel) and the proper
separation (lower panel) between the stellar centres in case the
initial Meudon shift is used (dashed line) and in case the initial
shift is set to zero (continuous line). The evolution equation for the
shift is the same for the two simulations.

It is clear that the coordinate orbit of the evolution started with
the Meudon shift has a noticeable amount of eccentricity (which
appears as large oscillations of the coordinate separation of the
stars during the inspiral), which is absent in the simulation in which
the shift is zero at the initial time. In addition, the adoption of a zero 
initial shift results into a larger initial violation of the $2$-norm
of the Hamiltonian constraint, which is however below $10^{-6}$ for
the typical resolution used.

As a final remark we note that the proper separations of the stars, the
maximum of the rest-mass density and other gauge-invariant quantities,
such as the gravitational waveforms, are instead very similar during
the inspiral phase.

\end{appendix}

\begin{acknowledgments}

  It is a pleasure to thank the group in Meudon (Paris) for producing
  and making available the initial data used in these calculations. We
  are also grateful to N.~Dorband, A.~Nagar, D.~Pollney and
  C.~Reisswig for help in the calculation of the gravitational waves
  and to E.~Schnetter and all the ``Carpet-developers'' for their help
  with AMR. We thank also R.~De Pietri, Y.~Eriguchi, G.~M.~Manca,
  M.~Shibata and S.~Yoshida for useful discussions and comments. The
  computations were performed on the clusters Peyote, Belladonna and
  Damiana of the AEI. This work was supported in part by the DFG grant
  SFB/Transregio~7 and by the JSPS Postdoctoral Fellowship For Foreign
  Researchers, Grant-in-Aid for Scientific Research (19-07803).

\end{acknowledgments}

\bibliography{aeireferences}

\end{document}